%% file: interior.tex
\documentclass{article}
\usepackage{latexsym}
\usepackage{graphics}
\usepackage{color}
\usepackage{amsmath, amsthm, amssymb}

\newtheorem{theorem}{Theorem}[section]

\newtheorem{corollary}[theorem]{Corollary}
\newtheorem{proposition}[theorem]{Proposition}

\newcommand{\tl}{\frac{\theta}{\lambda}}
\newcommand{\zn}{\frac{\zeta}{\nu}}

\begin{document}                        

\title{The interior of charged black holes and the problem of uniqueness
		in general relativity}              
		
\author{Mihalis Dafermos}                       
\maketitle   

\begin{abstract}
      We consider a spherically
symmetric double characteristic initial value problem
for the Einstein-Maxwell-(real) scalar field equations.
On the initial outgoing characteristic, the data
is assumed to satisfy the Price law decay
widely believed to hold on an event horizon
arising from the collapse of an asymptotically
flat Cauchy surface. We establish
that the heuristic mass inflation scenario put forth by Israel 
and Poisson is mathematically correct in the context of
this initial value problem. In particular, 
the maximal future development has a future boundary,
over which the spacetime is extendible as a $C^0$ metric,
but along which the Hawking mass blows up identically;
thus, the spacetime is inextendible as a $C^1$ metric. 
In view of recent results of the author in collaboration
with I.~Rodnianski, which 
rigorously establish the validity of Price's law as an upper
bound for the decay of scalar field hair, 
the $C^0$ extendibility result applies to the 
collapse of complete asymptotically flat spacelike initial data
where the scalar field is compactly supported. This shows
that under Christodoulou's $C^0$ formulation, the strong cosmic
censorship conjecture is false for this system.
\end{abstract}

 \tableofcontents

\section{Introduction}
The fact that the
initial value problem for the Einstein equations is locally
well-posed connects
general relativity to the classical 19th-century physics of fields,
and to the theory of equations of evolution in general. At first,
it may seem that this also casts the theory squarely into the sphere
of Newtonian determinism, the principle that initial data
determine uniquely the solution. In quasilinear hyperbolic
equations, however, such as the Einstein equations, where uniqueness
is controlled by the \emph{a priori} unknown
global geometry of the characteristics,
it is not immediate that one can pass from ``local'' to ``global'' 
determinism. In fact, it is well known that the Einstein equations
allow for smooth Cauchy horizons 
to emerge in 
evolution from complete initial data,
i.e.~uniqueness may fail \emph{without any loss of regularity 
in the solution}. This behavior indeed occurs in the celebrated 
Kerr solution; also in the spherically-symmetric
Reissner-Nordstr\"om solution, as can be seen most
easily by examining its so-called
\emph{Penrose diagram}\footnote{See~\cite{mi:mazi}.}:
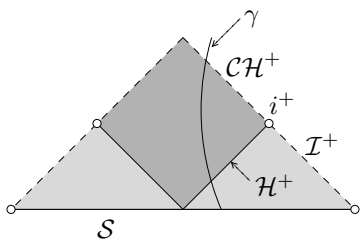
\[
\input{RN2.pstex_t}
\]
In more physical language, 
our ability to predict the future from initial conditions on the complete
Cauchy surface $\mathcal{S}$, in particular the fate of the 
observer depicted above by the timelike geodesic $\gamma$, fails,
as $\gamma$ ``exits'' the domain of dependence of $\mathcal{S}$
in finite proper time.
This failure, however, is not accompanied by
any sort of blow up, as measured by $\gamma$, which would indicate
that classical relativity should no longer apply.

The physical problems arising from this ``loss of determinism''
would be resolved if the phenomenon turned out to be unstable, 
i.e.~if for generic initial data,
uniqueness does indeed hold as long as the solution remains regular. 
The conjecture that this is the case, due to Penrose, goes by the name of 
\emph{strong cosmic censorship}. Definite formulations are given
in~\cite{mi:mazi, md:si, chr:givp, chru:uitl}.

The strong cosmic censorship 
conjecture was motivated by arguments of first
order perturbation theory, indicating that the Cauchy
horizon $\mathcal{CH}^+$ in a spherically symmetric Reissner-Nordstr\"om background
is indeed unstable.
The rigorous study of the stability of the Reissner-Nordstr\"om
Cauchy horizon in a nonlinear p.d.e.~setting was initiated
in \cite{md:si}. 
A certain spherically symmetric
characteristic initial value problem
was considered for the
Einstein-Maxwell-scalar field\footnote{The considerations 
leading to this particular system are discussed
at length in \cite{md:si, md:bb, bdim:scsbhi, brsm:bhs}.
It is the analogy between the repulsive
mechanism of charge and angular momentum which
makes it possible to
study the formation and instability of Cauchy
horizons while remaining in
the context of spherical symmetry. For the case where $F_{\alpha\beta}=0$,
the reader should consult~\cite{chr:ins, chr:mt, chr:bv, chr:sgsf}.} 
equations:
\begin{equation}
\label{Einstein-xs}
R_{\alpha\beta}-\frac12Rg_{\alpha\beta}=2T_{\alpha\beta},
\end{equation}
\begin{equation}
\label{Maxwell-xs}
F^{,\alpha}_{\alpha\beta}=0, F_{[\alpha\beta,\gamma]}=0,
\end{equation}
\begin{equation}
\label{wave-xs}
g^{\alpha\beta}\phi_{;\alpha\beta}=0,
\end{equation}
\begin{equation}
\label{em-xs}
T_{\alpha\beta}=\phi_{;\alpha}\phi_{;\beta}
-\frac12g_{\alpha\beta}\phi^{;\gamma}\phi_{;\gamma}
+F_{\alpha\gamma}F^{\gamma}_{\ \beta}-\frac14g_{\alpha\beta}F^{\gamma}_{\ \delta}
F^{\delta}_{\ \gamma},
\end{equation}
and it was proven
that the maximal development of initial data has future boundary
a Cauchy horizon over which the spacetime is extendible
as a $C^0$ metric. On the other hand, given an appropriate
additional assumption on the initial data, it was shown
that the Hawking mass blows up identically along this Cauchy horizon,
and thus the spacetime is inextendible as a $C^1$ metric.

The generic occurrence of a $C^1$-singular (but $C^0$-regular!)
Cauchy horizon in the interior of spherically symmetric charged
black holes arising from gravitational collapse
was first conjectured 
by W. Israel and E. Poisson \cite{ispo:isbh} on the 
basis of heuristic considerations; they termed this phenomenon
\emph{mass inflation}. Although several numerical
\cite{brsm:bhs, lb:sb} studies subsequently 
confirmed the mass inflation scenario,
it remained
somewhat controversial, as it is at odds with
the original picture put forth by Penrose, which had suggested
the generic occurrence
of a spacelike $C^0$-singular surface, terminating at $i^+$, rather than
a null $C^1$-singular Cauchy horizon.
 
The results of \cite{md:si} showed that in principle mass inflation
could occur for solutions of the Einstein-Maxwell-scalar field system;
the scenario was thus not
an artifice of the Israel-Poisson heuristics. The results
of~\cite{md:si}
could not show, however,
whether the scenario actually occurred
for the collapse of generic asymptotically flat 
spacelike initial data. 
The initial outgoing null characteristic 
considered in \cite{md:si}
was meant to represent the event horizon of a black hole
that had already formed. Moreover, to simplify the problem as much
as possible,  the assumptions imposed
on the event horizon
were very strong, in particular,
the event horizon was required \emph{not} to carry incoming
scalar field radiation. This implied that it geometrically
coincided with a Reissner-Nordstr\"om event horizon,
and thus was at the same time an apparent horizon, i.e.~it was
foliated by marginally trapped spheres.
The event horizon of a black hole arising from
gravitational collapse, however, will in general have a qualitatively
different structure from that of the Reissner-Nordstr\"om event horizon.
It will not be foliated by marginally trapped spheres,
and it will possess a radiating decaying scalar
field ``tail''. Heuristic analysis \cite{gpp:de1, lb:sbh, bo:lt, jb:gcc}
going back to Price \cite{rpr:ns},
together with more recent numerical results \cite{gpp:de, bo:lte, mc:bhsf},
have indicated that, in the case of massless scalar field matter,
this tail will decay polynomially with exponent $-3$,
with respect to a naturally defined advanced
time coordinate. This decay rate is widely known as
\emph{Price's law}.

In the present paper, we demonstrate that results analogous to those of
\cite{md:si} can be established for
data 
satisfying a weak form of the conjectured Price
law decay discussed above.\footnote{In particular, this
data includes the intial data considered in the numerical
studies \cite{brsm:bhs,lb:sb}.}
To state the theorems, it will be helpful to apply the
standard notation of Penrose diagrams. The reader 
can refer to~\cite{mi:mazi}. In particular
the labelled subsets $\mathcal{CH}^+$, $\mathcal{I}^+$, $i^+$, $\mathcal{H}^+$
acquire a well-defined meaning from their position in the diagram. 

The first main result of the paper is given by
\begin{theorem}
\label{eis9ew}
Fix constants $0<e<\varpi_+$, $C>0$, $p>\frac12$.
Define $r_+=\varpi_++\sqrt{\varpi_+^2-e^2}$, and fix a constant $r_0<r_+$.
Let $r$ and $\phi$ be functions defined on 
the union $\mathcal{C}_{out}\cup_{\{p\}}\mathcal{C}_{in}$
of two connected $C^\infty$
$1$-dimensional manifolds, each with boundary a point, 
identified topologically at the boundary point.
Parametrizing $\mathcal{C}_{out}$ by $[V,\infty)$, for a suitably large
$V$,
assume that $r$ and 
$\phi$ are $C^2$ functions in $\mathcal{C}_{out}$ 
with $r\ge r_0>0$, $\partial_vr\ge 0$,  
\[
\lim_{v\to\infty}r=r_+,
\] 
\begin{equation}
\label{avtitoumavva}
|\partial_v\phi|\le Cv^{-p},
\end{equation}
and, defining 
\[
\varpi(v)=-\frac12\int_{v}^\infty{r^2}(\partial_v\phi)^2dv+\varpi_+,
\]
assume
\[
\partial_vr=1-\frac{2\varpi}r+\frac{e^2}{r^2}.
\]
Parametrizing $\mathcal{C}_{in}$ by $[0,u_0)$ for $u_0<r_0$, assume that
$r$ and $\phi$ are $C^2$ in $\mathcal{C}_{out}$, and
that $\partial_ur=-1$, and
$|\partial_u\phi|\le\bar{C}$, for some
constant $\bar{C}$.
There exists a unique maximal
spherically symmetric globally hyperbolic
$C^2$ solution 
$\{\mathcal{M}, g, F_{\mu\nu}, \phi\}$ of the
Einstein-Maxwell-scalar field equations with two-dimensional
Lorentzian quotient manifold $\mathcal{Q}=\mathcal{M}/SO(3)$,
such that $\mathcal{C}_{out}\cup\mathcal{C}_{in}$ embeds into
$\mathcal{Q}$ as a double null boundary, with 
$J^+(p)\cap\mathcal{Q}=D^+(\mathcal{C}_{out}\cup\mathcal{C}_{in})\cap\mathcal{Q}$,
and such that the functions $r,\phi$ on $\mathcal{Q}$, and the constant $e$,
induced by the solution, restrict to
the prescribed values on $\mathcal{C}_{out}\cup\mathcal{C}_{in}$,
and the renormalized Hawking mass function $\varpi$ restricts to 
its prescribed value along $\mathcal{C}_{out}$.
Moreover, for some non-empty connected subset 
$\mathcal{C}'_{out}\subset\mathcal{C}_{out}$,
$D^+(\mathcal{C}'_{out}\cup\mathcal{C}_{in})\cap\mathcal{Q}$ has Penrose
diagram depicted below:
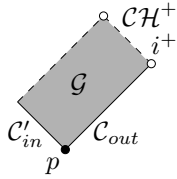
\[
\input{arxikoedw00.pstex_t}
\]
The function $r$ extends by monotonicity to $\mathcal{CH}^+$
such that 
\[
r(q)\to r_-=\varpi_+-\sqrt{\varpi_+^2-e^2}
\]
as $q\to i^+$ along
$\mathcal{CH}^+$.

If $p>1$, then  
the functions $r$, $\phi$, and the Lorentzian metric
$\bar{g}$ of $\mathcal{G}$ extend continuously to 
$\mathcal{G}\cup\mathcal{CH}^+$.
Thus,
$(\mathcal{M},g)$ can be extended to a larger
$(\tilde{\mathcal{M}},\tilde{g})$ where $\tilde{g}$ is $C^0$,
where $\tilde{\mathcal{M}}$ can be chosen spherically symmetric,
and such that its spherically symmetric quotient has a subset
with Penrose diagram depicted below:
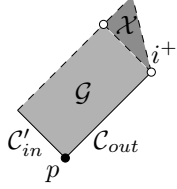
\[
\input{arxikoedw0.pstex_t}
\]
\end{theorem}

The second main result of the paper is 
\begin{theorem}
\label{allo9ewr}
Let $p>\frac12$, and $C,c,\epsilon>0$.
Consider data as in Theorem \ref{eis9ew}, where
in place of $(\ref{avtitoumavva})$, we assume
\[
0<cv^{-3p+\epsilon}\le|\partial_v\phi|\le Cv^{-p}
\]
for large enough $v$.
It follows that the Hawking mass, interpreted in the limit, blows up identically
along $\mathcal{CH}^+$,
in particular, extensions $(\tilde{\mathcal{M}},\tilde{g})$
depicted above
\emph{cannot} have $C^1$ metric.
\end{theorem}

When the above results were first obtained, the only motivation
for the assumptions was the heuristic and numerical evidence
mentioned above.
In view of recent work \cite{mi:mazi} of the author in collaboration
with I.~Rodnianski, however, it follows now that
the assumptions of Theorem \ref{eis9ew},
in fact the stronger bound, 
\begin{equation}
\label{pioduvato}
|\partial_v\phi|\le C_\epsilon v^{-3+\epsilon},
\end{equation}
for any $\epsilon>0$,
indeed hold on the event horizon of black holes arising
from the collapse of general complete spherically symmetric asymptotically flat
spacelike data
for which the charge is non-zero\footnote{The reader should note that in the case
of non-vanishing charge, complete 
spherically symmetric Cauchy data necessarily will have two
asymptotically flat ends, and this easily implies
that trapped surfaces occur and thus
an event horizon necessarily forms. See~\cite{md:sssts}.}, and
the scalar field is compactly supported\footnote{In fact, Theorem \ref{eis9ew}
 will apply to
the collapse of spacelike data more general than
those where the scalar field and its gradient have
compact support, \emph{cf.}~\cite{lb:nc}.} on the
initial hypersurface 
$\mathcal{S}$, provided that some additional assumption
is made ensuring that the black hole not be extremal in the 
limit.
In particular, the Lorentzian $2$-dimensional quotient $\mathcal{Q}$
of the future Cauchy development $\{\mathcal{M}, g, F_{\mu\nu}, \phi\}$
of such data contains
a subset $\mathcal{D}\cup\mathcal{G}$ with Penrose diagram depicted as the union of
the lighter-shaded regions below:
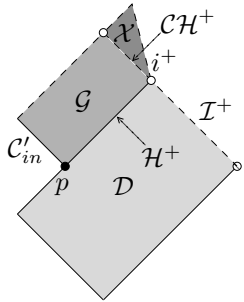
\[
\input{arxikoedw.pstex_t}
\]
In the above diagram, $\mathcal{D}$ is the region of~\cite{mi:mazi},
$\mathcal{H}^+$ is its event horizon,
and $\mathcal{C}'_{in}$ and $\mathcal{C}_{out}=\mathcal{H}^+\cap J^+(p)$ and
$\mathcal{G}$ are as in Theorem~\ref{eis9ew}.
It follows
that $\mathcal{D}\cup\mathcal{G}\cup\mathcal{CH}^+\cup\mathcal{X}$
represents the quotient manifold of a subset of a 
proper extension of $(\mathcal{M},g)$
to a larger $(\tilde{\mathcal{M}},\tilde{g})$, where $\tilde{g}$ is a
$C^0$ metric. The null curve $\mathcal{CH}^+$ is a 
subset of the quotient of the \emph{Cauchy
horizon} of $\mathcal{M}$ in $\tilde{\mathcal{M}}$. 
Thus, the question of the $C^0$ \emph{in}extendibility 
of the future Cauchy development of complete spherically symmetric
asymptotically
flat spacelike data is now completely resolved \emph{in the negative}
for the Einstein-Maxwell-scalar
field equations, i.e.~we have the following
Corollary of Theorem~\ref{eis9ew}
\begin{corollary}
Strong cosmic censorship, under the formulation 
of~\cite{chr:givp},
is false for the Einstein-Maxwell-scalar field equations under
spherical symmetry.
\end{corollary}

If the conditions of Theorem~\ref{allo9ewr} are satisfied on $\mathcal{H}^+$
for the collapse of generic spacelike initial data, then the $C^0$
extensions depicted above will not be $C^1$. A weaker version of
strong cosmic censorship would then be true. Retrieving the conditions
of Theorem~\ref{allo9ewr} on the event horizon of
the Cauchy development of an appropriate notion of
generic spacelike initial data remains thus an important open problem.

As for the question of Price law decay,
it should be noted that 
previously, Stalker and Machedon~\cite{st:pc} had obtained
bounds similar to $(\ref{pioduvato})$
for the wave equation on a \emph{fixed} Schwarzschild
or Reissner-Nordstr\"om background.

The techniques of this paper are refinements of methods
initiated in \cite{md:si}. Serious new difficulties arise, however, 
that affect both the qualitative picture and the estimates at 
our disposal. Since the event horizon and the apparent horizon no 
longer coincide, 
one must show that an apparent horizon (and consequently, a trapped
region) emerges in evolution. This can be accomplished by using 
the celebrated ``red-shift'' effect.
(See Section~\ref{rsrsec}. Note that the ``red-shift'' effect plays a fundamental
role in~\cite{mi:mazi} as well.)
Another difficulty is that since 
the initial data are not trapped or marginally trapped, monotonicity
special to trapped regions cannot be immediately applied. 
On the one hand, this implies that the pointwise estimate for 
one of the controlling quantities used heavily in \cite{md:si} is here lost,
and new arguments must be introduced to circumvent this. 
On the other hand, an additional argument is also necessary to ensure
that the monotonicity of the mass difference, crucial
to the proof of blow-up, holds in the trapped
region, as this depended on appropriate conditions being met
at the apparent horizon\footnote{Since in \cite{md:si}, the apparent
horizon coincided with the event horizon, these conditions could be
imposed directly as an assumption on initial data.}. 
This additional argument is accomplished again employing the
red-shift effect.

Besides the novel qualitative features discussed above, the
perturbations introduced here differ from those of \cite{md:si}
quantitatively, and this difference persists after the trapped
region has formed. We will discuss this further at the end of
Section~\ref{ivpsec}. In some sense, this renders the proof of
blow-up slightly easier but the proof of stability harder.
In particular, the so-called ``stable blue-shift region'' 
must be handled with considerably
more care. For this, certain $BV$ estimates for the scalar field will
prove useful.

Despite these differences, this paper confirms the qualitative
picture of the emergence of a ``stable red-shift'' region,
a ``no-shift'' region, and a sufficiently large ``stable blue-shift''
region, before the effects of instability become large. It thus supports
the thesis of \cite{md:si} 
that this picture may indeed represent
the interior of black holes that emerge from realistic gravitational
collapse.

The outline of this paper is as follows:
In Section \ref{sphsymsec} we present the spherically symmetric
Einstein-Maxwell-scalar field equations and in Section \ref{ivpsec} we
formulate precisely the initial value problem.
The definition of the maximal future development, 
together with an extension theorem involving the
area radius $r$, are stated in Section \ref{mddsec}. 
Section \ref{monsec} reviews some of the important monotonicity
properties of the equations that are central in the analysis. 
Section \ref{ovvsec} gives an overview of the main ideas employed in proving
both stability and blow-up, and more 
generally, in the analysis
of black hole interiors. Stability is then established
in the course of Sections \ref{rsrsec}, \ref{nsrsec}, \ref{sbssec}, \ref{bsbsec}, 
and \ref{c0esec}.
Blow-up is shown in Section \ref{minsec}.
Certain \emph{a priori} $BV$-estimates for the
scalar field $\phi$ are left to Section \ref{bvesec}. 
Finally, we give in the Appendix two simple causal constructions,
to which we will often make reference.

\section{The Einstein-Maxwell-scalar field equations in spherical
symmetry}
\label{sphsymsec}

Let $\mathcal{M}$ be a $C^3$ $4$-dimensional 
Hausdorff manifold, let $g$ be a $C^2$
time-oriented Lorentzian metric, let $F_{\alpha\beta}$ be
a $C^1$ anti-symmetric $2$-tensor defined on $\mathcal{M}$, and let $\phi$
be a $C^2$ function defined on $\mathcal{M}$. We say that
$\{\mathcal{M},g,F_{\alpha\beta},\phi\}$ is a solution to the
Einstein-Maxwell-scalar field equations if the equations
$(\ref{Einstein-xs})$--$(\ref{em-xs})$
are satisfied
pointwise
on $\mathcal{M}$,
where $R_{\alpha\beta}$ and $R$ denote the Ricci and scalar
curvature of $g_{\alpha\beta}$, respectively, and the Einstein summation 
convention has been applied. We refer to $F_{\alpha\beta}$ as the \emph{electromagnetic
field}, and $\phi$ as the \emph{scalar field}.

We say that a solution $\{\mathcal{M},g,F_{\alpha\beta},\phi\}$ of
the Einstein-Maxwell-scalar field equations is \emph{spherically symmetric}
if the group $SO(3)$ acts by isometry on $(\mathcal{M},g)$ preserving
$F_{\alpha\beta}$ and $\phi$.
We have the following:

\begin{proposition}
\label{megalnprotasn}
Let $\mathcal{Q}$ be a $C^3$ $2$-dimensional manifold, let
$\bar{g}$ be a $C^2$ time-oriented Lorentzian metric on $\mathcal{Q}$,
let $r$ and $\phi$ be $C^2$ functions, and let $e$ be a constant.
Assume $r$ to be strictly positive, and assume that there exist
future-directed global coordinates
$u$ and $v$ on
$\mathcal{Q}$ such that
\begin{equation}
\label{W-orism}
\bar{g}=-\Omega^2(du\otimes dv+dv\otimes du).
\end{equation}
(Such coordinates are known as null coordinates.)
Define a function $m$ by
\begin{equation}
\label{m-orism}
m=\frac{r}2(1-\bar{g}(\nabla r,\nabla r)),
\end{equation}
a function 
\begin{equation}
\label{mu-orism}
\mu=\frac{2m}r
\end{equation}
and
a function
\begin{equation}
\label{pi-orism}
\varpi=m+\frac{e^2}{2r}.
\end{equation}
Define 
\begin{equation}
\label{ruqu}
\partial_ur=\nu, 
\end{equation}
\begin{equation}
\label{rvqu}
\partial_vr=\lambda,
\end{equation}
\begin{equation}
\label{z-orism}
r\partial_u\phi=\zeta,
\end{equation}
\begin{equation}
\label{th-orism}
r\partial_v\phi=\theta.
\end{equation}
Assume $\nu<0$, and define a function
\begin{equation}
\label{kappaoris}
\kappa=-\frac14\Omega^{2}{\nu}^{-1}.
\end{equation}
Suppose 
\begin{equation}
\label{puqu}
\partial_u\varpi=\frac{1}{2}(1-\mu)\left(\zn\right)^2\nu,
\end{equation}
\begin{equation}
\label{pvqu}
\partial_v\varpi=\frac{1}{2}\kappa^{-1}\theta^2,
\end{equation}
\begin{equation}
\label{fdb1}
\partial_u\kappa=
\frac1{r}\left(\zn\right)^2\nu\kappa,
\end{equation}
\begin{equation}
\label{sign1}
\partial_u\theta=-\frac{\zeta\lambda}r,
\end{equation}
\begin{equation}
\label{sign2}
\partial_v\zeta=-\frac{\theta\nu}r.
\end{equation}
Let $\gamma$ denote the standard metric on $S^2$, 
let $\mathcal{M}$ denote the manifold $\mathcal{Q}\times S^2$,
let $\pi_1:\mathcal{M}\to\mathcal{Q}$, $\pi_2:\mathcal{M}\to S^2$
denote the standard projections, and define
\begin{equation}
\label{ansatz}
g=\pi_1^*\bar{g}+(\pi_1^*r)^2\pi_2^*\gamma
\end{equation}
\begin{equation}
\label{Fcharge}
F_{\alpha\beta}dx^\alpha\otimes dx^\beta=
\frac{e}{(\pi_1^*r)^2}\pi_1^*(\Omega^2 du\wedge dv).
\end{equation}
Then $\{\mathcal{M},g,F_{\alpha\beta}, \pi_1^*\phi\}$ is a spherically
symmetric solution to the Einstein-Maxwell-scalar field equations.
$\mathcal{Q}$ represents the space of group orbits, 
\begin{equation}
\label{Qdef}
\mathcal{Q}=\mathcal{M}/SO(3).
\end{equation}
By $(\ref{ansatz})$, $\pi_1^{-1}(q)$ is a spacelike sphere in $\mathcal{M}$,
for all $q\in\mathcal{Q}$.
The function $r$ can be interpreted geometrically
as 
\begin{equation}
\label{geoint}
r(p)=\sqrt{{\rm Area(\pi_1^{-1}(p))}/4\pi},
\end{equation}
and $m(p)$ as the \emph{Hawking mass} of the surface $\pi_1^{-1}(p)$. 
We shall call $\mu$ the \emph{mass-aspect function}, and
$\varpi$ the \emph{renormalized Hawking mass}.
The constant $e$ is called the \emph{charge} and can be retrieved
from $g$ and $\phi$ by the relation
\begin{eqnarray}
\label{giatoe}
\nonumber
T_{\alpha\beta}dx^\alpha\otimes dx^\beta&=&
\frac{e^2}{2(\pi_1^*r)^4}\pi_1^*\bar{g}+\frac{e^2}{(\pi_1^*r)^2}\pi_2^*\gamma\\
&&\hbox{}+Hess(\phi)
-\frac12\bar{g}(\nabla\phi,\nabla\phi)g.
\end{eqnarray}

Conversely now, let $\{\mathcal{M}, g, F_{\mu\nu}, \phi\}$ be a 
spherically symmetric $C^2$ solution of the Einstein-Maxwell-scalar field
system. Define $\mathcal{Q}$ by $(\ref{Qdef})$, assume that
for all $p$, $\pi_1^{-1}(p)$
is either a point or a spacelike sphere,
define $r$ by $(\ref{geoint})$, and assume that 
$\mathcal{Q}^{>0}=\mathcal{Q}\cap\{r> 0\}$
can be endowed with the structure of a $2$-dimensional $C^3$ Lorentzian manifold,
such that $\pi_1^{-1}(\mathcal{Q}^{>0})=
\mathcal{Q}^{>0}\times S^2$, with metric defined by
$(\ref{ansatz})$, where $\bar{g}$ is a $C^2$ Lorentzian metric
on $\mathcal{Q}^{>0}$. Finally, let $\mathcal{U}\subset\mathcal{Q}^{>0}$ be a
connected
open subset covered by future-directed global null coordinates $(u,v)$,
i.e.~such that $(\ref{W-orism})$ holds. In $\pi_1^{-1}(\mathcal{U})$, 
it follows that
the energy-momentum tensor $T_{\alpha\beta}$ 
can be written as $(\ref{giatoe})$ for some
constant $e\ne0$.
Define 
$m$, $\mu$, $\varpi$, $\nu$, $\lambda$, $\zeta$, $\theta$, 
$\kappa$ on $\mathcal{U}$ by equations $(\ref{m-orism})$--$(\ref{kappaoris})$, 
and assume $\nu<0$. Then it
follows that these functions are $C^1$ and the equations $(\ref{puqu})$--$(\ref{sign2})$ hold pointwise on 
$\mathcal{U}$.
\end{proposition}

\begin{proof}
See~\cite{md:sssts, md:si}.
\end{proof}

We note that $(\ref{W-orism})$ and $(\ref{m-orism})$
imply that
\[
\kappa(1-\mu)=\lambda,
\]
and $(\ref{fdb1})$, $(\ref{pvqu})$ and $(\ref{rvqu})$ then give
\begin{equation}
\label{nqu}
\partial_v\nu=\frac{2\nu\kappa}{r^2}
\left(\varpi-\frac{e^2}{r}\right).
\end{equation}
By equality of mixed partials, we have
\begin{equation}
\label{lqu}
\partial_u\lambda=\frac{2\nu\kappa}{r^2}
\left(\varpi-\frac{e^2}{r}\right).
\end{equation}
We can also derive the equation
\begin{equation}
\label{fdb2}
\partial_v\left(\frac{\nu}{1-\mu}\right)=
\frac{1}r\left(\tl\right)^2\lambda\left(\frac{\nu}{1-\mu}\right),
\end{equation}
which holds wherever $1-\mu\ne0$, $\lambda\ne0$.

In this paper, we will
construct a spherically symmetric solution to $(\ref{Einstein-xs})$--$(\ref{em-xs})$ by
solving a double characteristic initial value problem for the system
$(\ref{ruqu})$--$(\ref{sign2})$. 
The initial data will be described in the next section.

\section{The initial value problem}
\label{ivpsec}
Fix 
\[
0<e<\varpi_+<\infty,
\]
and define 
\[
r_+=\varpi_++\sqrt{\varpi_+^2-e^2}.
\]
(The constant $e$ represents the charge,
and $\varpi_+$, $r_+$ will represent the asymptotic
renormalized Hawking mass, and asymptotic area radius, respectively,
along $\mathcal{C}_{out}$.)
Choose an additional constant
\[
0<r_0<r_+.
\]
(This will represent a lower bound for $r$ along $\mathcal{C}_{out}$.)
Choose further constants 
\[
\Lambda>0,
\]
\[
p>\frac12.
\] 
(These will be related to the modulus and power-law decay rate of
the scalar field.)
We will need in addition a certain smallness assumption, which,
for fixed values of the above constants, we will derive
by restricting to a large $V$. To formulate this, define for
each $v$,
constants
\[
\tilde{M}(v)=\frac{\Lambda}{2(2p-1)}v^{-2p+1}+
\frac{\Lambda r_+}2v^{-2p},
\]
\[
\tilde{K}(v)=\frac{\Lambda e^2}{r_+r_0(2p-1)}v^{-2p+1},
\]
\begin{equation}
\label{c_1ori}
c_1(v)=\frac{e^2}{\frac{e^2}{r_+^2}+\frac52\tilde{M}(v)+2\tilde{K}(v)},
\end{equation}
\begin{equation}
\label{M1ori}
M_1(v)=1+\frac{e^2}{c_1^2}+\frac{\tilde{M}(v)+2\varpi_+}{c_1},
\end{equation}
and choose
$V>0$ 
to satisfy
\begin{equation}
\label{giatorkatw}
r_+-r_0\ge\frac{\Lambda}{2p-1}V^{-2p+1},
\end{equation}
\begin{equation}
\label{alloxipriv}
0<\varpi_+-\frac{e^2}{r_+}-2(\tilde{K}(V)+\tilde{M}(V)),
\end{equation}
\begin{eqnarray}
\label{trello2}
\nonumber
\left(V-2pr_+^2
\left(\varpi_+-\frac{e^2}{r_+}-2(\tilde{K}(V)+\tilde{M}(V))
\right)^{-1}\log V\right)^{-p-1}\cdot\\
\nonumber
\cdot(\log V)\left(2p^2r_+^2
\left(\varpi_+-\frac{e^2}{r_+}+\frac32(\tilde{K}(V)+\tilde{M}(V))
\right)^{-1}\right)+\\
+r_+pV^{-2p-1}\le V^{-p},
\end{eqnarray}
\begin{eqnarray}
\label{trello}
\nonumber
\frac{r_+^2}{e^2}\left(\frac32\tilde{M}(V)+2(\tilde{M}(V)+\tilde{K}(V))\right)
\\
\hbox{}
\le\frac{c_1(V)}{6M_1(V)r_{+}}\left(\varpi_+-\frac{e^2}{r_+}-2(\tilde{K}(V)+\tilde{M}(V))
\right).
\end{eqnarray}
In what follows, denote $\tilde{M}=\tilde{M}(V)$, $\tilde{K}=\tilde{K}(V)$,
$M_1=M_1(V)$, $c_1=c_1(V)$.

We are ready now to define initial data 
on $\mathcal{C}_{out}$ and $\mathcal{C}_{in}$, satisfying
the conditions of  Theorem~\ref{eis9ew}.
With the hindsight of that result, namely that $\mathcal{G}$
can clearly be covered by a null coordinate system $(u,v)$, of which
$\mathcal{C}_{out}$ and $\mathcal{C}_{in}$ are axes, we define,
given $0<u_0<r_0$, the set $\mathcal{K}(u_0)$ as the subset of 
the $(u,v)$-plane given by
\[
\mathcal{K}(u_0)=[0,u_0)\times[V,\infty),
\]
and set $\mathcal{C}_{out}=\{0\}\times[V,\infty)$ 
and $\mathcal{C}_{in}=[0,u_0)\times \{V\}$. 

On $\{0\}\times[V,\infty)$, we prescribe 
$\kappa$, $r$, 
$\varpi$, $\lambda$, $\theta$, $\phi$.
Since our problem--interpreted geometrically--has one
``dynamic degree of freedom'', the data should depend in some sense
on one ``free'' function. It is most convenient to make this function
$\lambda$.\footnote{The reader may
notice that this prescription of data is slightly different
from that given in the formulation of Theorem~\ref{eis9ew}. 
The equivalence of the two prescriptions follows
from a computation in the
proof of Theorem~\ref{sxesntwv2} below.}
We proceed as follows:
Set
\begin{equation}
\label{INITv1}
\kappa(0,v)=1,
\end{equation}
and define an arbitary
function 
\begin{equation}
\label{au9aireto}
\lambda(0,v)\in C^1\cap L^1,
\end{equation}
such that, defining 
the $C^2$ function $r(0,v)$,
by
\begin{equation}
\label{rdef}
r(0, v)=r_+-\int_{v}^{\infty}\lambda(0,\bar{v})d\bar{v},
\end{equation}
$\lambda$ and $r$ are
subject only to the condition that the following inequalities hold:
\begin{equation}
\label{newINIT}
0\le\lambda\le \Lambda v^{-2p},
\end{equation}
\begin{equation}
\label{newINIT2}
|\partial_v\lambda|\le \Lambda v^{-2p},
\end{equation}
\begin{equation}
\label{newINIT3}
\frac{\lambda(1-\lambda)}2-\frac{r\partial_v\lambda}2
-\frac{e^2}{2r^2}\lambda\ge0.
\end{equation}

(Note that it is easy to explicitly construct $\lambda$, $r$ satisfying
$(\ref{newINIT3})$: For instance, prescribing $\lambda$ satisfying
$(\ref{newINIT})$ and $(\ref{newINIT2})$ on
some interval $[V',\infty)$, and requiring in addition $\lambda$
to be monotonically nonincreasing, then 
$(\ref{newINIT3})$ is satisfied
in an interval $[V,\infty)$, for some $V\ge V'$,
because $r\to r_+$, $\lambda\to0$, as $v\to\infty$, and
$\frac{e^2}{r_+^2}<1$.)

The definition $(\ref{rdef})$ and inequality $(\ref{newINIT})$
clearly yield
\begin{equation}
\label{profavws*}
r_+-r(0,v)\le \frac{\Lambda}{2p-1}v^{-2p+1},
\end{equation}
and thus, in view of $(\ref{giatorkatw})$,
\begin{equation}
\label{newINIT4}
r(0,v)\ge r_0.
\end{equation}

Define $\varpi(0,v)$ by setting
\begin{eqnarray}
\label{pioris}
\varpi(0,v)&=&\frac12r(1-\kappa^{-1}\lambda)(0,v)
+\frac{e^2}{2r}(0,v)\\
\nonumber
&=&\frac12r(1-\lambda)(0,v)+\frac{e^2}{2r}(0,v).
\end{eqnarray}
From $(\ref{newINIT})$, $(\ref{profavws*})$ and $(\ref{pioris})$, we have
\begin{equation}
\label{birxi}
\varpi_+-\varpi(0,v)\le
\frac{\Lambda}{2(2p-1)}v^{-2p+1}+
\frac{\Lambda r_+}2v^{-2p}=\tilde{M}(v)\le\tilde{M}.
\end{equation}
Define $m$ by $(\ref{pi-orism})$ and $\mu$ by $(\ref{mu-orism})$. 
We have 
\begin{equation}
\label{exomev}
1-\mu(0,v)=(1-\mu)\kappa(0,v)=\lambda(0,v),
\end{equation}
and thus, by
$(\ref{newINIT})$, and $(\ref{newINIT2})$,
we have
\begin{equation}
\label{exomev3ava}
0\le 1-\mu\le \Lambda v^{-2p},
\end{equation}
\begin{equation}
\label{peftei}
|\partial_v(1-\mu)|\le \Lambda v^{-2p}.
\end{equation}

Differentiating $(\ref{pioris})$ with respect to $v$, and applying the
inequalities
$(\ref{newINIT3})$, $(\ref{newINIT})$ and $(\ref{peftei})$,
we obtain
\begin{equation}
\label{nova}
0\le\partial_v\varpi\le \frac12C^2v^{-2p}
\end{equation}
for some constant $C\ge0$, where $C=C(\Lambda, r_+)$.
We select $\theta(0,v)$ as a continuous
function such that $\theta^2(0,v)=2\partial_v\varpi(0,v)$.
We then have
\begin{equation}
\label{INITv2}
|\theta(0,v)|\le Cv^{-p}.
\end{equation}
Finally, choose a constant $\Phi_0$, and define
\begin{equation}
\label{orise}
\phi(0,v)=\Phi_0+\int_V^v\frac\theta{r}(0,\bar{v})d\bar{v}.
\end{equation}
(If $-\infty<\int_V^\infty\frac\theta{r}(0,\bar{v})d\bar{v}<\infty$,
we can choose $\Phi_0$ to be the negative of this integral.
With this choice $\phi(0,v)\to0$ as $v\to\infty$.)

%Given $\lambda,r,\varpi,\theta$ defined above
%along some ray $[V',\infty)$, 
%we set $\tilde{M}(V)=\varpi(0,V)-\varpi_0$,
%and $\tilde{L}(V)=\frac{e^2}r(0,V)-\frac{e^2}{r_+}$,
%and note that
%\begin{equation}
%\label{birxi}
%0<\varpi_0-\varpi\le\tilde{M},
%\end{equation}
%\begin{equation}
%\label{ikixi}
%0<\frac{e^2}r-\frac{e^2}{r_+}\le\tilde{K},
%\end{equation}
%and $\tilde{M},\tilde{K}\to0$ as $V\to\infty$. 

%Finally, our initial data on the $v$-axis will be
%the functions $\lambda$, $r$, $\varpi$, $\theta$, restricted
%to some ray $[V,\infty)$, where $V$ satisfies
%the additional conditions $(\ref{alloxi})$,
%$(\ref{trello2})$ and $(\ref{trello})$ to be imposed in
%Section 8 where they will be needed. 

Having determined initial data on $\{0\}\times[V,\infty)$, we turn to
$[0,u_0)\times\{V\}$. Again, as our problem has one degree of freedom,
our data should depend in some sense on the choice of one ``free'' function.
Let us make this function $\zeta$, i.e., define
an arbitrary bounded continuous function $\zeta(u,V)$. In particular,
\begin{equation}
\label{INITu1}
|\zeta(u,V)|\le \overline{C},
\end{equation}
for some constant $\overline{C}$.
To complete the data, set
\begin{equation}
\label{INITu2}
\nu(u,V)=-1.
\end{equation}

As noted in the introduction, 
our original motivation for these assumptions was the Price law
conjecture of~\cite{rpr:ns}. In view of~\cite{mi:mazi}, however,
we have the following
\begin{theorem}
\label{sxesntwv2}
Consider an asymptotically flat spherically symmetric spacelike initial data
set\footnote{See~\cite{mi:mazi}.}
for the Einstein-Maxwell-scalar field system where the scalar field
and its gradient are initially of compact support, let 
\[
\{\mathcal{M},g,F_{\mu\nu},\phi\}
\]
denote its (maximal) Cauchy development,
let $\mathcal{Q}=\mathcal{M}/SO(3)$, let 
$\mathcal{D}$ be the region of~\cite{mi:mazi},
and let $\mathcal{U}=\mathcal{Q}\cap J^+(\mathcal{D})$. 
Then $\mathcal{U}$ satisfies the assumptions of the second
part of Proposition \ref{megalnprotasn}.
Assume that the initial hypersurface is complete,
and, on its quotient in $\mathcal{Q}$, the inequality
$0<e<r$ holds. Then it 
follows that there exists a $u_0>0$ and a subset 
$\mathcal{G}'(u_0)\subset\mathcal{U}\subset\mathcal{Q}$,
covered by global null coordinates $(u,v)$, such that
$\mathcal{G}'(u_0)\subset \mathcal{K}(u_0)$,
\[
\{0\}\times[V,\infty)\subset\mathcal{G}'(u_0),
[0,u_0)\times\{V\}\subset\mathcal{G}'(u_0),
\] 
\[
\mathcal{G}'(u_0)=D^+(\{0\}\times[V,\infty)\cup[0,u_0)\times\{V\})
\cap\mathcal{Q},
\]
and such that,
defining $\Omega$ by $(\ref{W-orism})$ with respect to these coordinates,
defining $\nu$ by $(\ref{ruqu})$, $\kappa$ by $(\ref{kappaoris})$, 
$\theta$ by $(\ref{th-orism})$,
and $\zeta$ by $(\ref{z-orism})$, then $\kappa$, $r$,
$\lambda$, $\nu$, $\theta$, $\zeta$, $\phi$ satisfy 
$(\ref{INITv1})$--$(\ref{INITu2})$
on
$\{0\}\times[V,\infty)\cup[0,u_0)\times\{V\}$.
\end{theorem}
\begin{proof}
That $\mathcal{U}$ satisfies the conditions of Proposition~\ref{megalnprotasn},
including $\nu<0$,
follows from~\cite{md:sssts}.
In view of \cite{chge:givp}, the rest of the theorem
is really a simple exercise in change of coordinates.
Let $\mathcal{D}$ be the
region of~\cite{mi:mazi} satisfying
the conditions of the main theorem, and extend the original $(u,v)$
coordinate system on $\mathcal{D}$ defined in~\cite{mi:mazi}
to a global regular
null coordinate system on all of $\mathcal{Q}$. Recall
the event horizon $\mathcal{H}^+\subset\mathcal{D}$ 
given in $(u,v)$-coordinates by $\{U\}\times[1,\infty)$,
and the inequalities
\[
1\ge\kappa(U,v)\ge c,
\]
\[
|\theta|\le C_\omega v^{-\omega},
\]
\[
\lambda(v)\ge 0,
\]
for all $v\ge 1$, for some constants $c>0$, $\omega>1$, 
$C_\omega\ge 0$, and also recall that
\[
v\to\infty
\]
along $\mathcal{H}^+$,
and
\begin{equation}
\label{mazaorio}
\varpi(v)\to \varpi_+,
\end{equation}
as $v\to\infty$,
for a constant $\varpi_+>e$, and 
\begin{equation}
\label{r--orio}
r(v)\to r_+=\varpi_++\sqrt{\varpi^2-e^2}.
\end{equation}
Define a function $v^*$ on $J^+(U,1)\cap\mathcal{Q}$
by 
\[
v^*(u,v)=\int_1^v\kappa(U,\bar{v})d\bar{v}+1.
\]
It is clear that $v^*$ defines a new regular advanced time coordinate,
\[
c\le \frac{d {v^*}}{d v}\le 1,
\]
and thus
\[
c(v-1)\le{v^*}-1\le v-1.
\]
Define 
$\kappa^*=\frac14{\Omega^*}^2\nu^{-1}=
\left(\frac{\partial {v^*}}{\partial v}\right)^{-1}\kappa$,
$\theta^*=r\partial_{v^*}\phi=\left(\frac{\partial {v^*}}{\partial v}\right)^{-1}\theta$,
$\lambda^*=\partial_{v^*}r=\left(\frac{\partial {v^*}}{\partial v}\right)^{-1}\lambda$.
We have
\begin{equation}
\label{eiv1}
\kappa^*(U,v^*)=1,
\end{equation}
i.e.~
\begin{equation}
\label{eiv1*}
\lambda^*(U,v^*)=(1-\mu)(U,v^*)
\end{equation}
and
\begin{equation}
\label{apotoallo9}
|\theta^*(U,v^*)|\le c^{-1}C_\omega v^{-\omega}\le C{v^*}^{-\omega},
\end{equation}
for $C=c^{-1}C_\omega$. 
Finally, set $r_0=r(U,1)>0$.

In what follows, let us drop the $*$, and refer to the
new coordinate as $v$, $\lambda^*$ as $\lambda$, etc.
In general, now, given $r(U,v)\ge r_0$, $\varpi(U,v)$,
and a continuous $\theta(U,v)$, satisfying $(\ref{rvqu})$ and
$(\ref{pvqu})$, and such that $(\ref{mazaorio})$, $(\ref{r--orio})$, 
$(\ref{eiv1})$, $(\ref{eiv1*})$ hold, 
it follows from 
\[
1-\mu(U,v)=\int_v^\infty{\frac{\theta^2}
re^{\int_v^{\bar{v}}{\frac2{r^2}\left(\frac{e^2}r-
\varpi\right)}}d\bar{v}},
\]
that the inequalitites $(\ref{newINIT})$, $(\ref{newINIT2})$
and $(\ref{newINIT3})$ hold for large enough $\Lambda\ge 0$, where
$\Lambda=\Lambda(C, r_0, r_+)$.\footnote{In particular, this fact shows
the equivalence of the data set up in this section and 
the assumptions of Theorem~\ref{eis9ew}.}
Let $V$ satisfy $(\ref{giatorkatw})$--$(\ref{trello})$, 
consider the null ray $v^*=V$, and choose $U'$ so that
the $[U,U']\times\{V\}$ is
contained in $\mathcal{Q}\cap\{r>0\}$. (Such a $U'$ exists
because $\mathcal{Q}\cap\{r>0\}$ is open.)
Define a function $u^*$ on
\[
\mathcal{G}'=D^+([U,U')\times\{V\}\cup U\times[V,\infty))\cap \mathcal{Q}
\]
by 
\[
u^*(u,v)=-\int_U^u\nu(\bar{u},V)d\bar{u}.
\]
Since $\nu<0$, $u^*$ defines a regular
coordinate on $\mathcal{G}'$. Let $(u_0,V)$ denote in $(u^*,v)$-coordinates
the point $(U',V)$ in $(u,v)$ coordinates.

We can define $\nu^*(u^*,v)=\partial_{u^*}r$, 
$\zeta^*(u^*,v)=r\partial_{u^*}\phi$,
and we have that 
\[
\nu^*(u^*,V)=-1,
\]
\[
|\zeta^*(u^*,V)|\le\bar{C},
\]
for some $\bar{C}$, the latter inequality following simply from continuity
of $\zeta^*$ and compactness of $[0,u_0]\times\{V\}\subset\mathcal{Q}$.
 
Again, let us drop the $*$ from our coordinate, and write
$u$ for $u^*$, $\nu$ for $\nu^*$, etc. We shall refer
to $\mathcal{G}'$ as $\mathcal{G}'(u_0)$.
With the above definitions, $r$, $\lambda$, $\nu$,
$\theta$, $\zeta$, $\phi$, $\kappa$ satisfy  
$(\ref{ruqu})$--$(\ref{sign2})$ in $\mathcal{G}'(u_0)$, 
while they satisfy
$(\ref{INITv1})$--$(\ref{INITu2})$ on 
$[0,u_0)\times\{V\}\cup\{0\}\times[V,\infty)$,
with $e$, $\varpi_+$, $\Lambda$, $r_0$ as defined in the course of this proof,
and $p=\omega$. 
\end{proof}

The data considered in \cite{md:si}, 
expressed in the coordinates of the present paper,
are such that $\theta$ vanishes initially on the event horizon, 
and $\zeta$ decays to $0$
polynomially in $u$ on the ingoing segment as $u\to0$. 
Thus, the data defined in this section include
in particular the data of \cite{md:si}.\footnote{The more restricted
data of Section~\ref{minsec}, however, necessary to prove blow up,
do not contain the analogous restricted
data of Theorem 2 of \cite{md:si}.}
One should note that once the solution has evolved up to a spacelike
curve beyond the apparent horizon 
(for instance $\Gamma_{E}$
of Proposition~\ref{prwtnprotasn}), it would be difficult to decide,
on the basis of measurements
along that curve, whether in the initial data $\theta$ vanished
on the event horizon or merely decayed exponentially; this
is due to
the backscattering of radiation.
Thus, the data of \cite{md:si}
should be compared more generally
with data in which $\theta$ decays exponentially. 
This (exponential vs.~polynomial) is the important quantitative
difference 
between the data of \cite{md:si} and the data considered here.

\section{The maximal future development}
\label{mddsec}
Recall the subset 
$\mathcal{K}(u_0)$ of the $(u,v)$ plane defined in Section~\ref{ivpsec}.
In this section, unless otherwise noted, causal relations $J^+$, $J^-$, $D^+$ 
will refer
to the induced time-oriented Lorentzian metric $-(du\otimes dv+dv\otimes du)$ 
in $\mathcal{K}(u_0)$. 

Introducing the notation
\[
|\psi|^k_{(u,v)}=|\psi|_{C^k(J^-(u,v)\setminus(u,v))},
\]
we define the norm
\begin{eqnarray}
\label{normdef}
|(r,\lambda,\nu,\kappa,\varpi,\theta,\zeta, \phi)|^k_{(u,v)}
=\max\left\{|r^{-1}|^k_{(u,v)},|\lambda|^k_{(u,v)},|\nu|^k_{(u,v)},
\right.\\
\nonumber
\hbox{}\left.|\nu^{-1}|^{k-1}_{(u,v)},|\kappa|^{k-1}_{(u,v)}, |\kappa^{-1}|^{k-1}_{(u,v)},
|\varpi|^k_{(u,v)},|\theta|^{k-1}_{(u,v)},|\zeta|^{k-1}_{(u,v)}\right\}.
\end{eqnarray}
Standard techniques (see~\cite{chr:bv}) then yield the following 
\begin{theorem}
\label{standard}
Let $k$ be an integer $k\ge1$.
Let $r$, $\kappa$, $\lambda$, $\varpi$, $\theta$, $\phi$ be functions
on $\{0\}\times[V,\infty)$ satisfying $(\ref{INITv1})$--$(\ref{orise})$, and
let $\zeta$, $\nu$ be functions on $[0,u_0)\times\{V\}$
satisfying $(\ref{INITu1})$--$(\ref{INITu2})$, and assume that 
$\theta(0,v)\in C^{k-1}$, $\zeta(u,V)\in C^{k-1}$.
Then there exists a unique non-empty open set
\[
\mathcal{G}(u_0)\subset\mathcal{K}(u_0),
\]
and unique extensions of the functions
$r$, $\lambda$, $\nu$, $\kappa$, $\varpi$, $\theta$,
$\zeta$, $\phi$ to $\mathcal{G}(u_0)$
such that
\begin{enumerate}
\item
The functions satisfy $(\ref{ruqu})$--$(\ref{th-orism})$, 
$(\ref{puqu})$--$(\ref{sign2})$,
and 
\[
|(r,\lambda,\nu,\kappa,\varpi,\theta,\zeta)|^k_{(u,v)}<\infty,
\]
for all $(u,v)\in\mathcal{G}(u_0)$.
\item
$\mathcal{G}(u_0)$ is a past subset of $\mathcal{K}(u_0)$,
i.e.~$J^{-}(\mathcal{G}(u_0))\subset\mathcal{G}(u_0)$.
\item
For each $(u,v)\in\partial\overline{\mathcal{G}(u_0)}$,
we have
\[
|(r,\lambda,\nu,\kappa,\varpi,\theta,\zeta)|^1_{(u,v)}=\infty.
\]
\end{enumerate}
\end{theorem}
Here $\overline{\mathcal{G}(u_0)}$
denotes the closure of $\mathcal{G}(u_0)$ in the topology of $\mathcal{K}(u_0)$.
The collection
\begin{equation}
\label{suvo9}
\{\mathcal{G}(u_0), r, \lambda, \nu, \kappa, \varpi, \theta, \zeta, \phi\}
\end{equation}
is the so-called \emph{maximal future development} of the
initial data set. By property 3, it follows easily
that if $\theta(0,v)\in C^{\tilde{k}}$,
$\zeta(u,V)\in C^{\tilde{k}}$, for $\tilde{k}>1$, 
then $(\ref{suvo9})$ does not depend on
the value of $1\le k\le \tilde{k}$ we fix in Theorem~\ref{standard}. 
Thus, we need not refer explicitly to the value of $k$ in $(\ref{suvo9})$.
 
One should note that in this topology,
$\partial\overline{\mathcal{G}(u_0)}=\overline{\mathcal{G}(u_0)}\setminus\mathcal{G}(u_0)$
is empty in the case of the Reissner-Nordstr\"om solution, i.e.~it
includes neither the segment $\{u_0\}\times[V,\infty)$, 
nor the Cauchy horizon.

One can easily prove that the following result of \cite{md:si}
also holds for the initial data considered here:

\begin{theorem}
\label{epekt9e}
The function $r$ can be extended continuously
to a function on $\overline{\mathcal{G}(u_0)}$, such that $r(p)=0$
for all $p\in\partial\overline{\mathcal{G}(u_0)}$.
\end{theorem}

It follows from the above theorem that if 
$p\in\overline{\mathcal{G}(u_0)}$ and $r(p)>0$,
then $p\in\mathcal{G}(u_0)$. 
We shall make repeated use of this extension principle in this paper.

Defining $\bar{g}$ on $\mathcal{G}(u_0)$ by
$(\ref{W-orism})$, where $-\Omega^2=4\nu\kappa$,
it follows by Proposition~\ref{megalnprotasn} that 
from the collection $(\ref{suvo9})$ we can construct 
a spherically symmetric solution to $(\ref{Einstein-xs})$--$(\ref{em-xs})$. 
Let
us denote this solution by 
\begin{equation}
\label{mialusn}
\{\mathcal{M}', g', F'_{\mu\nu}, \phi'\}.
\end{equation}
The solution $(\ref{mialusn})$
can be proven to be the unique 
globally hyperbolic solution to the Einstein-Maxwell-scalar field
equations $(\ref{Einstein-xs})$--$(\ref{em-xs})$,
admitting the data, interpreted upstairs, as a suitable ``past boundary''.
Since ultimately, we are interested here in the characteristic initial value
problem only for data which arise from Cauchy data, we will not discuss
in detail the issue of uniqueness for this characteristic initial value problem when
interpreted upstairs. (See, however, \cite{rigid}.) 
For data which do arise from Cauchy data, we have, however, the following
\begin{proposition} 
Let $\{\mathcal{M}, g, F_{\mu\nu}, \phi\}$, $\mathcal{Q}$, $\mathcal{G}'(u_0)$
 be as in Theorem~\ref{sxesntwv2}, and let $\mathcal{G}(u_0)$ be
as in Theorem~\ref{standard}, applied to the functions 
on $\{0\}\times[V,\infty)\cup[0,u_0)\times\{V\}$ given by
Theorem~\ref{sxesntwv2}. Then
\[
\mathcal{G}'(u_0)=\mathcal{G}(u_0), 
\]
and the functions $r$, $\nu$, $\lambda$, $\nu$, $\kappa$,
$\theta$, $\zeta$, $\phi$ of Theorem~\ref{standard} coincide on 
$\mathcal{G}(u_0)$ with the functions defined in
Theorem \ref{sxesntwv2}.
\end{proposition}
\begin{proof}
This follows immediately from 
the uniqueness part of Theorem \ref{standard}, the maximality
of the Cauchy development, and 
the fact that, by the results of~\cite{md:sssts},
the set $\mathcal{G}'(u_0)\subset\mathcal{Q}^{>0}$ of Theorem~\ref{sxesntwv2}
satisfies the conditions of Proposition \ref{megalnprotasn}.
\end{proof}

It follows from the above Proposition that statements
about $(\ref{suvo9})$, in the special case where Theorem~\ref{standard} 
is applied to
data arising from Theorem~\ref{sxesntwv2}, 
are in fact statements about the unique solution 
to a spacelike Cauchy problem.
Moreover, in this case, 
\[
\mathcal{G}'(u_0)\subset \mathcal{Q}\setminus J^-(\mathcal{I}^+),
\]
i.e. $\mathcal{G}(u_0)$ can be viewed as a subset of the black hole
region of an asymptotically flat spacetime. In a slight abuse
of notation, we will often refer to $\mathcal{G}(u_0)$ as describing 
black hole region even when it does not arise as above.

In the sequel, 
we shall consider initial data as having been fixed once and for all.
Applying Theorem~\ref{standard}, we 
are given $(\ref{suvo9})$.
We note, however, the following. In the course of the paper, we will restrict
attention again and again
to sets
\begin{equation}
\label{tosuvolo}
D^+([0,U)\times\{V\}\cup\{0\}\times[V,\infty))\cap\mathcal{G}(u_0)
=\mathcal{G}(u_0)\cap\mathcal{K}(U)
\end{equation}
for smaller and smaller $U<u_0$.
This set $(\ref{tosuvolo})$, 
together with the restrictions of the functions $r$, $\nu$, $\lambda$,
$\kappa$, $\varpi$, $\theta$, $\zeta$, $\phi$, correspond
to the collection $(\ref{suvo9})$ that emerges 
from Theorem~\ref{standard} applied to the restriction of
the initial data to $[0,U)\times\{V\}\cup\{0\}\times[V,\infty)$.
There is thus no confusion if we denote the set $(\ref{tosuvolo})$ by
$\mathcal{G}(U)$.   

Given $U$, keep in mind that there will be
three distinct topological and
conformal structures which one might want to consider:
$(\mathcal{G}(U),-\Omega^2du dv)$, $(\mathcal{K}(U),-dudv)$, 
and the (closed) Penrose diagram $\overline{\mathcal{PD}}$ 
of the spacetime $\mathcal{G}(U)$.
The structure of $\mathcal{K}(U)$ is convenient for 
doing analysis, in particular, for
formulating Theorems~\ref{standard} and~\ref{epekt9e}.
The structure of the Penrose diagram 
$\mathcal{PD}$ is convenient for global causal geometric statements,
like saying that the spacelike 
curve $\gamma\subset\mathcal{K}(U)$ terminates at $i^+$ or parametrizing
what will be Cauchy horizon $\mathcal{CH}^+$.
Finally, the structure of $(\mathcal{G}(U),-\Omega^2dudv)$
is fundamental, as it is the structure of our spacetime itself. 
All three structures
coincide when restricted to $\mathcal{G}(U)$.

\section{Monotonicity}
\label{monsec}
Since $\nu<0$ on $[0,u_0)\times\{V\}$, it follows by definition 
of $(\ref{normdef})$ that 
\begin{equation}
\label{nmono}
\nu<0
\end{equation}
throughout $\mathcal{G}(u_0)$. 
In view of the fact that $r(0,v)\le r_+$, integration of $(\ref{nmono})$
in $u$ now gives
\begin{equation}
\label{*IKI*}
r\le r_+.
\end{equation}

In contrast to $\nu$, the sign of
$\lambda$ will change in evolution.
Define the so-called
\emph{trapped region} $\mathcal{T}$ by
\[
\mathcal{T}=\{(u,v)\in\mathcal{G}(u_0):\lambda(u,v)<0\},
\]
and the \emph{apparent horizon} $\mathcal{A}$ by
\[
\mathcal{A}=\{(u,v)\in\mathcal{G}(u_0):\lambda(u,v)=0\}.
\]
From $(\ref{normdef})$, the equation $\lambda=\kappa(1-\mu)$,
and the fact that $\kappa>0$ on $\{0\}\times[V,\infty)$,
it follows that $\kappa$ is positive in $\mathcal{G}(u_0)$, and 
 that $1-\mu(u,v)<0$ if and only if $(u,v)\in\mathcal{T}$,
and $1-\mu(u,v)=0$ if and only if $(u,v)\in\mathcal{A}$. 
%Moreover, $\mathcal{A}$, if non-empty, is achronal, and since
%$\lambda(0,v)\ge0$,
%$\mathcal{T}=J^+(\mathcal{A})$.

From $(\ref{fdb1})$ and $(\ref{nmono})$, we have
\begin{equation}
\label{F1}
\partial_u\kappa\le0.
\end{equation}
In $\mathcal{T}$, $\frac{\nu}{1-\mu}$ is also a positive
quantity, and 
\begin{equation}
\label{F2}
\partial_v\frac\nu{1-\mu}\le0,
\end{equation}
by $(\ref{fdb2})$
The above inequality implies that
if $(u,v)\in\mathcal{A}$, then $(u,v^*)\in\mathcal{A}\cup\mathcal{T}$,
for $v^*>v$, while if $(u,v)\in\mathcal{T}$, then
$(u,v^*)\in\mathcal{T}$ for $v^*>v$.

Note that $\frac{\nu}{1-\mu}$ clearly blows up 
identically on $\mathcal{A}$,
so this fraction cannot be immediately utilized as a controlling 
quantity in $\mathcal{T}$, except in the case
where $\mathcal{A}$ coincides with the event horizon.
(Compare with \cite{md:si} where one did have \emph{a priori} pointwise
bounds for $\frac{\nu}{1-\mu}$ throughout $\mathcal{G}(u_0)$, on account
of our special choice of initial data!)

Finally, in view of the above
and $(\ref{puqu})$ and $(\ref{pvqu})$, it is clear
that we have the following
additional monotonicity in $\mathcal{T}$:
\begin{equation}
\label{spemov}
\partial_v\varpi\ge0, \partial_u\varpi\ge0.
\end{equation}

We note here that the inequalities of this section,
with the exception of $(\ref{spemov})$, are quite general, and
depend only on the dominant energy condition.   This is nicely discussed
in \cite{chr:sgrf}. 
We shall return to the issue of monotonicity in Section~\ref{minsec},
where, in particular, we shall be able to prove, for appropriate initial
data, the inequality $\partial_u\partial_v\varpi\ge0$, in $\mathcal{T}$. 

\section{Remarks on the analysis of black holes}
\label{ovvsec}
Before beginning the study of our system, 
it will be useful to make a few preliminary remarks that will give
a taste of the special flavor of the analysis of 
$(\ref{puqu})$--$(\ref{sign2})$
in black hole regions. 
In addition to providing an outline
of the approach taken in the following sections, these remarks
may be relevant for understanding more
general charged or rotating black holes.

In view of Theorem~\ref{epekt9e}, to delimit the extent of the maximal
domain of development, it is enough to control (from below!) $r$. 
The function $r$ would indeed be controlled if
we could appropriately bound
its null derivatives, let us say $\nu$.
But integrating equation $(\ref{nqu})$, one deduces that it
is the following quantity that must then be controlled:
\begin{equation}
\label{apeiro;}
\int_{v_1}^{v_2}{\frac{2\kappa}{r^2}
\left(\varpi-\frac{e^2}r\right)dv}.
\end{equation}
In the Reissner-Nordstr\"om solution, where $\phi\equiv0$, $\theta\equiv0$,
$\zeta\equiv0$, it follows by $(\ref{fdb1})$ that 
\[
\partial_u\kappa=0,
\]
and thus, by $(\ref{INITv1})$, 
$\int_{v_1}^{v_2}\kappa dv=v_2-v_1$.
This means that as $v_2\to\infty$, the integral $(\ref{apeiro;})$
blows up in
the trapped region{\footnote{It is instructive to compare
with the domain of outer communications $\mathcal{D}$ of~\cite{mi:mazi}. 
There, as $v_2\to\infty$,
$r\to\infty$, and this keeps the integral bounded. In the 
trapped region, on the other hand, we have the \emph{a priori} 
upper bound $(\ref{*IKI*})$ for $r$.} for all values of $u$.
For fixed $u>0$, the sign of $\left(\varpi-\frac{e^2}r\right)$, however, becomes
negative as $v\to\infty$. Thus the effect of
$(\ref{apeiro;})$ in integrating $(\ref{nqu})$ is to make
$\nu$ tend to $0$ in the limit as $\mathcal{CH^+}$ is approached.
The negative sign for $\left(\varpi-\frac{e^2}r\right)$ is
thus ``favorable'' for controlling $r$.
We will call this the \emph{blue-shift} sign.

If we examine, however, the sign of $\left(\varpi-\frac{e^2}r\right)$
on $\{0\}\times[V,\infty)\cup[0,u_0)\times\{V\}$, we find that it is positive. 
This is the ``unfavorable'' sign for the control of $r$,
and we will call it the \emph{red-shift} sign.

From this point of view, key to the formation
of the Cauchy horizon is the fact that $\left(\varpi-\frac{e^2}r\right)$
changes sign in evolution from red-shift to
blue-shift\footnote{We see that in the case $e=0$,
such a change of sign is impossible; this explains
in particular why
the
Schwarzschild solution
behaves so differently from the
Reissner-Nordstr\"om solution.}, and moreover, that
this blue-shift region $\mathcal{B}$ is sufficiently large, so as
to ``compensate'' for the effect of the red-shift region $\mathcal{R}$, which
in the meantime has made $|\nu|$ large.

In order to understand better the significance of
$(\ref{apeiro;})$,
it is useful to separate out a region 
where $(\ref{apeiro;})$
can be uniformly bounded. We will call such regions
``no-shift'' regions, and denoted them $\mathcal{N}$.
The precise boundary of any such
region is a bit arbitrary, but
nonetheless, the concept is useful. 
The Reissner-Nordstr\"om solution can be decomposed
as follows:
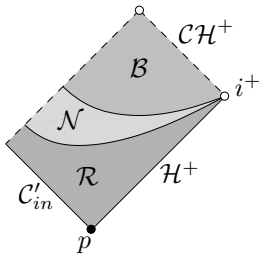
\[
\input{RNshiftveo.pstex_t}
\]

The above picture is our starting point for understanding
the evolution of the initial value problem considered
here. The partition into 
red-shift, no-shift, and blue-shift regions will be key
to understanding analytically the various quantities of
our system. 

We have already seen in the discussion above
the different effects on $r$ in each of the three regions. 
But these regions also have very different effects on $\varpi$.
We turn now to examine this point.

In the Reissner-Nordstr\"om solution, the scalar field
vanishes, and thus $\varpi$ is constant. 
With a scalar field present, however, 
$(\ref{puqu})$ implies that, all other things being equal, 
the \emph{smaller} the value of
$|\nu|$, the larger the value of $\partial_u\varpi$.
Thus, the red-shift sign, which was unfavorable for
the control of $r$, is \emph{favorable} for the control of $\varpi$,
while the blue-shift sign, which was favorable for
$r$, is \emph{unfavorable} for $\varpi$.

The fact that the blue-shift sign is ``unfavorable'' for
$\varpi$ is basically what led to the conjecture of strong
cosmic censorship in the first place.
We will return to this issue shortly. Let us first try to
understand the strategy employed
for proving the stability result, Theorem~\ref{eis9ew}.
 
As noted earlier, key to the Reissner-Nordstr\"om behavior of
$r$ is the blue-shift region. Yet, with the hindsight of our result,
namely that $\varpi\to\infty$ as $v\to\infty$, one
does \emph{not} have the blue-shift sign sufficiently
close to $\mathcal{CH}^+$ in the topology of the Penrose
diagram. The best one can hope for, then, is to have
at least a sufficiently
large piece of blue shift region $\mathcal{B}$, somewhere.
As we shall see, the whole situation is quite delicate. An outline of
the complete argument is as follows:

\begin{enumerate}
\item
In Section~\ref{rsrsec}, we shall define what will be a redshift region 
$\mathcal{R}=J^-(\Gamma_E)$, and we shall show 
that $\mathcal{R}$ contains an achronal apparent horizon $\mathcal{A}$. 
Using the red-shift sign of $(\ref{apeiro;})$,
we will be able to bound all quantities in
$\mathcal{R}$. The future boundary of $\mathcal{R}$ will
be a spacelike curve $\Gamma_E\subset\mathcal{T}$,
on which $\varpi-\frac{e^2}r=E>0$.
\item
Next, in Section~\ref{nsrsec}, we shall define what will be
a no-shift region $\mathcal{N}=J^+(\Gamma_E)\cap J^-(\Gamma_{-\xi})$.
Again, we shall be able to bound all quantities in $\mathcal{N}$,
using the uniform boundedness of $(\ref{apeiro;})$.
In particular, $\lambda\sim1$, $\kappa\sim1$, $\varpi\sim\varpi_+$. 
On the spacelike curve $\Gamma_{-\xi}$, we will have
$\varpi-\frac{e^2}r=-\xi<0$. The sign has turned blue-shift!
\item
In Section~\ref{sbssec}, we shall define what will be a blue-shift
region $\mathcal{B}_\gamma=J^+(\Gamma_{-\xi})\cap J^-(\gamma)$.
Understanding the $v$-dimensions of this region will be fundamental.
On the one hand, $\mathcal{B}_\gamma$ will be small enough to allow us
to control all quantities, by playing off the decay rate of the scalar
field with the the rate in which
$(\ref{apeiro;})$ blows up. In particular, $\kappa\sim 1$,
$\varpi\sim\varpi_+$. On the other hand, 
$\mathcal{B}_\gamma$
will be large enough so as to have a suitable
``moderating'' effect on $\lambda=\partial_vr$.\footnote{An important technical
point in this paper is how to relate $(\ref{apeiro;})$, which is 
$\log \nu(u,v_2)-\log \nu(u,v_1)$, with an analogous expression for
$\lambda$. It turns out that pointwise estimates are more natural for
$\lambda$ than for $\nu$ because of the stability of $\kappa=\lambda(1-\mu)$.}
Specifically, 
on the curve $\gamma$, we will show
\begin{equation}
\label{stagrngora}
\lambda(u|_{\gamma}(v),v)\le Cv^{-s}
\end{equation}
for an $s>1$. The significance of $s>1$ is that $(\ref{stagrngora})$ can
be integrated in $v$.
\item
We will show in Section~\ref{bsbsec} that
if there are regions in $J^+(\gamma)$
where the sign of $(\ref{apeiro;})$ becomes \emph{red-shift}, 
then $(\ref{apeiro;})$ is uniformly bounded in those
regions, i.e. $J^+(\gamma)$ can be decomposed into \emph{blue-shift}
and \emph{no-shift} regions. Either
way, a bound $(\ref{stagrngora})$ of the form
is preserved. This allows us to bound $r$ \emph{a priori} away from $0$  in
$\mathcal{G}(U)$, for sufficiently small $U$.
It will follow by Theorem~\ref{epekt9e} that $\mathcal{G}(U)=
\mathcal{K}(U)$. 
\item
Finally, estimating suitable quantities in $\mathcal{G(U)}$, 
we will be able to show in Section~\ref{c0esec}, 
that $\bar{g}$, $r$, and $\phi$ can be extended continuously
to $\mathcal{CH}^+$, completing the proof of Theorem~\ref{eis9ew}.
\end{enumerate}
The various curves and
regions referred to in the above outline
are depicted below:
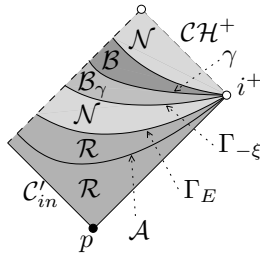
\[
\input{gevshiftveo.pstex_t}
\]

We now turn to discuss the blow-up of $\varpi$. Again, the key
is the blue-shift region, which tends to make $\varpi$ large. 
Yet, as noted earlier, the growing of $\varpi$
tends to limit the strength of, and eventually completely
destroy, the blue-shift region. It seems that we have to face
the uncertain competition between these two effects in the potentially unstable
part of the blue-shift region, $\mathcal{B}\cap J^+(\gamma)$. 
Our argument proceeds by contradiction. There are two steps:
\begin{enumerate}
\item
It is shown that 
for appropriate initial data, the following dichotomy holds:
Either the solution in $J^+(\gamma)$ is ``qualitatively
similar'' to the Reissner-Nordstr\"om solution in $\mathcal{B}$,
\emph{or} $\varpi$ blows up identically on $\mathcal{CH}^+$.
The phrase ``qualitatively similar'' means in particular
that the sign of $(\ref{apeiro;})$ should remain
blue-shift, $\varpi$ should extend to a finite-valued function
on $\mathcal{CH}^+$ with $\varpi(q)\to\varpi_+$ as $q\to i^+$,
and $\lambda$ should decay exponentially in $v$ for fixed $u$.
The arguments proving this dichotomy rely heavily on the
strong monotonicity properties proven in Section~\ref{minsec}.
\item
By step 1, if $\varpi$ does not blow up, we can assume that the solution in
$J^+(\gamma)$ is indeed ``qualitatively similar'' to Reissner-Nordstr\"om
in the sense described above. But this implies that the 
\emph{linear} mechanism for blow-up should apply.
Indeed, the blue-shift factor $(\ref{apeiro;})$ operates, 
making $\varpi$ large, 
contradicting 
in particular 
the assumption that $\varpi(q)\to\varpi_+$. Thus, $\varpi$ blows up
after all.
\end{enumerate}

The proof of step 2 can be considered a linear theory
argument\footnote{Indeed, this computation shows a form of blow up
for the  decoupled problem as well.} 
that should be compared with 
Chandrasekhar and Hartle \cite{ch:cchRN}, 
even though, with the help of the monotonicity proven
in Section~\ref{minsec}, we will be able to derive it almost immediately. 
In this sense, the
linear theory that was responsible for the original conjecture
of strong cosmic censorship
indeed finds its way in the present proof, 
albeit as a part of a contradiction argument.

\section{The red-shift region}
\label{rsrsec}
Recalling the inequality $(\ref{alloxipriv})$, one can 
choose a positive constant $E$ such that
\begin{equation}
\label{alloxi}
\varpi_+-\frac{e^2}{r_+}-2(\tilde{K}+\tilde{M})<
E<\varpi_+-\frac{e^2}{r_+}-\frac32(\tilde{K}+\tilde{M}).
\end{equation}
We will define our \emph{red-shift} region $\mathcal{R}\subset\mathcal{G}(u_0)$ 
by the relation:
\[
\mathcal{R}=\left\{(u,v)\in\mathcal{G}(u_0):\left(\varpi-\frac{e^2}r\right)(\tilde{u},
\tilde{v})>E
{\rm\ for\ all\ }(\tilde{u},\tilde{v})\in J^-(u,v)\right\}.
\]
Note that by the inequality $(\ref{birxi})$ and the inequality
\begin{eqnarray}
\label{ikixi}
\nonumber
0\le \left(\frac{e^2}{r}-\frac{e^2}{r_+}\right)(0,v)&=&
\frac{e^2(r_+-r)}{r_+r}(0,v)\\
&\le&\frac{\Lambda e^2}{r_0r_+(2p-1)}v^{-2p+1}\\
\nonumber
&=&\tilde{K}(v)\le \tilde{K},
\end{eqnarray}
it follows that
$\mathcal{R}$ is a non-empty subset of $\mathcal{G}(u_0)$.
By continuity,
$\mathcal{R}$ contains some open set (in the topology of $\mathcal{K}(u_0)$)
containing $\{0\}\times[V,\infty)$.
As $\mathcal{R}$ is clearly a past set, its future boundary
$\Gamma_{E}$, given by
\[
\Gamma_{E}=\overline{\mathcal{R}}\cap I^+(\mathcal{R})
\cap\mathcal{G}(u_0)\setminus\mathcal{R},
\]
is a (possibly empty) achronal curve.

Step 1 of the outline of Section \ref{ovvsec} is accomplished by the following 
\begin{proposition}
\label{prwtnprotasn}
For $U_1$ sufficiently small,
$\Gamma_{E}\cap\mathcal{G}(U_1)\ne\emptyset$ is spacelike and terminates 
at $i^+$. 
Moreover, on $\Gamma_{E}\cap\mathcal{G}(U_1)$, the equality
\[
\varpi-\frac{e^2}r=E
\]
holds
identically,
and
$\Gamma_{E}\cap\mathcal{G}(U_1)\subset\mathcal{T}$:
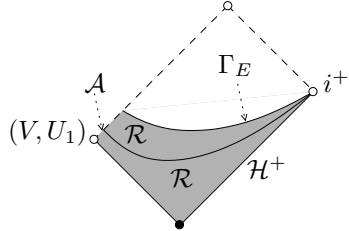
\[
\input{prop1veo.pstex_t}
\]
Finally, as depicted in the above Penrose diagram, $\mathcal{A}\cap\mathcal{G}(U_1)$
is achronal, and 
\[
\mathcal{T}\cap\mathcal{G}(U_1)=I^+(\mathcal{A})
\cap\mathcal{G}(U_1).
\]
\end{proposition}

\begin{proof}
The proof will proceed by a bootstrap argument. We first define a region
$\tilde{\mathcal{R}}$ as the set of points $(u,v)\in\mathcal{G}(u_0)$ such that  
the following
estimates hold for all $(\tilde{u},\tilde{v})\subset J^-(u,v)$:
\begin{equation}
\label{BS1}
\left(\varpi-\frac{e^2}r\right)(\tilde{u},\tilde{v})>E,
\end{equation}
\begin{equation}
\label{BS2}
r(\tilde{u},\tilde{v})> \frac{c_1}2,
\end{equation}
\begin{equation}
\label{BS3}
|1-\mu(\tilde{u},\tilde{v})|<2M_1,
\end{equation}
\begin{equation}
\label{booteva}
\left|\frac\zeta\nu\right|(\tilde{u},\tilde{v})
<\frac{3\tilde{C}r_+^2}{Ec_1} \tilde{v}^{-p},
\end{equation}
\begin{equation}
\label{bootduo}
|\theta(\tilde{u},\tilde{v})|<\tilde{C}\tilde{v}^{-p},
\end{equation}
for $c_1$, $M_1$ given by  $(\ref{c_1ori})$ and $(\ref{M1ori})$,  and $\tilde{C}$
defined by
\begin{equation}
\label{miaepilogn}
\tilde{C}=\max\left\{  
\frac{2c_1E}{r_+^2}\overline{C}
e^{V\frac{E}{r_+^2}}e^{-p}\left(\frac{pr_+^2}{E}\right)^{p}
,3C \right\}.
\end{equation}

In $\overline{\tilde{\mathcal{R}}}\cap\mathcal{G}(U_1)$,
we shall be able to retrieve, in fact improve the bounds
$(\ref{booteva})$, $(\ref{bootduo})$, 
$(\ref{BS2})$, and $(\ref{BS3})$, for $U_1$ 
sufficiently small. 
A simple continuity
argument will then imply
that $\tilde{\mathcal{R}}\cap\mathcal{G}(U_1)=\mathcal{R}\cap\mathcal{G}(U_1)$.

The estimates $(\ref{booteva})$, $(\ref{bootduo})$ thus
obtained will easily
lead to the remaining conclusions of the proposition.

Before giving the details of the proof, it might be useful to
point out in advance some of the main points.
We will see that our bootstrap assumptions above imply
\begin{equation}
\label{masdivei}
e^{
\int_V^v{\frac{2\kappa}{r^2}\left(\varpi-\frac{e^2}r\right)}}
\ge e^{Hv},
\end{equation}
for some positive constant
$H$.  Integrating the equation
$(\ref{integrating})$, and playing off the above exponential factor
$(\ref{masdivei})$, one can transfer 
the polynomial decay $(\ref{bootduo})$ of $\theta$ in $v$ into similar
decay $(\ref{booteva})$ of $\frac{\zeta}{\nu}$. 
This is the heart of the \emph{red-shift} technique.

As far as retrieving the bootstrap assumptions is concerned, we
note in advance three sources for a smallness factor:
\begin{enumerate}
\item
The assumptions $(\ref{giatorkatw})$--$(\ref{trello})$ on $V$.
\item
Restricting to $\mathcal{G}(U_1)$ for small
$U_1$, and then noting that $\int_0^u d\tilde{u}\le U_1$.
\item
Noting that our bootstrap assumptions imply $\int_0^{u}\nu(\bar{u},v) d\bar{u}$
is small.
\end{enumerate}

We now proceed with the proof in detail. Consider a point
$(u,v)\in\overline{\tilde{\mathcal{R}}}\cap\mathcal{G}(u_0)$. By continuity,
the bounds
$(\ref{booteva})$, $(\ref{bootduo})$, 
$(\ref{BS2})$, and $(\ref{BS3})$ clearly hold in $J^-(u,v)$,
where the strict inequalities are replaced by non-strict inequalities.
Let us first note that $(\ref{BS2})$ provides a bound
\begin{equation}
\label{nfirst}
\left|\int_0^u{\nu(\bar{u},v) d\bar{u}}\right|\le r_+-\frac{c_1}2,
\end{equation}
as we have $\nu<0$ and $r(0,v)\le r_+$.
We can derive a different bound for the same quantity
as follows: In view of the bounds $(\ref{BS2})$, $(\ref{BS3})$,
and $(\ref{F1})$, we have that
\begin{eqnarray*}
0\le\frac{2\kappa}{r^2}
\left(\varpi-\frac{e^2}{r}\right)&=&
\frac{2\kappa}{r^2}\left(-\frac r2(1-\mu)+\frac r2-\frac{e^2}{2r}\right)\\
&\le& r_0^{-2}r_+(2M_1+1).
\end{eqnarray*}
Thus, integrating
$(\ref{nqu})$, in view of $(\ref{INITu2})$,
it follows that $-\nu\le e^{r_0^{-2}r_+(2M_1+1)v}$ and consequently,
\begin{equation}
\label{nsecond}
\left|\int_0^u{\nu(\bar{u},v)d\bar{u}}\right|\le ue^{r_0^{-2}r_+(2M_1+1)(v-V)}.
\end{equation}
Integrating the equation $(\ref{fdb1})$ with initial condition
provided by $(\ref{INITv1})$, using
$(\ref{booteva})$ one obtains the bound 
\[
\kappa(u,v)\ge 
e^{-\frac{18\tilde{C}^2r_+^2}{c_1E^2}v^{-2p}
\left|\int_0^u\nu(\tilde{u},v)d\tilde{u}\right|}.
\]
Choosing $V_1$ such that
\[ 
\frac{9\tilde{C}^2r_+^2}{E^2}V_1^{-2p}(r_+-c_1/2)
\le\max\left\{c_1(\log 4)^{-1},\frac{\tilde{M}}{4M_1}\right\},
\]
and then choosing $U_{1,1}$ so that
\[
\frac{9\tilde{C}^2r_+^2}{E^2}
U_{1,1}e^{r_0^{-2}r_+(2M_1+1)(V_1-V)}
\le\max\left\{c_1(\log 4)^{-1},\frac{\tilde{M}}{4M_1}\right\},
\]
it follows that for 
$(u,v)\in\overline{\tilde{\mathcal{R}}}\cap\mathcal{G}(U_{1,1})$ we have
\begin{equation}
\label{nowhave}
\kappa(u,v)>\frac12.
\end{equation}
Now, integrating $(\ref{puqu})$ and using $(\ref{booteva})$
together with our bound $(\ref{nfirst})$
for $\left|\int{\nu du}\right|$, we deduce
\begin{equation}
\label{topi}
|\varpi(u,v)-\varpi(0,v)|\le \frac{18\tilde{C}^2r_+^2}{E^2}M_1v^{-2p}
(r_+-c_1/2).
\end{equation}
On the other hand, $(\ref{nsecond})$ implies
\[
|\varpi(u,v)-\varpi(0,v)|\le \frac{18\tilde{C}^2r_+^2}{E^2}M_1ue^{Av}.
\]
Thus, in 
$\overline{\tilde{\mathcal{R}}}\cap\mathcal{G}(U_{1,1})$,
\begin{equation}
\label{palabo}
-\frac32\tilde{M}+\varpi_+\le\varpi\le\frac12\tilde{M}+\varpi_+. 
\end{equation}
This estimate
together with $(\ref{BS1})$ yields
\begin{equation}
\label{beraber}
r(u,v)\ge\frac{e^2}{\frac{e^2}{r_+^2}+\frac52\tilde{M}+2\tilde{K}},
\end{equation}
\begin{equation}
\label{asteraki}
|1-\mu|(u,v)\le1+\frac{e^2}{c_1^2}+\frac{\tilde{M}+2\varpi_+}{c_1},
\end{equation}
which, in view of
$(\ref{M1ori})$ and $(\ref{c_1ori})$, improves $(\ref{BS2})$ and $(\ref{BS3})$.

We are left with improving $(\ref{booteva})$
and $(\ref{bootduo})$.
Integrating the equation
\begin{equation}
\label{integrating}
\partial_v\left(\frac\zeta\nu\right)=
-\frac\theta{r}-\left(\frac\zeta\nu\right)\frac{2\kappa}{r^2}
\left(\varpi-\frac{e^2}r\right)
\end{equation}
yields
\[
\frac\zeta\nu(u,v)=
\int_V^v{-\frac{\theta}r(u,\bar{v})e^{-\int_{\bar{v}}^v{
\frac{2\kappa}{r^2}\left(\varpi-\frac{e^2}r\right)}}d\bar{v}}+
\frac\zeta\nu(u,V)e^{-\int_V^v{\frac{2\kappa}{r^2}\left(
\varpi-\frac{e^2}r\right)}}
\]
and thus
\[
\left|\frac\zeta\nu(u,v)\right|
\le
\int_V^v{\left|\frac{\theta}r(u,\bar{v})\right|e^{-\int_{\bar{v}}^v{
\frac{2\kappa}{r^2}\left(\varpi-\frac{e^2}r\right)}}d\bar{v}}
+
\left|\frac\zeta\nu(u,V)\right|
e^{\int_V^v{\frac{2\kappa}{r^2}\left(
\frac{e^2}r-\varpi\right)}}.
\]
It follows from $(\ref{nowhave})$ and our bootstrap assumptions
$(\ref{BS1})$ and $(\ref{BS2})$ that
\begin{equation}
\label{cokmuhim}
\frac{2\kappa}{r^2}\left(\varpi-\frac{e^2}r\right)
\ge
\frac{E}{r_+^2}.
\end{equation}
Thus we have in fact
\begin{equation}
\label{haveinfact}
\left|\frac\zeta\nu(0,v)\right|
\le
\int_V^v{\frac{\tilde{C}\bar{v}^p}{c_1}e^{-(v-\bar{v})\frac{E}{{r_+}^2}}
d\bar{v}}
+
\overline{C}e^{V\frac{E}{{r_+}^2}}e^{-v\frac{E}{{r_+}^2}}.
\end{equation}
Integrating by parts the term
\[
\int_V^v{\frac{\tilde{C}\bar{v}^{-p}}{c_1}e^{-(v-\bar{v})\frac{E}{{r_+}^2}}
d\bar{v}},
\]
and noting that it is positive one obtains
\begin{eqnarray}
\label{sxedov1}
\nonumber
\int_V^v{\frac{\tilde{C}\bar{v}^{-p}}{c_1}e^{-(v-\bar{v})\frac{E}{{r_+}^2}}
}
&\le&
\frac{r_+^2\tilde{C}v^{-p}}{c_1E}\\
&&\hbox{}-
\int_V^v{\frac{r_+^2(-p)\tilde{C}\bar{v}^{-p-1}}{c_1E}
e^{-(v-\bar{v})\frac E{r_+^2}}d\bar{v}}.
\end{eqnarray}
Now,
\begin{eqnarray}
\label{sxedov2}
\nonumber
&&\int_V^v{\frac{r_+^2p\tilde{C}\bar{v}^{-p-1}}{c_1E}
e^{-(v-\bar{v})\frac{E}{r_+^2}}d\bar{v}}
\\
\nonumber
&=&\int_V^{v-\frac{2pr_+^2}{E}\log v}
{\frac{r_+^2p\tilde{C}\bar{v}^{-p-1}}{c_1E}
e^{-(v-\bar{v})\frac{E}{r_+^2}}d\bar{v}}\\
\nonumber
&&+\int_{v-\frac{2pr_+^2}{E}\log v}^v
{\frac{r_+^2p\tilde{C}\bar{v}^{-p-1}}{c_1E}
e^{-(v-\bar{v})\frac{E}{r_+^2}}d\bar{v}}\\
\nonumber
&\le&\frac{pr_+^2\tilde{C}V^{-p-1}}{c_1E} v^{-2p}
+\frac{2pr_+^2}{E}(\log v)\frac{r_+^2p}{c_1E}\tilde{C}(
v-\frac{2pr_+^2}{E}\log v)^{-p-1}\\
&\le&\frac{r_+^2\tilde{C}v^{-p}}{c_1E},
\end{eqnarray}
where the last inequality requires $(\ref{trello2})$.
Since the second term of $(\ref{haveinfact})$
can be bounded by $\frac{r_+^2\tilde{C}v^{-p}}{2c_1E}$,
it follows from $(\ref{miaepilogn})$,
$(\ref{haveinfact})$, $(\ref{sxedov1})$ 
and $(\ref{sxedov2})$ that
$(\ref{booteva})$ can be improved.

The equation $(\ref{sign1})$ can be rewritten as
\begin{equation}
\label{astduo}
\partial_u\theta=-\zn\frac{\nu\lambda}r.
\end{equation}
Moreover, the bound $(\ref{F1})$ together with $(\ref{asteraki})$
imply that $\lambda\le M_1$. Integration of $(\ref{astduo})$
yields
\begin{equation}
\label{2hac}
|\theta|(u,v)\le
Cv^{-p}+\frac{3\tilde{C}r_+^2M_1}{c_1^2E}v^{-p}\left|\int_0^u{\nu}(\bar{u},v)
d\bar{u}\right|.
\end{equation}
To improve $(\ref{bootduo})$ we will need 
yet another bound on $\int\nu$. A simple computation,
using $(\ref{BS1})$ and $(\ref{alloxi})$
yields
\[
r-r_+\ge\frac{-rr_+}{e^2}\left(
2\tilde{K}+2\tilde{M}+(\varpi-\varpi_+)\right),
\]
and since $r-r_+\le0$, as $\lambda(0,v)\ge0$ and $\nu<0$,
this gives, using $(\ref{palabo})$,
\[
-\int_0^u\nu\le r_+-r\le
\frac{r_+^2}{e^2}\left(\frac32\tilde{M}+
2(\tilde{M}+\tilde K)\right).
\]
Thus, by $(\ref{trello})$
and the second term on the right hand side of
$(\ref{miaepilogn})$,
it follows from $(\ref{2hac})$ that
$(\ref{bootduo})$ is also improved.

Since 
\[
J^-\left(\overline{\tilde{\mathcal{R}}}\right)\cap\mathcal{G}(U_{1,1})
\subset
\overline{J^-\left(\tilde{\mathcal{R}}\right)}\cap\mathcal{G}(U_{1,1})
=
\overline{\tilde{\mathcal{R}}}\cap\mathcal{G}(U_{1,1}),
\]
the above improved estimates show that
\[
\overline{\tilde{\mathcal{R}}}\cap\mathcal{G}(U_{1,1})\cap \mathcal{R}
\subset
\tilde{\mathcal{R}}\cap\mathcal{G}(U_{1,1}).
\]
Thus, $\tilde{\mathcal{R}}$ is an open and closed subset in the topology of
$\mathcal{R}$, and since $\mathcal{R}$ is connected, it follows that
\[
\tilde{\mathcal{R}}=\mathcal{R}.
\]

Let us suppose the curve $\Gamma_{E}$
is either empty or does not terminate at $i^+$. Then there is a 
$\tilde{u}\le U_{1,1}$ such that $\mathcal{G}(\tilde{u})$ 
does not contain any point on this curve, i.e.~such that
$\mathcal{G}(\tilde{u})=\mathcal{R}\cap\mathcal{G}(\tilde{u})$.
In particular, from the bound $(\ref{BS2})$ it follows
that $r\ge c>0$ on $\partial\overline{\mathcal{G}(\tilde{u})}$, where
the boundary is taken as a subset of $\mathcal{K}(\tilde{u})$,
and thus, by
Theorem \ref{epekt9e}, 
$\partial\overline{\mathcal{G}(\tilde{u})}=\emptyset$ and consequently
\begin{equation}
\label{avtifasn!!}
\mathcal{R}\cap\mathcal{G}(\tilde{u})=\mathcal{G}(\tilde{u})=\mathcal{K}(\tilde{u}).
\end{equation}
On the other hand, integration of
$(\ref{nqu})$ with the bound $(\ref{cokmuhim})$
yields
\[
-\nu\ge e^{\frac{E}{r_+^2}(v-V)}
\]
in $\mathcal{R}$.
Integrating the above inequality in $u$ gives
\begin{equation}
\label{avtifasn!!!}
r(u,v)\le r(0,v)-ue^{\frac{E}{r_+^2}(v-V)}.
\end{equation}
Fixing $u<\tilde{u}$, it follows from $(\ref{avtifasn!!})$
that $(u,v)\in R$ for all $v$.
Taking $v$ sufficiently large, $(\ref{avtifasn!!!})$
contradicts $(\ref{BS2})$. Thus $\Gamma_{E}\ne\emptyset$
and terminates at $i^+$.

To show that $\Gamma_{E}$
is spacelike, with
\begin{equation}
\label{toshow}
\varpi-\frac{e^2}r=E,
\end{equation}
we note that if this is not the case,
then there exists a constant-$v$ null component $N\subset \Gamma_{E}$
with past endpoint at $q$, with $\varpi-\frac{e^2}r=E$ at $q$.
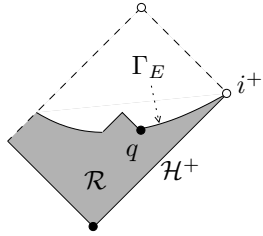
\[
\input{prop1bveo.pstex_t}
\]
This follows for otherwise one could clearly extend $J^-(\Gamma_E)$
to a larger past set such that the estimates of $\mathcal{R}$ are satisfied,
i.e.~$\mathcal{R}\not\subset J^-(\Gamma_E)$.

Thus, to show the desired property of $\Gamma_{E}$,
it suffices to show that
at any point $q$ on $\Gamma_{E}$ such that
$(\ref{toshow})$ holds, $\Gamma_{E}$ is in fact spacelike
at $q$.

We note first that since $\Gamma_{E}$ is achronal
and terminates at $i^+$, given any $\tilde{v}$, there exists
a $0<\tilde{u}$ 
such that $u\le \tilde{u}$ implies $v|_{\Gamma_E}(u)\ge\tilde{v}$.\footnote{See
the Appendix for an explanation of this notation.}
Thus, by $(\ref{topi})$ and $(\ref{nova})$, 
since
\begin{equation}
\label{supernova}
|\varpi(u,v)-\varpi_+|\le\hat{C}v^{-2p},
\end{equation}
it follows that given $\delta>0$,
we can choose $0<U_{1,2}\le U_{1,1}$ small enough so that
$\varpi(u,v)-\varpi_+\ge -\delta$.
Thus at $q$, the identity
$(\ref{toshow})$ together with $(\ref{alloxi})$
implies 
\begin{eqnarray*}
r_+-r&\ge&\frac{rr_+}{e^2}\left(\frac32(\tilde{K}+\tilde{M})
+(\varpi-\varpi_+)\right)\\
&\ge&\frac{c_1r_+}{e^2}\left(\frac32(\tilde{K}+\tilde{M})-\delta\right)\\
&\ge&\delta,
\end{eqnarray*}
where the last inequality follows for sufficiently small $\delta$.
The above inequality and $(\ref{supernova})$ imply 
$1-\mu<-L<0$ on $\Gamma_{E}\cap\mathcal{G}(U_{1,3})$ for some $L>0$,
$0<U_{1,3}\le U_{1,2}$.
The inequality $(\ref{nowhave})$ then implies that $-\lambda>L/2$.
Thus, from the identities
\[
\partial_u\left(\varpi-\frac{e^2}r\right)
=\nu\left(\frac{e^2}{r^2}+\frac12\left(\zn\right)^2(1-\mu)\right),
\]
\[
\partial_v\left(\varpi-\frac{e^2}r\right)
=\lambda\left(\frac{e^2}{r^2}+\frac12\theta^2\kappa^{-1}\right),
\]
it follows that for $u\le U_{1,4}\le U_{1,3}$,
\[
\partial_u\left(\varpi-\frac{e^2}r\right)<0,
\partial_v\left(\varpi-\frac{e^2}r\right)<0,
\]
where $U_{1,4}$ has been chosen so that $\left|\zn\right|$
and $|\theta|\kappa^{-1}$ are suitably small on 
$\Gamma_{E}\cap\mathcal{G}(U_{1,4})$,
in view of the bounds $(\ref{booteva})$ and $(\ref{bootduo})$.
Thus $\Gamma_{E}\cap\mathcal{G}(U_1)$ 
is spacelike and satisfies the requirements
of the proposition, for $U_1=U_{1,4}$. 

Finally, we remark that in $\mathcal{R}\cap\mathcal{G}(U_1)$, 
$\lambda$ is strictly monotone decreasing
as a function of $u$. Thus, 
\[
\{\lambda=0\}\cap \mathcal{R}\cap\mathcal{G}(U_1)=
\mathcal{A}\cap \mathcal{R}\cap
\mathcal{G}(U_1)
\]
is a graph over the event horizon. From the properties outlined in Section
\ref{monsec}, it now follows that $\mathcal{A}\cap \mathcal{R}\cap\mathcal{G}(U_1)$ 
is in fact an achronal curve
terminating at $i^+$. The final statement of the proposition follows
immediately. \end{proof}

\section{A ``no-shift'' region}
\label{nsrsec}
Proposition \ref{prwtnprotasn} has successfully brought us into the
trapped region $\mathcal{T}$. The importance of the large contribution
of the red-shift factor
in the analysis is clear.  
In this section, we shall investigate a region 
$\mathcal{N}\subset J^+(\Gamma_{E})$, where $(\ref{apeiro;})$ will be bounded.
This will thus be a \emph{no-shift} region, in the terminology of Section~\ref{ovvsec}.
In $\mathcal{N}$, the sign of $\varpi-\frac{e^2}r$ will change
from red-shift to blue-shift. $\mathcal{N}$ will thus bring
us into the blueshift region $\mathcal{B}$ which we shall investigate
in the next section.

Let $r_-$ denote the smaller root $r_-=\varpi_+-
\sqrt{\varpi_+^2-e^2}$ of the quadratic equation
$r^2-2\varpi_+r+e^2=0$.\footnote{This is the constant
value of $r$ on the Reissner-Nordstr\"om Cauchy horizon.} 
Choosing $\xi$ such that
\begin{equation}
\label{periergoasteri}
E>-\xi>\varpi_+-\frac{e^2}{r_-},
\end{equation}
we define our \emph{no-shift} region $\mathcal{N}$ as
follows:
\begin{eqnarray*}
\mathcal{N}&=&\left\{(u,v)\in J^+(\Gamma_{E}))
\cap\mathcal{G}(U_1):\right.\\
&&\hbox{}\left.
\left(\varpi-\frac{e^2}r\right)(\tilde{u},\tilde{v})
>-\xi{\rm\ for\ all\ }(\tilde{u},\tilde{v})\in J^-(u,v)\right\}
\end{eqnarray*}
Since $\mathcal{N}$ 
is clearly a past subset of $J^+(\Gamma_{E})$, 
its future boundary $\Gamma_{-\xi}$ in $\mathcal{G}(U_1)$
\[
\Gamma_{-\xi}=\overline{\mathcal{N}}\cap I^+(\mathcal{N})\cap\mathcal{G}(U_1)
\setminus\mathcal{N}
\]
is a (possibly empty) achronal curve.

We have
\begin{proposition}
\label{deutprotasn}
For $0<U_2$ sufficiently small,
$\Gamma_{-\xi}\cap\mathcal{G}(U_2)\ne\emptyset$, and
is a nonempty spacelike curve terminating at $i^+$. 
Moreover, on $\Gamma_{-\xi}\cap\mathcal{G}(U_2)$,
the equality
\[
\varpi-\frac{e^2}r=-\xi
\]
holds
identically.
\end{proposition}

\begin{proof}
Again, we prove this Proposition by a bootstrap argument.
We consider a region $\tilde{\mathcal{N}}$ defined as
the set of 
$(u,v)\in \mathcal{N}$ such that for all $(\tilde{u},\tilde{v})
\in J^-(\tilde{u},\tilde{v})\cap \mathcal{N}$,
the inequalities
\begin{equation}
\label{nBS4}
\tilde{v}-v|_{\Gamma_E}(\tilde{u})< H,
\end{equation}
\begin{equation}
\label{nBS5}
\varpi(\tilde{u},\tilde{v})<\Pi
\end{equation}
hold, for constants $H$, $\Pi$ to be determined below.
By the previous proposition, $\tilde{\mathcal{N}}$ is nonempty, 
provided $\Pi>\varpi_+$.
We will show that in $\overline{\tilde{\mathcal{N}}}\cap \mathcal{N}
\cap\mathcal{G}(U_2)$,
for small enough $U_2$, assumptions
$(\ref{nBS4})$ and $(\ref{nBS5})$ can in fact
be improved, for appropriate choice of $H$, $\Pi$.
It will follow
by a continuity argument that 
$\tilde{\mathcal{N}}\cap\mathcal{G}(U_2)=\mathcal{N}\cap\mathcal{G}(U_2)$.
The bounds obtained in $\tilde{\mathcal{N}}$ will easily imply the
assertion of the proposition.

Again, before giving the details of the proof,  it might
be useful to point out some of the key points. 
The bootstrap
assumption $(\ref{nBS4})$ together with $(\ref{F1})$
implies in particular
a bound for $\int\kappa dv$, when the integral is taken in $\tilde{\mathcal{N}}$,
hence the \emph{no-shift} property of $\mathcal{N}$.
Estimates for the scalar field derivatives $\theta$ and
$\zeta$ in this section are derived from
the Propositions of Section \ref{bvesec}; a discussion of these
is deferred till then. These estimates and
the boundedness of $\int\kappa dv$ then allow us
to control all other quantities, in particular
improving $(\ref{nBS5})$. The bootstrap assumption $(\ref{nBS4})$
is improved by noting that $|1-\mu|$ and $\kappa$
can be bounded below away from $0$, and thus 
\[
v-v|_{\Gamma_E}\sim\int_{v|_{\Gamma_E}}^v
{\kappa}\sim\int\lambda
\]
can be controlled by our upper bound on $r$.

Now for the proof in detail.
Note that the inequality
\begin{equation}
\label{3ava}
\varpi-\frac{e^2}{r}>-\xi
\end{equation}
together with $(\ref{nBS5})$ yield
\begin{equation}
\label{rbd}
r\ge e^2(\xi+\Pi)^{-1}.
\end{equation}

Consider $(u,v)\in\overline{\tilde{\mathcal{N}}}\cap \mathcal{N}$. 
Forming the characteristic rectangle
\[
J^+(u|_{\Gamma_E}(v),v|_{\Gamma_E}(u))\cap J^-(u,v)
\]
depicted
in the Penrose diagram below,
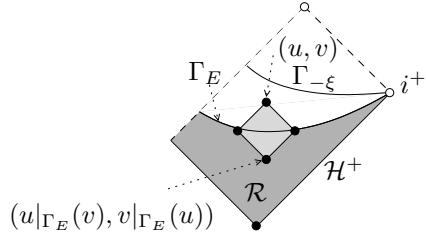
\[
\input{prop2veo.pstex_t}
\]
it is clear that $(\ref{nBS4})$ and $(\ref{nBS5})$ hold in
$J^+(u|_{\Gamma_E}(v),v|_{\Gamma_E}(u))\cap J^-(u,v)$, 
with the $\le$ sign replacing
the strict inequalities.
We claim that
\begin{eqnarray}
\label{tapia}
\nonumber
\int_{v|_{\Gamma_E}(u)}^v{|\theta|(u,\bar{v})d\bar{v}}
+\int_{u|_{\Gamma_E}(v)}^u{|\zeta|(\bar{u},v)d\bar{u}}\\
\nonumber
\hbox{\ }\le C_1(\xi,\Pi)\left(
\int_{v|_{\Gamma_E}(u)}^v{|\theta|(u|_{\Gamma_E}(v),\bar{v})d\bar{v}}
+\int_{u|_{\Gamma_E}(v)}^u{|\zeta|(\bar{u},v|_{\Gamma_E}(u))d\bar{u}}\right)\\
\hbox{\ }\le C_2(\xi,\Pi,H)v^{-p}.
\end{eqnarray}
The first of the above inequalities follows
from Proposition \ref{Estimates} of Section 13, to
be proved later,
while the second follows since, by $(\ref{booteva})$ and $(\ref{bootduo})$,
\[
\int_{v|_{\Gamma_E}(u)}^v|\theta|(u|_{\Gamma_E}(v),\bar{v})d\bar{v}
\le H\tilde{C}(v-H)^{-p}
\]
and
\begin{eqnarray*}
\int_{u|_{\Gamma_E}(v)}^u
{\left|\zn\right|(-\nu)(\bar{u},v|_{\Gamma_E}(\bar{u}))d\bar{u}}
&\le& \frac{3\tilde{C}r_+^2}{c_1E}(v-H)^{-p}\\
&&\hbox{}\cdot
\int_{u|_{\Gamma_E}(v)}^u{-\nu(\bar{u},v|_{\Gamma_E}(\bar{u}))d\bar{u}}\\
&\le& \hat{C}v^{-p}.
\end{eqnarray*}
Integrating $(\ref{integrating})$ with $(\ref{tapia})$, and applying
again $(\ref{booteva})$, one
obtains
\begin{eqnarray}
\label{yineyildiz}
\nonumber
\left|\zn\right|(u,v)&\le& \left(C_2v^{-p}(\xi+\Pi)e^{-2}
+\left|\zn\right|(u,v|_{\Gamma_{E}}(u))\right)
e^{\int_{v|_{\Gamma_{E}}}^v{2\xi
(\xi+\Pi)^2e^{-4}}}\\
&\le&C_3(\xi,\Pi,H)v^{-p}.
\end{eqnarray}
Since we have a bound $|1-\mu|\le C_4(\xi,\Pi)$,
integrating $(\ref{puqu})$, in view
of $(\ref{supernova})$, we obtain a bound
\begin{equation}
\label{posoakoma;}
\varpi-\varpi_+\le C_5(\xi,\Pi,H)v^{-2p}.
\end{equation}
In particular, $(\ref{posoakoma;})$
improves $(\ref{nBS5})$ once $u$ is restriced so as $u<U_{2,1}$ for some 
$U_{2,1}>0$, as long
as $\Pi$ is chosen so that $\Pi>\varpi_+$. (Recall that
since $\Gamma_{E}$ is spacelike and terminates at $i^+$,
given $\tilde{v}$, there exists $\tilde{u}$ such that
$u<\tilde{u}$ implies $v>\tilde{v}$ for $(u,v)\in J^+(\Gamma_{E})$.)
Thus, there is $c(\xi,E)>0$ such that
for $u<U_{2,2}$ for some $0<U_{2,2}\le U_{2,1}$, $(\ref{posoakoma;})$,
$(\ref{periergoasteri})$, and
$(\ref{3ava})$ together imply that
\begin{equation}
\label{LLL}
|1-\mu|(u,v)>c.
\end{equation}
Integrating $(\ref{fdb1})$ yields
\[
\kappa\ge\frac12e^{-\tilde{C}_3(\xi,\Pi,H)v^{-2p}},
\]
and hence
we can select $0<U_2\le U_{2,2}$ so that for $u\le U_2$ we also have
\[
\kappa\ge\frac13.
\]
This yields
\[
-\int_{v|_{\Gamma_{E}}(u)}^v\lambda(u,\bar{v})d\bar{v}\ge 
c\int_{v|_{\Gamma_{E}}(u)}^v\kappa(u,\bar{v})d\bar{v}
\ge\frac{c}3(v-v|_{\Gamma_{E}}(u)).
\]
If $H$ was selected so that $H>\frac3cr_+$, the above inequality
improves the bootstrap assumption $(\ref{nBS4})$.
Thus, since
\[
J^-\left(\overline{\tilde{\mathcal{N}}}\right)\cap \mathcal{N}
\subset
\overline{J^-(\tilde{\mathcal{N}})}\cap \mathcal{N}
=
\overline{\tilde{\mathcal{N}}}\cap \mathcal{N},
\]
we have shown
\begin{equation}
\label{kleisto}
\overline{\tilde{\mathcal{N}}}
\cap \mathcal{N}\cap\mathcal{G}(U_2)\subset{\tilde{\mathcal{N}}}.
\end{equation}
Since $\tilde{\mathcal{N}}$ is clearly open, $(\ref{kleisto})$ implies
that $\tilde{\mathcal{N}}$ is both open and closed as a subset of $\mathcal{N}$ in
the latter set's induced topology. Thus, since $\mathcal{N}$ is connected,
we have that
\[
\tilde{\mathcal{N}}\cap\mathcal{G}(U_2)=\mathcal{N}\cap\mathcal{G}(U_2).
\]
A similar argument as in the proof of Proposition \ref{prwtnprotasn} now
verifies that $\Gamma_{-\xi}$ is spacelike
and $\varpi-\frac{e^2}r=-\xi$ holds identically
on $\Gamma_{-\xi}$.
\end{proof}

It should be noted that one can derive a pointwise bound
for $\tl$ on $\Gamma_{-\xi}$, in analogy to $(\ref{yineyildiz})$.
Note first that $(\ref{LLL})$ implies 
\begin{equation}
\label{profaves}
|\lambda|>c/3
\end{equation}
in $\mathcal{N}$. Now
$(\ref{sign1})$ together with $(\ref{profaves})$ imply
\[
|\theta(u,v)|\le|\theta(u|_{\Gamma_{E}}(v),v)|
+C_6\int_{u|_{\Gamma_{E}}(v)}^{u}{|\zeta(\bar{u},v)|
d\bar{u}}.
\]
Thus, by $(\ref{tapia})$
we obtain
\begin{equation}
\label{toavalogo}
\left|\tl\right|\le c|\theta|\le C_7(\xi,E)v^{-p}
\end{equation}
on $\Gamma_{-\xi}$.

\section{The stable blue-shift region}
\label{sbssec}
In the red-shift region $\mathcal{R}$ and the ``no-shift'' region
$\mathcal{N}$, the analysis has been relatively simple,
in $\mathcal{R}$ because the unbounded factor 
$(\ref{apeiro;})$
appears with a favorable sign for controlling $\varpi$, 
while in $\mathcal{N}$ because $(\ref{apeiro;})$ is
uniformly bounded. The next region we will consider is more delicate; it 
is the first in which we will have to
deal with an unbounded 
$(\ref{apeiro;})$  
carrying
the unfavorable (as to controlling $\varpi$)
``blue-shift'' sign.

The next proposition will
refer to a curve $\gamma\subset\mathcal{G}(U_2)$
defined by the relation
\begin{equation}
\label{gammadef}
v|_\gamma-v|_\Gamma=\alpha\log v|_\gamma
\end{equation}
where $\alpha$ is some positive constant.
As this curve can be written $v|_\gamma=f(v|_\Gamma)$,
where $f(v)$ is the inverse of $x-\alpha\log x$,
and $f'>0$ for large enough $v$, it 
follows that $v|_\gamma$ decreases
in $u$.
Thus $\gamma$ is easily seen to be spacelike.
We have:

\begin{proposition}
\label{sta9mpleprot}
For $0<U_3$ sufficiently small, and for suitable choice of $\alpha$
in $(\ref{gammadef})$,
we have $\gamma\cap\mathcal{K}(U_3)\subset\mathcal{G}(U_3)$, 
$-\lambda\le av^{-s}$ on 
$\gamma\cap\mathcal{G}(U_3)$
for some $s>1$ and some $a$, and $r(u,v|_\gamma(u))
\to r_-$ as $u\to0$. 
\end{proposition}

\begin{proof}
We define the region $\mathcal{B}_\gamma$ by
\[
\mathcal{B}_\gamma=J^+(\Gamma_{-\xi})
\cap J^-(\gamma)
\]
and the region $\tilde{\mathcal{B}}_\gamma$ to be
the set of all $(u,v)\in \mathcal{B}_\gamma\cap\mathcal{G}(U_2)$ for which
for the following inequalities hold for all
$(\tilde{u},\tilde{v})\in J^-(u,v)\cap \mathcal{B}_\gamma$:
\begin{equation}
\label{neBS2}
-\xi-\epsilon<\left(\varpi-\frac{e^2}r\right)(\tilde{u},\tilde{v})<-\xi+\epsilon,
\end{equation}
\begin{equation}
\label{neBS3}
r_--\epsilon<r(\tilde{u},\tilde{v})<r_-+\epsilon.
\end{equation}
The choice of $\xi$, $\epsilon$ will be decided
in the course of the proof. In particular, $\xi$ will be chosen
such that $\tilde{\mathcal{B}}_\gamma$ is non-empty; for this it suffices
that 
on $\Gamma_{-\xi}$,
\begin{equation}
\label{upo91}
r<r_-+\epsilon.
\end{equation}
On the other hand, $\xi$ will be such that 
\begin{equation}
\label{upo92}
-\xi-\epsilon<\varpi_+-\frac{e^2}{r_-}.
\end{equation}
In $\overline{\tilde{\mathcal{B}}_\gamma}\cap\mathcal{G}(U_3)$, 
for sufficiently small $U_3$,
we shall be able to estimate all quantities.
In particular, we shall be able to improve $(\ref{neBS2})$
and $(\ref{neBS3})$; by a continuity argument and
Theorem \ref{epekt9e}, this will give
\begin{equation}
\label{mperdemevo}
\tilde{\mathcal{B}}_\gamma\cap\mathcal{G}(U_3)=
\mathcal{B}_\gamma\cap\mathcal{G}(U_3)=
\tilde{B}_\gamma\cap\mathcal{K}(U_3).
\end{equation}
Our estimates in $\tilde{\mathcal{B}}_\gamma$ will thus apply up until the
curve $\gamma$ (proving this will be part 1), 
and this will allow us to deduce a bound for
$\lambda$ (proving this will be part 2).
The bound for $r$ will then follow easily.

We begin with part 1. We will prove $(\ref{mperdemevo})$ for
all choices of $\xi$, $\epsilon$ satisfying $(\ref{upo91})$
on $\Gamma_{-\xi}$,
and $(\ref{upo92})$.
Our situation is somewhat opposite to that
of the previous theorem: For here we are given, by $(\ref{gammadef})$,
the maximum $v$-dimension of
the region $\mathcal{B}_\gamma$, whereas there the $v$-dimension was a bootstrap
assumption that we had to improve. Very briefly, the bounds
for $\theta$ and $\zeta$ given by Proposition \ref{Estimates},
together with $(\ref{gammadef})$,
after suitable restriction of $\alpha$, give
us the smallness necessary to retrieve $(\ref{neBS2})$ and
$(\ref{neBS3})$, as well as to prove that $\kappa\sim1$.

Consider a point $(u,v)\in \overline{\tilde{\mathcal{B}}_\gamma}
\cap \mathcal{G}(U_2)$ and the 
characteristic rectangle 
\[
J^+(u|_{\Gamma_{-\xi}}(v),v|_{\Gamma_{-\xi}}(u))\cap J^-(u,v)
\]
depicted below:
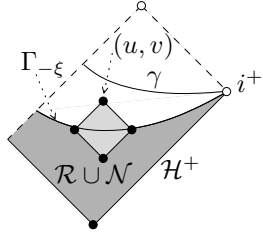
\[
\input{prop3veo.pstex_t}
\]
The estimates $(\ref{neBS2})$, $(\ref{neBS3})$ hold on 
$J^+(\Gamma_{-\xi})\cap J^-(u,v)$, where
the $\le$ sign replaces the strict inequalities. 
We wish to determine bounds on 
\[
\int_{v|_{\Gamma_{-\xi}}(u)}^v{|\theta|(u,\bar{v})d\bar{v}}
\] 
and
\[
\int_{u|_{\Gamma_{-\xi}(v)}}^u{|\zeta|(\bar{u},v)d\bar{u}}. 
\]
By Proposition \ref{Estimates}, in view
of the bound $(\ref{neBS3})$,
we have
\begin{eqnarray}
\label{wehave}
\int_{v|_{\Gamma_{-\xi}}(u)}^v{|\theta|(u,\bar{v})d\bar{v}}
+\int_{u|_{\Gamma_{-\xi}(v)}}^u{|\zeta|(\bar{u},v)d\bar{u}}
\\
\nonumber
\hbox{\ }\le C_1
\left(\int_{v|_{\Gamma_{-\xi}}(u)}^v{|\theta|(u|_{\Gamma_{-\xi}}(v)
,\bar{v})d\bar{v}}
+\int_{u|_{\Gamma_{-\xi}(v)}}^u{|\zeta|(\bar{u},v|_{\Gamma_{-\xi}}(u))d\bar{u}}\right),
\end{eqnarray}
where $C_1=C_1(r_--\epsilon)$.
Now,
\begin{eqnarray*}
\int_{v|_{\Gamma_{-\xi}}(u)}^v{|\theta|(u|_{\Gamma_{-\xi}}(v)
,\bar{v})d\bar{v}}&\le&\int_{v|_{\Gamma_{-\xi}}(u)}^v{Cv^{-p}}\\
&\le& C_2(v-\log{(v+\log v^\alpha)^\alpha})^{-p}\log{v^\alpha}\\
&\le& 
C_3(C_2,\alpha)v^{-p}\log v^\alpha.
\end{eqnarray*}
On the other hand, 
\begin{eqnarray*}
\int_{u|_{\Gamma_{-\xi}(v)}}^u{|\zeta|(\bar{u},v|_{\Gamma_{-\xi}}(u))d\bar{u}}
&\le&
\int_{u|_{\Gamma_{-\xi}(v)}}^u{\left|\zn\right|(-\nu)
(\bar{u},v|_{\Gamma_{-\xi}}(u))d\bar{u}}\\
&\le& C_4\left(v|_{\Gamma_{-\xi}}(u)\right)^{-p}\le C_5v^{-p}.
\end{eqnarray*}
It follows then from $(\ref{wehave})$
that 
\begin{equation}
\label{KAIauto}
\int_{v|_{\Gamma_{-\xi}}(u)}^v{|\theta|(u,\bar{v})d\bar{v}}
+\int_{u|_{\Gamma_{-\xi}(v)}}^u{|\zeta|(\bar{u},v)d\bar{u}}
\le
C_6v^{-p}\log v^\alpha.
\end{equation}

Integrating the equation $(\ref{integrating})$ and using the bounds 
proved immediately above, together with $(\ref{yineyildiz})$,
yields the inequality
\begin{eqnarray}
\label{exeiakoma}
\nonumber
\sup_{\bar{u}\in\left[u|_{\Gamma_{-\xi}}(v),u\right]}\left|\zn(\bar{u},v)\right|
&\le& \left(\frac{C_6}{r_--\epsilon}
v^{-p}\log v+C_7v^{-p}\right)e^{\frac{2(\xi+\epsilon)}{(r_--\epsilon)^2}
\log v^\alpha}\\
&\le& C_8v^{-p+\frac{2(\xi+\epsilon)}{(r_--\epsilon)^2}\alpha}
\log v^\alpha.
\end{eqnarray}

From this, it is quite easy to improve the bounds $(\ref{neBS2})$ and
$(\ref{neBS3})$ at $(u,v)$. 
Choose $\alpha=\alpha(\xi,\epsilon)$ so that
\begin{equation}
\label{epilogn1}
-p+\frac{(\xi+\epsilon)}{(r_--\epsilon)^2}\alpha<0,
\end{equation}
and
integrate $(\ref{puqu})$,
written
as
\[
\partial_u\varpi=-\frac12(1-\mu)\left|\zn\right||\zeta|,
\]
along $\left[u|_{\Gamma_{-\xi}}(v),u\right]\times\{v\}$.
Given $\epsilon_2>0$, by 
our bounds $(\ref{KAIauto})$ and
$(\ref{exeiakoma})$, and 
an upper bound\footnote{This bound does not depend on $\xi$, $\epsilon$,
as long as $\epsilon$ is small enough.} on $1-\mu$ following from $(\ref{neBS2})$ and
$(\ref{neBS3})$,
%\[
%\varpi\le C_9(\xi,\epsilon),
%\]
we obtain
\begin{equation}
\label{togetherwith}
\varpi(u,v)\le\varpi_++\epsilon_2,
\end{equation}
so long as $u\le U_{3,1}$ for some
$0<U_{3,1}(\epsilon_2,\ldots)$. 
But now, the inequality
$1-\mu<0$ yields
\[
(r-(\varpi-\sqrt{\varpi^2-e^2}))
(r-(\varpi+\sqrt{\varpi^2-e^2}))<0,
\]
and since the second factor is negative,
\[
r\ge\varpi-\sqrt{\varpi^2-e^2}.
\]
For $\epsilon_2$ small enough, this improves $(\ref{neBS3})$,
and then, applying this better bound for $r$ together with the bound
$(\ref{togetherwith})$,
improves $(\ref{neBS2})$, in view of the restriction
$(\ref{upo92})$.

Since 
\[
J^-\left(\overline{\tilde{\mathcal{B}}_\gamma}\right)
\cap J^+(\Gamma_{-\xi})
\subset
\overline{\tilde{\mathcal{B}}_\gamma},
\]
we
have in fact shown that
\[
\overline{\tilde{\mathcal{B}}_\gamma}\cap\mathcal{G}(U_{3,1})
\subset
\tilde{\mathcal{B}}_\gamma.
\]
Thus $\tilde{\mathcal{B}}_\gamma\cap\mathcal{G}(U_{3,1})$ is an open and closed set
in the topology of $\mathcal{B}_\gamma\cap\mathcal{G}(U_{3,1})$. 
Since the latter set is connected, it follows
that
\[
\tilde{\mathcal{B}}_\gamma\cap\mathcal{G}(U_{3,1})=\mathcal{B}_\gamma
\cap\mathcal{G}(U_{3,1}).
\]
Now, with respect to the topology of $\mathcal{K}(U_{3,1})\cap J^+(\Gamma_{-\xi})$,
we have
\[
\partial\overline{\mathcal{B}_\gamma\cap\mathcal{G}(U_{3,1})}
\subset(\gamma\cap\mathcal{K}(U_{3,1}))\cup\partial\overline{\mathcal{G}(U_{3,1})}.
\]
On the other hand, since $(\ref{neBS3})$ holds on 
$\overline{\mathcal{B}_\gamma\cap\mathcal{G}(U_{3,1})}$, with the $\le$
replacing the strict inequality, it follows by Theorem
\ref{epekt9e} that 
\[
\partial\overline{\mathcal{B}_\gamma\cap\mathcal{G}(U_{3,1})}
\cap
\partial\overline{\mathcal{G}(U_{3,1})}=\emptyset,
\]
and thus 
\[
\partial\overline{\mathcal{B}_\gamma\cap\mathcal{G}(U_{3,1})}
\subset\gamma\cap\mathcal{K}(U_{3,1}).
\]
Consequently, we have
\[
B_\gamma\cap\mathcal{G}(U_{3,1})=
B_\gamma\cap\mathcal{K}(U_{3,1}),
\]
i.e.~$\gamma\cap\mathcal{K}(U_{3,1})\subset\mathcal{G}(U_{3,1})$, and
the estimates
$(\ref{neBS2})$ and $(\ref{neBS3})$ hold in 
$\mathcal{B}_\gamma\cap\mathcal{K}(U_{3,1})$.

Integrating $(\ref{fdb1})$ and using $(\ref{exeiakoma})$ now 
yields\footnote{Note that, the bounds on $\zn$ in
$J^{-}(\Gamma_{-\xi})$ show that for $u\le U_{3,2}'$,
$\kappa\ge1-\frac{\epsilon}2$ in that region.}
that for $u\le U_{3,2}$, for some $0<U_{3,2}\le U_{3,1}$,
\begin{equation}
\label{oexomev}
\kappa\ge1-\epsilon.
\end{equation}
This concludes part 1.

Part 2 requires considerable care. As it is $\lambda$ which we desire
to bound in absolute value on $\gamma$, we certainly need
a lower bound on 
\begin{equation}
\label{posot1}
\int_{u|_{\Gamma_{-\xi}}(v)}^u\frac{\nu}{1-\mu}(\bar{u},v)d\bar{u},
\end{equation}
a quantity we have not as of yet
controlled.\footnote{It is $\lambda$ on $\gamma$, 
and not $\nu$, that we estimate, because
it is $\lambda$, and not $\nu$, that we control pointwise
on $\Gamma_{-\xi}$.} What we have, however, in view of $(\ref{oexomev})$,
is a lower bound on 
\begin{equation}
\label{posot2}
\int_{v|_{\Gamma_{-\xi}}(u)}^v \kappa(u,\bar{v})d\bar{v}.
\end{equation}
It turns out that $(\ref{posot1})$ and $(\ref{posot2})$ are
in fact very related, but
this relation only becomes apparent by comparing with
a special ``zigzag''-like curve which is contained in 
$J^+(\Gamma_E)\cap J^-(\Gamma_{-\xi})$, 
where $1-\mu$ is bounded
away from zero. Exploiting such a curve, we will
obtain a lower bound for $(\ref{posot1})$ from
the lower bound for $(\ref{posot2})$. As we shall see,
there is little room for loss in this argument,
as $\alpha$ is constrained already by 
$(\ref{epilogn1})$. Indeed, our margin of error
is precisely $\delta=p-\frac12$.

We begin now the proof of part 2.
Our first task is to relate $(\ref{posot1})$ and $(\ref{posot2})$.
Refer to the Penrose diagram below:
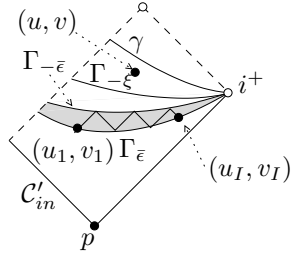
\[
\input{konta.pstex_t}
\]
Given a small $\overline\epsilon>0$, 
we fix the two curves\footnote{These two curves are given by
Proposition \ref{deutprotasn}, where $\overline\epsilon$ and $-\overline\epsilon$
take the place of $-\xi$.} 
$\Gamma_{\overline\epsilon}$
and $\Gamma_{-\overline\epsilon}$, and given 
\[
(u,v)\in
\mathcal{B}_\gamma\cap\mathcal{G}(U_{3,2}),
\]
we consider the $(\Gamma_{\overline\epsilon}, 
\Gamma_{-\overline\epsilon},u,v)$-zigzag\footnote{See the Appendix
for an explanation of this notation.}:
\[
\bigcup_{i=1}^{I-1}\{u_i\}\times[v_i,v_{i+1}]\cup[u_{i+1},u_i]\times\{v_{i+1}\}
\]
We have
\begin{equation}
\label{biryildiz}
1-\hat{C}\overline\epsilon
\le\frac{\sum_{i=1}^{I-1}\int_{v_i}^{v_{i+1}}\lambda(u_i,v) dv}
{\sum_{i=1}^{I-1}\int_{u_{i+1}}^{u_i}\nu(u,v_{i+1}) du}\le
1+\hat{C}\overline\epsilon,
\end{equation}
where $\hat{C}>0$ is a constant independent of $\overline\epsilon$.
To prove $(\ref{biryildiz})$, note first that in 
$J^+(\Gamma_{\overline\epsilon})\cap 
J^-(\Gamma_{\overline\epsilon})\cap
\mathcal{G}(U_{3,2})$
\begin{equation}
\label{notefirst}
1-\varpi_+^2e^{-2}-\hat{\epsilon}
<
1-\mu
<
1-\varpi_+^2e^{-2}+\hat{\epsilon},
\end{equation}
with $\hat\epsilon\to0$ as $\overline\epsilon\to0$. Thus, for
sufficiently small $\overline\epsilon$,
\begin{eqnarray}
\label{aptnmia}
\nonumber
-\sum_{i=1}^{I-1}\int_{v_i}^{v_{i+1}}\lambda(u_i,\bar{v}) d\bar{v}
&\ge&\frac12\left(\varpi_+^2e^{-2}-1\right)
\sum_{i=1}^{I-1}\int_{v_i}^{v_{i+1}}\kappa(u_i,\bar{v}) d\bar{v}\\
\nonumber
&\ge&\frac12\left(\varpi_+^2e^{-2}-1\right)
\int_{v|_{\Gamma_{-\overline\epsilon}}(u)}^{
v|_{\Gamma_{-\xi}}(u)}\kappa(u,\bar{v}) d\bar{v}\\
\nonumber
&\ge&\frac16\left(\varpi_+^2e^{-2}-1\right)
\left(v|_{\Gamma_{-\xi}}(u)-
v|_{\Gamma_{-\overline\epsilon}}(u)\right)\\
&\ge&h>0,
\end{eqnarray}
for some $h>0$ that can be chosen independent of $\overline\epsilon$,
and $\xi$, once one restricts to sufficiently small 
$\overline\epsilon$
and $\frac{e^2}{r_-}-\varpi_+-\xi$.

On the other hand,
\begin{eqnarray}
\label{aptnvalln}
\nonumber
\left|-\sum_{i=1}^{I-1}\int_{u_{i+1}}^{u_i}\nu(\bar{u},v_{i+1}) d\bar{u}
+\sum_{i=1}^{I-1}\int_{v_i}^{v_{i+1}}\lambda(u_i,\bar{v}) d\bar{v}\right|\\
\nonumber
\hbox{\ }\le\sup_{x,y\in J^+(\Gamma_{\overline\epsilon})\cap
J^-(\Gamma_{-\overline\epsilon})}{(r(x)-r(y))}\\
\hbox{\ }\le C\overline{\epsilon},
\end{eqnarray}
for some $C$ independent of $\overline\epsilon$.
Dividing $(\ref{aptnvalln})$ by 
\begin{equation}
\label{diairesn}
\sum_{i=1}^{I-1}\int_{u_{i+1}}^{u_i}(-\nu)(\bar{u},v_{i+1}) d\bar{u},
\end{equation}
noting
that, by addition of $(\ref{aptnmia})$ and $(\ref{aptnvalln})$,
an inequality similar to $(\ref{aptnmia})$ applies to
$(\ref{diairesn})$,
we obtain $(\ref{biryildiz})$.

The bounds $(\ref{notefirst})$ now give
\begin{eqnarray}
\label{ikiyildiz}
\nonumber
\int_{v|_{\Gamma_{-\overline\epsilon}}(u)}^{v}
\kappa (u,\bar{v})d\bar{v}&\le&
\sum_{i=1}^{I-1}\int_{v_i}^{v_{i+1}}\kappa(u_i,\bar{v}) d\bar{v}\\
\nonumber
&\le&
\frac{-1}{1-\varpi_+^2e^{-2}+\hat{\epsilon}}
\sum_{i=1}^{I-1}\int_{v_i}^{v_{i+1}}(-\lambda)(u_i,\bar{v}) d\bar{v}\\
\nonumber
&\le&\frac{1-\varpi_+^2e^{-2}-\hat{\epsilon}}
{1-\varpi_+^2e^{-2}+\hat{\epsilon}}
\sum_{i=1}^{I-1}\int_{v_i}^{v_{i+1}}\kappa(u_i,\bar{v}) d\bar{v}\\
&\le&
\frac{1}{1-\epsilon}
\frac{1-\varpi_+^2e^{-2}-\hat{\epsilon}}
{1-\varpi_+^2e^{-2}+\hat{\epsilon}}
\int_{v|_{\Gamma_{-\overline\epsilon}}(u)}^{v}
\kappa(u,\bar{v})d\bar{v},
\end{eqnarray}
where the last inequality follows from $(\ref{oexomev})$.
If we can show
\begin{equation}
\label{ifwe}
\sum_{i=1}^{I-1}\int_{u_{i+1}}^{u_i}(-\nu)(\bar{u},v_{i+1}) d\bar{u}\sim
\int_{u|_{\Gamma_{-\overline\epsilon}}(v)}^{u}
\frac{\nu}{1-\mu}(\bar{u},v)d\bar{u},
\end{equation}
then it will follow
from $(\ref{ikiyildiz})$ and $(\ref{biryildiz})$ that
\begin{equation}
\label{thenit}
\int_{v|_{\Gamma_{-\overline\epsilon}}(u)}^{v}
\kappa(u,\bar{v})d\bar{v}\sim
\int_{u|_{\Gamma_{-\overline\epsilon}}(v)}^u
\frac{\nu}{1-\mu}(\bar{u},v)d\bar{u}.
\end{equation}
For this, we must bound
\[
\int_{v|_{\Gamma_{-\bar\epsilon}}(u)}^v
{\left(\frac\theta\lambda\right)^2(-\lambda)(u,\bar{v})d\bar{v}}.
\]
This is done by a similar procedure.
Note that we do have a bound
\[
\int_{u|_{\Gamma_{-\overline\epsilon}}(v)}^{u}
\frac{\nu}{1-\mu}(\bar{u},v)d\bar{u}\le 
\frac{-1}{1-\varpi_+^2e^{-2}+\hat{\epsilon}}
\sum_{i=1}^{I-1}\int_{u_{i+1}}^{u_i}(-\nu)(\bar{u},v_{i+1}) d\bar{u},
\]
and thus, by $(\ref{biryildiz})$ and $(\ref{notefirst})$,
\begin{equation}
\label{suvepws}
\sum_{i=1}^{I-1}\int_{u_{i+1}}^{u_i}\frac\nu{1-\mu}(\bar{u},v_{i+1}) d\bar{u}\le 
\frac{1}{1-\hat{C}\overline\epsilon}
\frac{1-\varpi_+^2e^{-2}-\hat{\epsilon}}
{1-\varpi_+^2e^{-2}+\hat{\epsilon}}
\sum_{i=1}^{I-1}\int_{v_i}^{v_{i+1}}\kappa(u_i,\bar{v}) d\bar{v}.
\end{equation}
Integrating the equation
\begin{equation}
\label{toallo}
\partial_u\frac\theta\lambda=
-\frac\zeta{r}-\left(\frac\theta\lambda\right)\frac\nu{1-\mu}\frac2{r^2}
\left(\varpi-\frac{e^2}r\right),
\end{equation}
using $(\ref{suvepws})$, $(\ref{KAIauto})$, $(\ref{neBS3})$,
and $(\ref{toavalogo})$, we obtain
\[
\left|\frac\theta\lambda\right|(u,v)\le\left(
Cv^{-p}\log v^\alpha+C_8v^{-p}\right)e^{A\log v^\alpha}\le 
\tilde{C}v^{-p+A\alpha}\log v^\alpha
\]
and thus, using also $(\ref{KAIauto})$,
\begin{eqnarray*}
\int_{v|_{\Gamma_{-\bar\epsilon}}(u)}^v
{\left(\frac\theta\lambda\right)^2(-\lambda)(u,\bar{v})d\bar{v}} &\le& 
\left(\sup_{\mathcal{N}\cup \mathcal{B}_\gamma}
\left|\tl\right|\right)
\int_{v|_{\Gamma_{-\bar\epsilon}}(u)}^v
{|\theta|(u,\bar{v})d\bar{v}}\\
&\le&{\tilde C}'
v^{-2p+A\alpha}(\log v^\alpha)^2,
\end{eqnarray*}
where 
\[
A=\frac{2(\xi+\epsilon)}{(r_--\epsilon)^2}
\frac{1}{1-\hat{C}\overline\epsilon}
\frac{1-\varpi_+^2e^{-2}-\hat{\epsilon}}
{1-\varpi_+^2e^{-2}+\hat{\epsilon}},
\]
and $\tilde{C}'=\tilde{C}'(\xi)$
with $\tilde{C}'\to\infty$ as $\xi\to\frac{e^2}{r_-}-\varpi_+$.
Choosing $\xi$, $\epsilon$, $\overline\epsilon$, $\hat{\epsilon}$,
$\alpha$ so that in addition to $(\ref{epilogn1})$ we also
have
\begin{equation}
\label{epilogn2}
A\alpha<2p,
\end{equation}
we can choose $U_{3,3}\le U_{3,2}$ so that
\[
e^{\int_{v|_{\Gamma_{-\bar\epsilon}}(u)}^v
{\frac1r\left|\frac\theta\lambda\right|^2(-\lambda)(u,\bar{v})d\bar{v}}}
\le1+\epsilon
\]
for $(u,v)\in J^{-}(\gamma)\cap J^+(\Gamma_{-\bar\epsilon})\cap\mathcal{G}(U_{3,3})$.

Integrating now $(\ref{fdb2})$, we obtain $(\ref{ifwe})$;
specifically,
\begin{eqnarray*}
\int_{u|_{\Gamma_{-\overline\epsilon}}(v)}^u
\frac{\nu}{1-\mu}(u,\bar{v})du&\le&
\sum_{i=1}^{I-1}\int_{u_{i+1}}^{u_i}\frac\nu{1-\mu}(\bar{u},v_{i+1}) d\bar{u}\\
&\le&
\frac{-1}
{1-\varpi_+^2e^{-2}+\hat{\epsilon}}
\sum_{i=1}^{I-1}\int_{u_{i+1}}^{u_i}(-\nu)(\bar{u},v_{i+1}) d\bar{u}\\
&\le&\frac{1-\varpi_+^2e^{-2}-\hat{\epsilon}}
{1-\varpi_+^2e^{-2}+\hat{\epsilon}}
\sum_{i=1}^{I-1}\int_{u_{i+1}}^{u_i}\frac{\nu}{1-\mu}(\bar{u},v_{i+1}) d\bar{u}\\
&\le&(1+\epsilon)
\frac{1-\varpi_+^2e^{-2}-\hat{\epsilon}}
{1-\varpi_+^2e^{-2}+\hat{\epsilon}}
\int_{u|_{\Gamma_{-\overline\epsilon}}(v)}^{u}
\frac{\nu}{1-\mu}(\bar{u},v)d\bar{u},
\end{eqnarray*}
and thus
$(\ref{thenit})$;
in particular, for $(u,v)\in\gamma$, we have
\begin{eqnarray}
\label{kietsi}
\nonumber
\int_{u|_{\Gamma_{-\overline\epsilon}}(v)}^u
\frac{\nu}{1-\mu}&\ge&
\frac{1}{1+\epsilon}
\frac{1-\varpi_+^2e^{-2}+\hat{\epsilon}}
{1-\varpi_+^2e^{-2}-\hat{\epsilon}}
\frac{1}{1+\hat{C}\overline\epsilon}\sum_{i=1}^{I-1}\int_{v_i}^{v_{i+1}}\kappa
(u_i,\bar{v}) d\bar{v}\\
&\ge&
\frac{1}{1+\epsilon}
\frac{1-\varpi_+^2e^{-2}+\hat{\epsilon}}
{1-\varpi_+^2e^{-2}-\hat{\epsilon}}
\frac{1}{1+\hat{C}\overline\epsilon}(1-\epsilon)\alpha\log{v}.
\end{eqnarray}
On the other hand,
by our bound $(\ref{LLL})$ which holds in
particular in 
$J^+(\Gamma_{-\overline{\epsilon}})\cap
J^-(\Gamma_{-\xi})\cap\mathcal{G}(U_{3,3})$,
and our bounds on $r$, we have
\begin{equation}
\label{kialliws}
\int_{u|_{\Gamma_{-\overline\epsilon}}(v)}^
{u|_{\Gamma_{-\xi}}(v)}
\frac{\nu}{1-\mu}(\bar{u},v)d\bar{u}
\le\hat{a},
\end{equation}
where $\hat{a}=\hat{a}(\xi)$.
Putting $(\ref{kietsi})$ and $(\ref{kialliws})$ together
gives, for $(\bar{u},\bar{v})\in\gamma$,
\begin{equation}
\label{autoeivai}
\int_{u|_{\Gamma_{-\xi}}(\bar{v})}^{\bar{u}}
\frac{\nu}{1-\mu}\ge
\frac{1}{1+\epsilon}
\frac{1-\varpi_+^2e^{-2}+\hat{\epsilon}}
{1-\varpi_+^2e^{-2}-\hat{\epsilon}}
\frac{1}{1+\hat{C}\overline\epsilon}(1-\epsilon)\alpha\log{v}-\hat{a}.
\end{equation}

Integrating $(\ref{lqu})$ from $\Gamma_{-\xi}$ to $\gamma$, 
in view of the bound $(\ref{profaves})$
on $\Gamma_{-\xi}$, $(\ref{autoeivai})$, $(\ref{neBS2})$ and
$(\ref{neBS3})$,
 we now obtain that for $(u,v)\in
\gamma$,
\begin{equation}
\label{sump}
-\lambda(u,v)\le 
ae^{-A(1-\hat{C}\overline\epsilon)\frac{\xi-\epsilon}{\xi+\epsilon}
\left(\frac{r_+-\epsilon}{r_++\epsilon}\right)^2(1-\epsilon)\log v^{\alpha}},
\end{equation} 
where 
\[
a=ce^{\hat{a}2\frac{\xi+\epsilon}{(r_--\epsilon)^2}}.
\]
Since $p>\frac12$, we could have selected our quantities
so that in addition to $(\ref{epilogn1})$ and $(\ref{epilogn2})$,
we also have
\[
A\alpha(1-\hat{C}\overline\epsilon)
\frac{\xi-\epsilon}{\xi+\epsilon}
\left(\frac{r_+-\epsilon}{r_++\epsilon}\right)^2(1-\epsilon)
>1
\] 
For such a choice, we obtain
\begin{equation}
\label{kalolambda}
-\lambda\le av^{-s}
\end{equation}
for an $s>1$, as desired.
To show the statement about $r$, first note
that $\lambda\to0$ on $\gamma$, together with
$(\ref{oexomev})$,
implies $1-\mu\to0$ on $\gamma$. On the other hand,
$(\ref{exeiakoma})$ and $(\ref{puqu})$ imply that on $\gamma$,
$\varpi\to\varpi_+$. Since $1-\mu=1-\frac{2\varpi}r+\frac{e^2}r$,
and $r$ is bounded away from $r_+$, it follows that
$r\to r_-$ on $\gamma$, as $v\to\infty$. The proposition is thus proven
with $U_3=U_{3,3}$.
\end{proof}

\section{Beyond the stable blue shift region}
\label{bsbsec}
We are now ready to prove
\begin{theorem} 
\label{pera-r}
There exists a $U_4>0$ such that 
$\mathcal{G}(U_4)=\mathcal{K}(U_4)$. 
Moreover, $r$ can be extended
by monotonicity to a function on $\mathcal{CH}^+$ in the topology of the
Penrose diagram of $\mathcal{G}(U_4)$,
by
\begin{equation}
\label{stokwsu}
r_{\mathcal{CH}^+}(u)=\lim_{v\to\infty} r(u,v).
\end{equation}
In the limit,
\[
\lim_{u\to0} r_{\mathcal{CH}^+}(u)=r_-.
\] 
\end{theorem}

\begin{proof}
Suppose the theorem is false. This implies that 
given sufficiently small $\epsilon>0$, then 
for all $U'>0$,
there exist points $p\in\mathcal{G}(U')$ such that
\begin{equation}
\label{avdevisxuei}
r(p)=r_--\epsilon.
\end{equation}

To see how the falseness of the theorem implies
$(\ref{avdevisxuei})$, consider first the case where
for all $U'>0$, we have $\mathcal{G}(U')\ne\mathcal{K}(U')$.
It follows that $\partial\overline{\mathcal{G}(U')}\ne\emptyset$,
where the boundary is taken in the topology of $\mathcal{K}(U')$,
and thus, by Theorem \ref{epekt9e}, there exist points $p\in\mathcal{G}(U')$ 
such that $r(p)$ is arbitrarily close to $0$,
in particular, for $\epsilon<r_-$, satisfying $(\ref{avdevisxuei})$.
On the other hand, if for some $U'>0$, we have
$\mathcal{G}(U')=\mathcal{K}(U')$, then it is clear by monotonicity,
that a function $r_{\mathcal{CH}^+}$ can be defined by $(\ref{stokwsu})$,
for $0<u<U'$, and $r_{\mathcal{CH}^+}$ is non-increasing
in $u$. Moreover,
the limit $\lim_{u\to0}{r_{\mathcal{CH}^+}}$ exists
and is clearly less than or equal to $r_-$. If
$\lim_{u\to0}{r(u,\infty)}<r_-$, then for some $\epsilon,u>0$, 
$r_{\mathcal{CH}^+}(u)<r_--2\epsilon$, and this 
implies $(\ref{avdevisxuei})$. 

Assuming now $(\ref{avdevisxuei})$,
and restricting $u<U_{4,1}$ for some $U_{4,1}>0$, by the previous
proposition it follows that
$p\in J^+(\gamma)$. If we can show that
$(\ref{kalolambda})$ continues to hold in 
\[
\mathcal{U}=J^+(\gamma)\cap\{r\ge r_--\epsilon\},
\]
modulo a constant, i.e.~if we can show that
\begin{equation}
\label{modulo}
-\lambda\le Cv^{-s},
\end{equation}
for some $C$,
then we easily arrive at a contradiction.
Indeed, restricting to $\mathcal{G}(U_{4,2})$, for some
$0<U_{4,2}\le U_{4,1}$, so that 
$r_-+\epsilon/2\ge r(u,v|_\gamma(u))>r_--\epsilon/2$,
one computes for $(u,v)\in\mathcal{U}$,
\[
r(u,v)=r(u,v|_\gamma(u))+\int_{v|_\gamma(u)}^v{\lambda(u,\bar{v}) d\bar{v}}\ge
r_--\epsilon/2-\tilde{C}v^{-s+1}.
\]
For $u<U_{4,3}$, for some $0<U_{4,3}\le U_{4,2}$ 
we have that $\tilde{C}v^{-s+1}<\epsilon/3$ in $J^+(\gamma)$.
Thus 
\[
r(u,v)\ge r_--5\epsilon/6
\]
in $\mathcal{U}\cap\mathcal{G}(U_{4,3})$, 
and thus $\mathcal{U}\cap\mathcal{G}(U_{4,3})$ 
cannot contain points satisfying $(\ref{avdevisxuei})$,
a contradiction.

Thus it remains to show that
$(\ref{modulo})$ holds in $\mathcal{U}\cap\mathcal{G}(U_{4,3})$.

We partition $\mathcal{U}\cap\mathcal{G}(U_{4,3})$ into three subregions
$\mathcal{U}_1$, $\mathcal{U}_2$, $\mathcal{U}_3$, where
\[
\mathcal{U}_1=\{p\in\mathcal{U}\cap\mathcal{G}(U_{4,3}):\varpi-\frac{e^2}r<0\},
\]
\[
\mathcal{U}_2=\{p\in\mathcal{U}\cap\mathcal{G}
(U_{4,3}):\varpi(p)<\varpi_++C_1\}\setminus\mathcal{U}_1,
\]
and
\[
\mathcal{U}_3=\{p\in\mathcal{U}\cap\mathcal{G}(U_{4,3}):\varpi(p)\ge\varpi_++C_1\},
\]
for a sufficiently large $C_1$ to be determined later.

In $\mathcal{U}_1$, 
\[
\frac{2}{r^2}\frac{\nu}{1-\mu}\left(\varpi-\frac{e^2}r\right)<0.
\]
In $\mathcal{U}_2$, we compute
\begin{eqnarray*}
1-\mu&=&1-\frac{2\varpi}r+\frac{e^2}{r^2}\\
&=&1-\frac{2}{r}\left(\varpi-\frac{e^2}{r}\right)-\frac{e^2}{r^2}\\
&\le&1-\frac{e^2}{(r_-+\epsilon/2)^2}<0,
\end{eqnarray*}
for small enough $\epsilon$. Together with this
lower bound for $|1-\mu|$, we note that we have an upper bound
for $\left|\varpi-\frac{e^2}{r^2}\right|$. 
Thus, 
\begin{eqnarray*}
\int_{\mathcal{U}_2\cap[0,U_{4,3}]\times\{v\}}
{\left|\frac{2}{r^2}\frac{\nu}{1-\mu}\left(\varpi-\frac{e^2}r\right)\right|
(u,v)du}&<&
C_2\int_{\mathcal{U}_2\cap[0,U_{4,3}]\times\{v\}}(-\nu)(u,v)du\\
&\le& C_3,
\end{eqnarray*}
where $C_2$ and thus $C_3$ depend on $C_1$.
In $\mathcal{U}_3$, we compute
\begin{eqnarray*}
\frac{\frac{e^2}r-\varpi}{1-\frac{2\varpi}{r}+\frac{e^2}{r^2}}
&=&\frac{e^2-r\varpi}{r-2\varpi+\frac{e^2}r}\\
&=&\frac{r\varpi-e^2}{2\varpi-\left(\frac{e^2}r+r\right)}.
\end{eqnarray*}
For $C_1$ big enough, independent of $v$,
we have $\frac{e^2}r+r<\varpi$, and $e^2<r\varpi$,
and thus,
\[
\left|\frac{r\varpi-e^2}{2\varpi-\left(\frac{e^2}r+r\right)}\right|
<\frac{r\varpi}{\varpi}=r.
\]
This yields
\begin{eqnarray*}
\int_{\mathcal{U}_3\cap[0,U_{4,3}]\times\{v\}}
{\left|\frac{2}{r^2}\frac{\nu}{1-\mu}\left(\varpi-\frac{e^2}r\right)\right|
(u,v)du}\\
\hbox{\ \ }<2(r_--\epsilon)^{-1}
\int_{\mathcal{U}_3\cap[0,U_{4,3}]\times\{v\}}(-\nu(u,v) du)\\
\hbox{\ \ }
\le 2r_+(r_--\epsilon)^{-1}.
\end{eqnarray*}
Thus we have
\[
\int_{\mathcal{U}\cap[0,U_{4,3}]\times\{v\}}
{\frac{2}{r^2}\frac{\nu}{1-\mu}\left(\varpi-\frac{e^2}r\right)(u,v)du}<
C_3+2r_+(r_--\epsilon)^{-1},
\]
and finally, integrating $(\ref{lqu})$
with this bound, we infer that $(\ref{modulo})$ indeed applies
in $\mathcal{U}\cap\mathcal{G}(U)_{4,3}$. 
The theorem is proven with $U_4=U_{4,3}$.
\end{proof}

The proof of the above theorem implies in particular that
\[
\mathcal{U}\cap\mathcal{G}(U_4)=J^+(\gamma)\cap\mathcal{G}(U_4).
\] 
Moreover, $\mathcal{U}_2\cup\mathcal{U}_3$ can be understood
as a no-shift region, since, one can easily see in a similar fashion
that
\[
\sum_{i=2}^3\int_{\mathcal{U}_i\cap\{u\}\times[V,\infty)}\left|
\frac{2\kappa}{r^2}\left(
\varpi-\frac{e^2}r\right)\right|(u,v)dv<C,
\]
for some constant $C$ independent of $u$.
of the stable blue-shift region either remains blue-shift, or
is a no-shift region, just as claimed in Section \ref{ovvsec}.

\section{$C^0$ extension of the metric}
\label{c0esec}
By the monotonicity $(\ref{F1})$ of $\kappa$,
we can choose $U_5$ so that 
either
\begin{equation}
\label{miaupo9esn}
\int{\kappa(u,v)}<\infty
\end{equation}
for all $U_5>u>0$, or the above integral is infinite,
again
for all $U_5>u>0$. 
The Reissner-Nordstr\"om solution belongs to the
latter case. We will produce in Section \ref{minsec}
a large class of solutions for which the former case holds.

In this section we will show that under the assumption
that $p>1$,
Theorem \ref{pera-r} implies that the metric
can be extended continuously beyond the Cauchy
horizon. 

\begin{theorem}
Assume that in the initial data, $p>1$. 
There exists a $2$-dimensional manifold $\tilde{\mathcal{G}}$, with $C^0$ metric
$\tilde{g}$, and $C^0$ functions $\tilde{r}$ and $\tilde{\phi}$ defined
on $\tilde{\mathcal{G}}$, 
with Penrose diagram 
depicted below
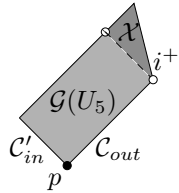
\[
\input{arxikoedw5.pstex_t}
\]
such that $(\mathcal{G}(U_5),\bar{g})$ embeds isometrically 
into $(\tilde{\mathcal{G}},\tilde{g})$ 
with image depicted above, and 
such that $\tilde{r}$ and $\tilde{\phi}$ restricted to 
$\mathcal{G}$
coincide with $r$ and $\phi$. If $(\ref{miaupo9esn})$ holds, then
$\tilde{g}$ can be chosen non-degenerate.
\end{theorem}
\begin{proof}
We will omit the proof in the
case where $(\ref{miaupo9esn})$ does not hold.\footnote{If
the conditions of the theorem of the next section hold,
then $(\ref{miaupo9esn})$ necessarily holds.}
Assume then $(\ref{miaupo9esn})$. To prove the theorem,
it suffices to show that coordinates
can be chosen 
downstairs so that $r$, $-\Omega^2=4\kappa\nu$ and $\phi$
extend to continuous functions up to $\mathcal{CH}^+$,
in the topology of the Penrose diagram minus the point
$i^+$, and to show that $-\Omega^2<0$.

The idea of this proof is to use the estimates of Section 14, which
by our bound on $r$ apply up to the Cauchy horizon,
to show that $r$, $\nu$, and $\kappa$ can be
continuously extended to the Cauchy horizon. The assumption
on $p$ is necessary to assure that $\int_v^\infty|\theta|\to0$
as $v\to\infty$. 

Redefine the $v$ coordinate so that
$\kappa(U_4,v)=1$. By assumption $(\ref{miaupo9esn})$, this
coordinate system has finite $v$-range $[V,\bar{V})$, and the 
Cauchy horizon $\mathcal{CH}^+$
is parametrized by $(0,U_4)\times\{\bar{V}\}$.

Let $\gamma$ be as before. Note that $\nu$ is unaffected by the
above change of
coordinates.
We proceed to show that $\nu$ can be extended to a continuous
function on
$(0,U_5)\times[V,\bar{V}]$
by setting 
\begin{equation}
\label{epektasn}
\nu(u,\bar{V})=\lim_{v\to\bar{V}}\nu(u,v).
\end{equation}

To show that the right hand side of $(\ref{epektasn})$ exists,
it suffices to show that
\begin{equation}
\label{stomndev}
\int_v^{\bar{V}}{
\frac{2\kappa}{r^2}\left(\varpi-\frac{e^2}r\right)(u,\bar{v})d\bar{v}}\to0
\end{equation}
as $v\to\bar{V}$. 

Fix a sufficiently large constant $C_1$ and let 
\[
\mathcal{U}_1=\{\varpi< C_1\}\cap[0,U_5)\times[V,\bar{V}),
\]
\[
\mathcal{U}_2=\{\varpi\ge C_1\}\cap[0,U_5)\times[V,\bar{V}).
\]
We have
\[
\left|\frac{2}{r^2}\left(\varpi-\frac{e^2}r\right)\right|\le C_2,
\]
in $\mathcal{U}_1$,
and
\[
\left|\frac{1}{1-\mu}\frac{2}{r^2}\left(\varpi-\frac{e^2}r\right)\right|\le C_2
\]
in $\mathcal{U}_2$, for some $C_2$, as long as $C_1$ is sufficiently large.
Thus,
\begin{eqnarray}
\label{peper}
\nonumber
\left|\int_v^{\bar{V}}\frac{2\kappa}{r^2}\left(\varpi-\frac{e^2}r\right)(u,\bar{v})d\bar{v}
\right|&\le&
\int_{\mathcal{U}_1\cap\{u\}\times[v,\bar{V})}{
\left|\frac2{r^2}\left(\varpi-\frac{e^2}r\right)\right|\kappa d\bar{v}}
\\
\nonumber
&&\hbox{}+
\int_{\mathcal{U}_2\cap\{u\}\times[v,\bar{V})}
\left|\frac{1}{1-\mu}\frac{2}{r^2}\left(\varpi-\frac{e^2}r\right)\right|
|\lambda| d\bar{v}\\
&\le&
\tilde{C}\int_{\mathcal{U}_1\cap\{u\}\times[v,\bar{V})}\kappa(u,\bar{v})d\bar{v}\\
&&\hbox{}+
\tilde{C}\int_{\mathcal{U}_2\cap\{u\}\times[v,\bar{V})}(-\lambda)(u,\bar{v})d\bar{v}.
\end{eqnarray}
By $(\ref{miaupo9esn})$,
and the fact that
\begin{equation}
\label{nonin}
\int_{\mathcal{U}_2\cap\{u\}\times[v,\bar{V})}\kappa(u,\bar{v})d\bar{v} 
\end{equation}
is non-increasing in $u$, it
is clear that the
right hand side of $(\ref{peper})$ is uniformly bounded on
any set $[u_1,u_2]\times[V,\bar{V})$, where $0<u_1\le u_2<U_5$. 
Moreover, for fixed $u$,
\[
\int_{\mathcal{U}_1\cap\{u\}\times[v,\bar{V})}(-\lambda)+
\int_{\mathcal{U}_2\cap\{u\}\times[v,\bar{V})}\kappa\to0
\]
as $v\to\bar{V}$. Thus $\nu(u,\bar{V})$ is well defined and satisfies
$A(u)<\nu(u,\bar{V})<-a(u)<0$, for constants $A\ge a>0$ depending on $u$, where
$A$ and $a$ are uniformly bounded above and below on compact subsets of $(0,U_5)$.

Note that by the inequalities $\partial_v\varpi\ge0$, 
$\partial_vr\le0$, we can extend
$\varpi$ and $r$ by monotonicity to functions
defined on $(0,U_5)\times[V,\bar{V}]$, where
$\varpi(\bar{V},U)$ may take values in the
extended real numbers. 
The result of the previous paragraph shows in particular
that $r$ is
a continuous function in $(0,U_5)\times[V,\bar{V}]$.
To see this, let $(u,\bar{V})$ be a point on the Cauchy
horizon, and let $\epsilon>0$ be given. By monotonicity,
it is clear that there exists a $\bar{v}<\bar{V}$ such that
$|r(u,\bar{V})-r(u,v')|<\frac{\epsilon}2$ for all $v'\ge\bar{v}$. 
On the other hand, given any $[u_1,u_2]$ containing $u$,
since $|\nu|\le C$ in $[u_1,u_2]\times(V,\bar{V}]$ for some $C$,
it follows that for $|u'-u|<\frac{\epsilon}{2C}$,
\begin{eqnarray*}
|r(u',v')-r(u,\bar{V})|&\le&|r(u',v')-r(u,v')|+|r(u,v')-r(u,\bar{V})|\\
&<&\left|\int_{u'}^u{\nu}\right|+\frac{\epsilon}2\\
&\le&C\left(\frac{\epsilon}{2C}\right)+\frac\epsilon2\le\epsilon.
\end{eqnarray*}
Continuity of $r$ follows immediately.

The function $\nu$ can also easily be seen to
be continuous on $(0,U_5)\times[V,\bar{V}]$.
First we note that by continuity of $r$,
\[
\int_{\mathcal{U}_1\cap\{u\}\times[v,\bar{V})}(-\lambda)=r(\bar{V},u)-r(v,u)\to0
\]
\emph{uniformly} in $u$, for $u\in [u_1,u_2]$, where $0<u_1\le u_2\le U_5$.
On the other hand, since $(\ref{nonin})$ is non-decreasing in $u$, it also
converges \emph{uniformly} in $u$ on compact subsets of $(0, U_5)$. Thus, the
convergence in $(\ref{stomndev})$ is also uniform on compact subsets. 
We have then that $\nu(u,\bar{V})$ is continuous in $u$, as $\nu$ is a uniform limit
of continuous functions $\nu(u,v_i)$, for $v_i\to\bar{V}$.
Again by the uniform convergence, it follows that $\nu(u,v)$ is
continuous in $(0,U_5)\times[V,\bar{V}]$.

Having extended $\nu$ to a continuous non-zero function,
to extend $\Omega$ it suffices to extend $\kappa$.
First we shall show that $\frac{\zeta}{\nu}$
extends to a continuous function on $(0,U_5)\times[V,\bar{V}]$.

Note that Propostion \ref{Estimates}, to be shown in
Section \ref{bvesec}, together with the assumption on
$p>1$ imply that 
\[
\int_{v_1}^{v_2}{|\theta|(u,\bar{v})d\bar{v}}<C
\]
and
\begin{equation}
\label{kovteuoume...}
\int_{u_1}^{u_2}{|\zeta|(\bar{u},v)d\bar{u}}<C,
\end{equation}
for a uniform constant $C$, for all $u_1,u_2,u\in(0,U_5)$,
$v_1,v_2,v\in[V,\bar{V}]$.
Fixing, as before, an interval $[u_1,u_2]$, given $\epsilon$,
we can choose
$v<\bar{V}$ large enough so that
\[
\int_{v}^{\bar{V}}{|\theta|(u_1,\bar{v})d\bar{v}}<\epsilon.
\]
On the other hand, we can also choose $v$ so that
\[
\int_{v}^{\bar{V}}{\frac{-\lambda}{r}(u,\bar{v})d\bar{v}}<\epsilon
\]
for all $u\in[u_1,u_2]$, since $r$ is continuous up to the
Cauchy horizon.
Applying $(\ref{kovteuoume...})$ and the
estimate $(\ref{analogous})$ of Section 14 for
small enough $\epsilon$ one obtains 
\[
\int_{v}^{\bar{V}}{|\theta|(u,\bar{v})d\bar{v}}<2\epsilon.
\]
for $u\in[u_1,u_2]$.
 
What we have just shown is that
\begin{equation}
\label{molistwra}
\int_{v}^{\bar{V}}{|\theta|(u,\bar{v})d\bar{v}}\to0
\end{equation}
uniformly in $u$ on compact subsets of $(0,U_5)$.
In view also of the uniformity of the convergence
of $(\ref{stomndev})$, discussed above, integrating the equation
\[
\partial_v\left(\zn\right)=
-\frac\theta{r}-\left(\zn\right)
\frac{2\kappa}{r^2}\left(\varpi-\frac{e^2}r\right),
\]
we obtain that
$\zn$ extends to a continuous function on
$(0,U_5)\times[V,\bar{V}]$.
Defining 
\[
\kappa(u,\bar{V})=e^{\int_u^U
{\left(\zn\right)^2\frac\nu{r}(u,\bar{V})du}},
\]
it is clear that this defines a continuous extension of
$\kappa$ to $(0,U_5)\times[V,\bar{V}]$.
From $(\ref{molistwra})$, the continuous extendibility
of $\phi$ to $(0, U_5)\times[V,\bar{V}]$ follows easily.
The theorem is thus proven.
\end{proof}

Defining now $\tilde{\mathcal{M}}$ from $\tilde{\mathcal{Q}}$
and $\tilde{r}$, as in Proposition~\ref{megalnprotasn},
we obtain easily the rest of Theorem~\ref{eis9ew}.

\section{Mass inflation}
\label{minsec}

In this section, it is shown that for  a large
class of data, the Hawking mass $m$ blows up
identically on the event horizon.

The additional condition that we will impose on initial data
is that there exist positive constants $V_1$ and $c$ such that 
\begin{equation}
\label{yenidurum}
v>V_1\Rightarrow
\left|\theta(0,v)\right|\ge cv^{-\tilde{p}},
\end{equation}
for a $\tilde{p}$ satisfying $3p>\tilde{p}>p>\frac12$.\footnote{To 
explicitly construct such data, one replaces $0$ by an appropriate
$c'v^{-2\tilde{p}}$ on the right hand side of $(\ref{newINIT3})$,
and then notes that if $\lambda$ is chosen monotonically
decreasing, with $\lambda\ge c''v^{-2\tilde{p}}$, then this new
condition is ensured for $v\ge V$.}
This class seems to include
the initial data considered in numerical work \cite{brsm:bhs}.
Without loss of generality, we will assume that
$(\ref{yenidurum})$ is true without the absolute values,
i.e.
\begin{equation}
\label{yenidurum'}
v>V_1\Rightarrow
\theta(0,v)\ge cv^{-\tilde{p}}.
\end{equation}

Note that the above assumptions imply in particular that
$\mathcal{A}\cap\{u=0\}=\emptyset$. 

\begin{proposition}
For initial data satisfying $(\ref{yenidurum'})$, where
$p<\tilde{p}<3p$,
it follows that on $\mathcal{A}\cap\mathcal{G}(U_6)$ for small enough $U_6$,
we have that $\zeta>0$, and $\theta>c'v^{-\tilde{p}}$, for some $c'>0$.
\end{proposition}

\begin{proof} For this, we must revisit the proof of
Proposition \ref{prwtnprotasn}. From $(\ref{newINIT})$,
and the bounds from Proposition \ref{prwtnprotasn} on the
sign of $\varpi-\frac{e^2}r$, it follows by integrating $(\ref{lqu})$
that 
\[
\lambda\le\tilde{C}v^{-2p}
\]
in $J^-(\mathcal{A})\cap\mathcal{G}(U_1)$, 
for some $\tilde{C}>0$.
It will be useful to keep in mind in this
proof that since $\mathcal{A}$ is
achronal, terminates at $i^+$,
and does not intersect the event horizon, it follows
that given $\tilde{v}$, one can always choose a $\tilde{u}>0$ so that
$v\ge\tilde{v}$ on $\mathcal{A}\cap\mathcal{G}(\tilde{u})$.

From our bounds on
$\left|\zn\right|$ derived in Proposition \ref{prwtnprotasn}, 
and condition $(\ref{yenidurum})$,
it follows by integration of
\[
\partial_u\theta=-\frac{\zeta\lambda\nu}{\nu r}
\]
that we can select 
$E$, and $U_{6,1}$
so that $\theta>cv^{-\tilde{p}}-\tilde{c}v^{-3p}$ in
$\mathcal{G}(U_{6,1})\cap J^-(\Gamma_{E})$.
Thus for $\tilde{p}<3p$, we can select $0<U_{6,2}\le U_{6,2}$
so that $\theta>c'v^{-\tilde{p}}$ on 
$J^-(\mathcal{A})\cap J^+(0,V_2)$
for large enough $V_2$.
Moreover, in $J^-(\mathcal{A})\cap J^+(0,V_2)$ we also have
\[
B(v-v^*)<\int_{v^*}^{v}
{\frac{2\kappa}{r^2}\left(\varpi-\frac{e^2}r\right)}
\]
for some $B>0$. Integrating $(\ref{integrating})$ now gives
(note there are no absolute values on the left) that 
\[
\zn\le-\int_{V_2}^v{c'{\tilde{v}}^{-\tilde{p}}
e^{-B(v-\tilde{v})}d\tilde{v}}+\overline{C}e^{-B(v-V_2)},
\]
where $\overline{C}=\sup_u{\left|\zn\right|(u,V_2)}$.
The first term on the left is greater than, or equal
to $c''v^{-\tilde{p}}$ for some $c''>0$,\footnote{For this, note
that the $v^{-p}$ on the right hand side of $(\ref{sxedov2})$ 
can be replaced by
a lesser power of $v$ for large enough $V_2$.} and thus, for small enough 
$0<U_6\le U_{6,2}$
one obtains that $\zn<0$ on $\mathcal{A}\cap\mathcal{G}(U_6)$, and
consequently $\zeta>0$. \end{proof}

\begin{proposition}
\label{fieldmonot}
If $\tilde\gamma\subset\mathcal{T}\cup\mathcal{A}$ is an
achronal curve and
$\zeta>0$, $\theta>0$ on $\tilde\gamma$,
then $\zeta>0$ and $\theta>0$ in
the future domain of dependence $D^+(\tilde\gamma)\cap\mathcal{G}(u_0)$ 
of $\tilde\gamma$. 
Moreover, in $D^+(\tilde\gamma)\cap\mathcal{G}(u_0)$ we also have
$\partial_v\zeta>0$, and $\partial_u\theta>0$.
\end{proposition}

\begin{proof} Note first that 
$D^+(\tilde\gamma)\cap\mathcal{G}(u_0)\subset\mathcal{T}$. 
If the first statement of the
proposition is false, there
must exist a point $(u,v)\in
D^+(\tilde\gamma)\cap\mathcal{G}(u_0)$
such that $\zeta>0$ and $\theta>0$ in 
$D^+(\tilde\gamma)\cap J^-(u,v)\setminus (u,v)$,
but one of these inequalities fails at $(u,v)$, 
i.e.~either $\zeta(u,v)=0$
or $\theta(u,v)=0$.
But
\[
\zeta(u,v)=\zeta(u,v|_{\tilde{\gamma}}(u))
+\int_{v|_{\tilde{\gamma}}(u)}^{v}{-\frac{\theta\nu}r
(u,\bar{v})d\bar{v}}>0,
\]
and
\[
\theta(u,v)=\theta(u|_{\tilde{\gamma}}(v),v)
+\int_{u|_{\tilde{\gamma}}(v)}^u
{-\frac{\zeta\lambda}r(\bar{u},v)d\bar{u}}
>0,
\]
a contradiction. 

The second statement of the proposition now follows immediately
from $(\ref{sign1})$ and $(\ref{sign2})$.
\end{proof}

\begin{corollary}
\label{massmonot}
Under the assumptions of Proposition \ref{fieldmonot},
$\partial_u\partial_v\varpi\ge0$ in $D^+(\tilde\gamma)\cap\mathcal{G}(u_0)$.
In particular, given a characteristic rectangle 
$\mathcal{X}=J^+(\bar{u},\bar{v})\cap J^-(u,v)$ with $\mathcal{X}\subset
D^+(\tilde\gamma)$
it follows that
\[
\varpi(u,v)-\varpi(\bar{u},v)\ge\varpi(u,\bar{v})-\varpi(\bar{u},\bar{v}).
\]
\end{corollary}

\begin{proof} Differentiating $(\ref{pvqu})$ we obtain
\begin{eqnarray*}
\partial_u\partial_v\varpi&=&\partial_u\left(\frac12\kappa^{-1}
\theta^2\right)\\
&=&\frac12\partial_u\kappa^{-1}\theta^2
+\kappa^{-1}\theta\partial_u\theta\ge0
\end{eqnarray*}
by the previous Proposition and the monotonicity $(\ref{F1})$ satisfied
by $\kappa$.
The second statement of the theorem follows by integration of
this inequality in $\mathcal{X}$. \end{proof}

Note that for $\tilde\gamma=\mathcal{A}\cap\mathcal{G}(U_1)$,
$D^+(\tilde\gamma)\cap\mathcal{G}(u_0)=J^+(\mathcal{A}\cap\mathcal{G}(U_1))$.

We can now state the theorem of this section.
\begin{theorem}
For initial data satisfying $(\ref{yenidurum})$, $\varpi$
blows up identically in the limit on the Cauchy horizon $\mathcal{CH}^+$ for
$0<u<U_4$, 
i.e.
\[
\lim_{v\to\infty}\varpi(u,v)=\infty.
\]
\end{theorem}

\begin{proof} The general two-step structure
of this proof was outlined in Section~\ref{ovvsec}. But in fact,
all the work is in step 1. Step 2, as we shall see,
is basically line $(\ref{part2})$.

Recall that the goal of step 1 is to prove that
if the mass does not blow up identically, then the
spacetime ``looks like'' Reissner-Nordstr\"om, in the sense
discussed in Section~\ref{ovvsec}.

The outline of step 1 is roughly as follows.
Let $\varpi_{\mathcal{CH}^+}(u)=\varpi(u,\bar{V})$ be the extended
real number valued function defined in Section~\ref{c0esec}, where $\bar{V}$ denotes
the coordinate of that section. (In this section, we will find it convenient
to use our original coordinates where the Cauchy horizon corresponds
to $v=\infty$.)
First
we will show that either
\begin{equation}
\label{ya}
\liminf_{u\to0}\varpi_{\mathcal{CH}^+}(u)=\varpi_+
\end{equation}
or 
\begin{equation}
\label{ya2}
\varpi_{\mathcal{CH}^+}=\infty
\end{equation}
identically.

The proof of the theorem thus reduces to showing 
that the assumption $(\ref{ya})$
leads to a contradiction. The rest of part 1 will 
derive more and more precise statements about the geometry of
the solution from
$(\ref{ya})$.
In particular, $(\ref{ya})$ shows that in $J^+(\gamma)
\cap\mathcal{G}(U')$, for sufficiently
small $U'$, in view of Theorem~\ref{pera-r}, we have that 
\begin{equation}
\label{masftavei}
\varpi-\frac{e^2}r<-A
\end{equation}
for some $A>0$.
This then allows us to prove that
\begin{equation}
\label{fragmevo;}
\int_v^\infty{\left(\tl\right)^2\lambda(u,\bar{v}) d\bar{v}}
\end{equation}
must remain bounded uniformly in $u$ for 
$(u,v)\in J^+(\gamma)\cap\mathcal{G}(U')$.
Such an estimate together with our zig-zag argument familiar from
Proposition \ref{sta9mpleprot} will give
us an estimate on estimate $\int\frac{\nu}{1-\mu}$. This in turn will
give as an estimate on $\lambda$ by integration of $(\ref{lqu})$.
 
Step 2 then consists of using this estimate on $\lambda$ together 
with our lower bound on $\theta$ that derives from Proposition
\ref{fieldmonot} to contradict the boundedness of $(\ref{fragmevo;})$.

We now give the details of the proof.
Suppose
\begin{equation}
\label{assume4now}
\liminf_{u\to0}\varpi_{\mathcal{CH}^+}>\varpi_+.
\end{equation}
It follows that there exits a $U_{7,1}>0$, $\epsilon>0$ such that
$\varpi_{\mathcal{CH}^+}>\varpi_++\epsilon$ for $u\le U_{7,1}$.
On the other hand, since 
\[
\lim_{v\to\infty}\varpi(u|_\gamma(v),v)=\varpi_+,
\]
there exists a $U_{7,2}$, such that
\[
\varpi(u|_\gamma(v),v)<\varpi_++\frac\epsilon2.
\]
In particular,
given a point $(u_0,v_0)\in\gamma$ with $u_0\le\min(U_{7,1},U_{7,2})$,
there exists a point $(u_0,v_1)$, $v_1>v_0$ such that
\[
\varpi(u_0,v_1)-\varpi(u_0,v_0)=\frac\epsilon3.
\]
Given $(u_i,v_{i+1})$, we can define 
\[
(u_{i+1},v_{i+1})=(u|_\gamma(v_{i+1}),v_{i+1}),
\]
and $(u_{i+1}, v_{i+2})$ to be such that
\begin{equation}
\label{stoepomevo}
\varpi(u_{i+1},v_{i+2})-\varpi(u_{i+1},v_{i+1})=\frac\epsilon3.
\end{equation}
This is depicted in the Penrose diagram below:
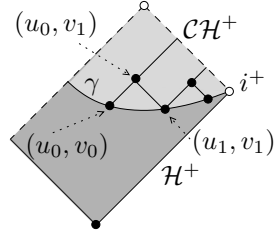
\[
\input{step1veo.pstex_t}
\]
Corollary \ref{massmonot} now
implies that 
\[
\varpi(u_k,v_i)-\varpi(u_k,v_k)\ge\frac{(i-k)\epsilon}3\to\infty,
\]
as $i\to\infty$. Thus, in view of the fact that $\partial_u\varpi\ge0$,
the assumption $(\ref{assume4now})$ leads to the conclusion
that $\varpi_{\mathcal{CH}^+}=\infty$ identically. If the proposition is 
false, it follows
that we must have $(\ref{ya})$.

For the bound on $(\ref{fragmevo;})$, we argue similarly. 
Since as discussed
above, $(\ref{ya})$ implies
$(\ref{masftavei})$ in $J^+(\gamma)\cap\mathcal{G}(U_{7,3})$ 
for some $U_{7,3}$, 
it follows that $\partial_u(-\lambda)\le0$
in $J^+(\gamma)\cap\mathcal{G}(U_{7,3})$.
Since by Proposition \ref{fieldmonot} we also have 
\[
\partial_u(\theta^2)=2\theta\partial_u\theta\ge0,
\]
it follows that
\[
-\int_{v_1}^{v_2}{\frac{\theta^2}\lambda(u,v)dv}
\]
is a non-decreasing function in $u$, provided $(u,v_1)\in 
J^+(\gamma)\cap\mathcal{G}(U_{7,3})$.

Suppose first that there exists a sequence of $u_i\to0$ such that
\begin{equation}
\label{allnavtifasn}
\int_{v|_\gamma(u_i)}^\infty{\frac{\theta^2}{(-\lambda)}(u_i,v)dv}=\infty.
\end{equation}
By the non-decreasing property
just proved, $(\ref{allnavtifasn})$ is clearly true when
$u_i$ is replaced by any $u<u_0$.
Now integrating $(\ref{nqu})$, it follows that $\nu$ extends to 
the function $0$ on the Cauchy horizon. 
Our assumption on the finiteness of $\varpi_{\mathcal{CH}^+}(u)$,
together with $(\ref{allnavtifasn})$,
clearly implies that 
\[
\left(1-\frac{2\varpi_{\mathcal{CH}^+}}{r_{\mathcal{CH}^+}}+\frac{e^2}{r_{\mathcal{CH}^+}^2}
\right)(u)=0.
\]
For otherwise, by the inequalities
$\partial_v\varpi\ge0$, $\partial_ur\ge0$, there would exist for each
$u$ a constant $V^*(u)$ such that $1-\mu(u,v)<-c'$ for 
$v\ge V^*(u)$. Integrating $(\ref{pvqu})$ in $\{u\}\times[V^*(u),\infty)$
would give $\varpi_{\mathcal{CH}^+}(u)=\infty$.  
But now, $\nu_{\mathcal{CH}^+}=0$ implies that
$r_{\mathcal{CH}^+}$ and thus $\varpi_{\mathcal{CH}^+}$ also is constant in $u$.
But $u_2>u_1$ implies
\[
\varpi_{\mathcal{CH}^+}(u_2)-\varpi_{\mathcal{CH}^+}(u_1)\ge
\varpi(u_2,v)-\varpi(u_1,v)>0,
\]
where the first inequality folows Corollary
\ref{massmonot} and the second from Proposition \ref{fieldmonot},
after integration of $(\ref{puqu})$.
So we arrive at a contradiction, and thus, the integral
on the left hand side of $(\ref{allnavtifasn})$ is finite for all $u$.

We will now show that 
\begin{equation}
\label{unifbnded}
\int_{v|_\gamma(u)}^\infty{\frac{\theta^2}\lambda(u,\bar{v})d\bar{v}},
\end{equation}
is uniformly bounded in $u$, in fact, tends to $0$ as $u\to0$.
For if not, since $(\ref{unifbnded})$
is a monotone quantity, for every $\epsilon>0$
there exists a $0<U_{7,4}\le U_{7,3}$ such that $u<U_{7,4}$ implies
\[
-\int_{v|_\gamma(u)}^\infty{\frac{\theta^2}\lambda(u,\bar{v})d\bar{v}}\ge\epsilon.
\]
Defining a sequence $(u_i,v_i)$ as before, but now
where $(\ref{stoepomevo})$ is replaced by the condition
\[
-\int_{v_1}^{v_2}{\frac{\theta^2}\lambda(u,v)dv}=\frac\epsilon3
\]
we obtain by our monotonicity $(\ref{allnavtifasn})$, and thus
a contradiction.

To bound $\int\frac{\nu}{1-\mu}$,
fix a point $(u,v)\in J^+(\gamma)\cap\mathcal{G}(U_{7,4})$,
and consider the 
$(\Gamma_{\overline{\epsilon}},\Gamma_{-\overline\epsilon}, u,
v)$-zigzag:
\[
\bigcup_{i=1}^{I-1}\{u_i\}\times[v_i,v_{i+1}]\cup[u_{i+1},u_i]\times\{v_{i+1}\}
\]
Let $J\le I$ the first value such that $u_J\le u|_{\gamma}(v)$,
and redefine $u_J$ to be $u|_{\gamma}(v)$.
Refer to the figure below:
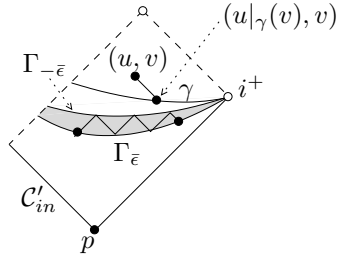
\[
\input{konta2.pstex_t}
\]
We have that
\begin{eqnarray*}
\int_{u|_\gamma(u)}^{u}{\frac{\nu}{1-\mu}(\bar{u},v)d\bar{u}}
&\sim&
\sum_{i=1}^{J-1}
\int_{u_{i+1}}^{u_i}{\frac{\nu}{1-\mu}(\bar{u},v_{i+1})d\bar{u}}\\
&\sim&
\sum_{i=1}^{J-1}
\int_{u_{i+1}}^{u_i}{-\nu (\bar{u},v_{i+1})d\bar{u}}\\
&\sim&
\sum_{i=1}^{J-1}
\int_{v_{i}}^{v_{i+1}}{-\lambda (u_i,\bar{v}) d\bar{v}}\\
&\sim&
\sum_{i=1}^{J-1}
\int_{v_{i}}^{v_{i+1}}{\frac{\lambda}{1-\mu}(u_i,\bar{v})d\bar{v}}\\
&\sim&v-(\alpha\log v+H).
\end{eqnarray*}
Integrating now $(\ref{lqu})$ from $\gamma$, in view of the above
bound and $(\ref{masftavei})$
it follows that 
\begin{equation}
\label{teleutaiobnd}
-\lambda\le c(u)e^{-\tilde{c}Av}
\end{equation}
where $\tilde{c}$
derives from the above $\sim$. We need not explicitly estimate the constants. 

This completes step 1.
For step 2, there is very little to do.
Applying
Proposition \ref{fieldmonot} together with $(\ref{teleutaiobnd})$, 
it follows that 
\begin{eqnarray}
\label{part2}
\nonumber
\int_{v|_\gamma(u)}^{\infty}{\left(\tl\right)^2(-\lambda)(u,\bar{v})d\bar{v}}&=&
\int_{v|_\gamma(u)}^{\infty}{\theta^2(-\lambda^{-1})(u,\bar{v})d\bar{v}}\\
&\ge&\int_{v|_\gamma(u)}^{\infty}{c\bar{v}^{-2\tilde{p}}e^{\tilde{c}A\bar{v}}
(u,\bar{v})d\bar{v}}=\infty,
\end{eqnarray}
which contradicts the fact proven above that this integral
is finite. Thus, $(\ref{ya})$ does not hold, so we must
indeed have $(\ref{ya2})$; the Theorem is proven. \end{proof}

If $\theta$ initially decays exponentially,
but there is also an appropriate exponential lower bound on $\theta$,
a similar theorem can in fact be proven.
To obtain the best results,
one has to be slightly more careful with the constants
connected to the $\sim$ in the chain of estimates above.
Compare with Section 10 or \cite{md:si}.

Heuristic analysis~\cite{gpp:de1, rpr:ns, lb:sbh, bo:lt} and 
numerical studies~\cite{gpp:de, bo:lte} suggest that the ``non-oscillatory''
behavior that we have assumed in the limit $(\ref{yenidurum})$ is
indeed to be expected, at least for the neutral massless
scalar field considered
here. As this is not likely to be true for more general matter,
and in any case has not been proved even for this matter, it would
be nice if assumption $(\ref{yenidurum})$ could in fact be weakened.

\section{$BV$ estimates for $\phi$}
\label{bvesec}

We prove in this section certain \emph{a priori}
estimates for the $L^1$ norms of $\theta$ and $\zeta$ in
null directions. The first result is 

\begin{proposition}
\label{or9ogwvio}
Consider a characteristic rectangle 
\[
[u_1,u_2]\times[v_1,v_2]\subset\mathcal{G}(u_0).
\]
Assume 
\[
\sup_{v_1\le v\le v_2}\int_{u_1}^{u_2}{|\nu|r^{-1}(u,v)du}<\delta_1
\]
and 
\[
\sup_{u_1\le u\le u_2}\int_{v_1}^{v_2}{|\lambda|r^{-1}(u,v)dv}<\delta_2,
\] 
for some $\delta_1>0$, $\delta_2>0$, $\delta_1\delta_2<1$. 
Then 
\begin{eqnarray}
\label{L1}
\nonumber
\int_{u_1}^{u_2}{|\zeta|(u,v_2)du}+\int_{v_1}^{v_2}{|\theta|(u_2,v)dv}
&\le&
C(\delta_1,\delta_2)\left(\int_{u_1}^{u_2}{|\zeta|(u,v_1)du}\right.\\
&&\hbox{}+\left.
\int_{v_1}^{v_2}{|\theta|(u_1,v)dv}\right).
\end{eqnarray}
\end{proposition}

\begin{proof} Let $(u,v)\in J^+(u_1,v_1)\cap J^-(u_2,v_2)$
and consider the quantities 
\[
Z(u,v)=\sup_{v_1\le\tilde v\le v}
|\zeta|(u,\tilde{v}),
\] 
and
\[
\Theta(u,v)=\sup_{u_1\le\tilde u\le u}|\theta|
(\tilde{u},v).
\]
We will derive in fact estimates for 
$\int_{u_1}^{u_2}{Z(u,v)du}$ and $\int_{v_1}^{v_2}{\Theta(u,v) dv}$.

Applying absolute values to the equation $(\ref{sign2})$,
and noting that $\partial_vZ\le\partial_v|\zeta|$ almost everwhere,
we obtain after integration that
\[
Z(u,v)\le\int_{v_1}^v{\frac{|\nu||\theta|}r(u,\bar{v})d\bar{v}}+Z(u,v_1).
\]
Thus, it follows that
\begin{eqnarray*}
\int_{u_1}^{u_2}{Z(u,v)du}&\le&
\int_{u_1}^{u_2}{\int_{v_1}^v{\frac{|\nu||\theta|}r(u,\bar{v})d\bar{v}}du}
+\int_{u_1}^{u_2}
{Z(u,v_1)du}\\
&=&\int_{v_1}^v{\int_{u_1}^{u_2}{\frac{|\nu||\theta|}r(u,\bar{v})du}
d\bar{v}}+\int_{u_1}^{u_2}{Z(u,v_1)du}\\
&\le&\int_{v_1}^v{\left(\sup_{u_1\le\tilde{u}\le u_2}
|\theta|(\tilde{u},\bar{v})\right)\int_{u_1}^{u_2}{\frac{|\nu|}rdu}d\bar{v}}
+\int_{u_1}^{u_2}{Z(u,v_1)du}\\
&\le&\delta_1\int_{v_1}^v{\Theta(u_2,\bar{v})d\bar{v}}+\int_{u_1}^{u_2}{Z(u,v_1)du}.
\end{eqnarray*}

Similarly,
one obtains
\[
\int_{v_1}^{v_2}{\Theta(u,v)dv}\le\int_{v_1}^{v_2}{\Theta(u_1,v) dv} +
\delta_2\int_{u_1}
^{u}{Z(\bar{u},v)d\bar{u}}
\]
and thus
\[
\int_{u_1}^{u_2}{Z(u,v_2)du}\le \int_{u_1}^{u_2}{Z(u,v_1)du}+
\delta_1\int_{v_1}^{v_2}{\Theta(u_1,v) dv}+\delta_1\delta_2 
\int_{u_1}^{u_2}{Z(u,v_2)du},
\]
which gives
\[
\int_{u_1}^{u_2}{Z(u,v_2)du}\le\frac{1}{1-\delta_1\delta_2}\int_{u_1}^{u_2}{Z(u,v_1)du}+
\frac{\delta_1}{1-\delta_1\delta_2}\int_{v_1}^{v_2}{\Theta(u_1,v) dv}.
\]
An analogous estimate for $\int_{v_1}^{v_2}{\Theta(u_2,v)dv}$ follows immediately:
\begin{eqnarray}
\label{analogous}
\nonumber
\int_{v_1}^{v_2}{\Theta(u_2,v)du}
&\le&\frac{1}{1-\delta_1\delta_2}\int_{v_1}^{v_2}{\Theta(u_1,v) dv}\\
&&\hbox{}+
\frac{\delta_2}{1-\delta_1\delta_2}\int_{u_1}^{u_2}{Z (u,v_1)du},
\end{eqnarray}
giving the proposition. \end{proof}

We would like to prove a version of Proposition~\ref{L1} when 
$\delta_1\delta_2$ is assumed finite, but not small.
We must be careful, however,
as the quantity
\[
\int{\sup_{\tilde{u}\le u}\lambda(\tilde{u},v)dv}
\]
is \emph{not} bounded.

The following theorem will be proven by showing
that an arbitrary characteristic rectangle in $\mathcal{G}(U_1)$
can be partitioned into $N^3$ rectangles, each of which satisfy
the assumptions of Proposition~\ref{L1}, where $N^3$ depends only on
pointwise bounds for $r$.

\begin{proposition}
\label{Estimates}
Consider 
characteristic rectangle 
\[
\mathcal{X}=[u_1,u_2]\times[v_1,v_2]\subset\mathcal{G}(U_1).
\]
We have
\begin{eqnarray*}
\int_{u_1}^{u_2}{|\zeta|(u,v_2)du}+\int_{v_1}^{v_2}{|\theta|(u_2,v)dv}
&\le&
C(\delta_1,\delta_2)\left(\int_{u_1}^{u_2}{|\zeta|(u,v_1)du}\right.\\
&&\hbox{}+\left.
\int_{v_1}^{v_2}{|\theta|(u_1,v)dv}\right).
\end{eqnarray*}
where $C=C(r^{-1}(u_1,v_1),r^{-1}(u_2,v_2),r_+)$.
\end{proposition}

\begin{proof}
Choose $\delta<1$. For an arbitrary function $f$,
Define
\[
V(f)=\sup_{x,y\in \mathcal{X}}{|\log f(x)-\log f(y)|}.
\]
We have
\[
V(r)\le\left|\log\min\{r(s),r(q)\}-\log r_+\right|.
\]
Partition $\{u_2\}\times[v_1,v_2]$ into $N_0$ segments
$\{u_2\}\times[\hat{v}_i,\hat{v}_{i+1}]$
such that
\[
\int_{\hat{v}_i}^{\hat{v}_{i+1}}{r^{-1}|\lambda|(u_2,v) dv}<\frac16\delta,
\]
for each $i$, $\hat{v}_1=v_1$, $\hat{v}_{N_0}=v_2$.
Since $\mathcal{A}\cap\mathcal{G}(U_1)$ is achronal,
it follows that if $\mathcal{A}\cap\{u_2\}\times[v_1,v_2]\ne\emptyset$,
then this intersection is either a point, or a null segment. In these cases,
arrange so
that this point or both endpoints of this null segment are included
as points $(u_2,\hat{v}_j)$ in the partition.
In view of the fact that the sign of $\lambda$ can change only once
on $\{u_2\}\times[v_1,v_2]$, it follows that 
$N_0$ can be chosen
\[
N_0<N=[12V(r)\delta^{-1}]+3.
\]

Since $\nu<0$, one can 
partition now, for each $i$, the segment
$[u_1,u_2]\times\{\hat{v}_i\}$ into $N_i$ subsegments
$[\hat{u}_{i,j},\hat{u}_{i,j+1}]\times\{\hat{v}_i\}$ with
$\hat{u}_{i,1}=u_1$, $\hat{u}_{i,N_i}=u_2$,
and
\[
\int_{\hat{u}_{i,j}}^{\hat{u}_{i,j+1}}{r^{-1}|\nu|(u,\hat{v}_i)du}<\frac16\delta
\]
for all $i$, $j$,
and with the extra condition that if 
$\mathcal{A}\cap[u_1,u_2]\times\{\hat{v}_i\}\ne\emptyset$,
then its endpoints (again, this set is either a point or a null segment)
are included in the partition as
$(\hat{u}_{i,j},\hat{v}_i)$,
for some $j$. Clearly, $N_i\le N$.

Since $\lambda$ changes sign at most 
once on $\{\hat{u}_{i,j}\}\times[\hat{v}_i,\hat{v}_{i+1}]$, we can partition
each of these segments
into $N_{ij}\le N$ subsegments $\{\hat{u}_{i,j}\}\times[\hat{v}_{i,k},\hat{v}_{i,k+1}]$
such that
\[
\int_{\hat{v}_{i,k}}^{\hat{v}_{i,k+1}}{r^{-1}|\lambda|(\hat{u}_{i,j},v)dv}
<\frac16\delta,
\]
for all $i$, $j$, $k$,
and such that, again,
if $\{\hat{u}_{i,j}\}\times[\hat{v}_{i,k},\hat{v}_{i,k+1}]\cap\mathcal{A}$ is
nonempty, then its endpoints are included as vertices in the partition.

Let us denote 
\[
\mathcal{X}_{ijk}=[\hat{u}_{i,j},\hat{u}_{j+1}]\times[\hat{v}_{i,k},\hat{v}_{i,k+1}].
\]
We have $\mathcal{X}=\cup\mathcal{X}_{ijk}$, and there
are at most $N^3$ rectangles in the collection.
For each $\mathcal{X}_{ijk}$, one of the following three statements
holds:
\begin{enumerate}
\item
$\mathcal{X}_{ijk}\subset\mathcal{T}\cup\mathcal{A}$ 
\item
$\mathcal{X}_{ijk}\subset\mathcal{G}(u_0)\setminus\mathcal{T}$.
\item
$(\hat{u}_{i,j+1},\hat{v}_{i,k})\in\mathcal{A}$
and $(\hat{u}_{i,j},\hat{v}_{i,k+1})\in\mathcal{A}$.
\end{enumerate}

For $f$ a function defined on $\mathcal{X}_{ijk}$ 
define
\[
V_{ijk}(f)= 
\sup_{x,y\in\mathcal{X}_{ijk}}{|\log f(x) -\log f(y)|}.
\]
Now, for those $\mathcal{X}_{ijk}\subset\mathcal{T}\cup\mathcal{A}$, 
it follows that 
\begin{eqnarray*}
\sup_{\hat{v}_{i,k}\le v\le\hat {v}_{i,k+1}}
\int_{\hat{u}_{i,j}}^{\hat{u}_{i,j+1}}{|\nu|r^{-1}(u,v)du}
&=&-\sup_{\hat{v}_{i,k}\le v\le\hat {v}_{i,k+1}}
\int_{\hat{u}_{i,j}}^{\hat{u}_{j+1}}
{\nu r^{-1}(u,v)du}\\
&\le& V_{ijk}(r)
\end{eqnarray*}
and
\begin{eqnarray*}
\sup_{\hat{u}_{i,j}\le u\le\hat{u}_{i,j+1}}
\int_{\hat{v}_{i,k}}^{\hat{v}_{i,k_1}}{|\lambda|r^{-1}(u,v)dv}&=&
-\sup_{\hat{u}_{i,j}\le u\le\hat{u}_{i,j+1}}
\int_{\hat{v}_{i,k}}^{\hat{v}_{i,k+1}}
{\lambda r^{-1}(u,v)dv}\\
&\le& V_{ijk}(r).
\end{eqnarray*}
But, by the monotonicity $\partial_v{r}\le0$, $\partial_u{r}<0$
in $\mathcal{T}\cup\mathcal{A}$, we have
\[
V_{ijk}(r)=\log r(\hat{u}_{i,j},\hat{v}_{i,k})-\log r(\hat{u}_{i,j+1},\tilde{v}_{i,k+1}).
\]
By construction of the rectangles it immediately follows that
\begin{eqnarray*}
\log r(\hat{u}_{i,j},\hat{v}_{i,k})-\log r(\hat{u}_{i,j+1},\hat{v}_{i,k+1})
&=&
\int_{\hat{u}_{i,j}}^{\hat{u}_{i,j+1}}{|\nu|r^{-1}(u,\hat{v}_{i,k+1}}du)\\
&&\hbox{}+
\int_{\hat{v}_{i,k}}^{\hat{v}_{i,k+1}}{|\lambda|r^{-1}(\hat{u}_{i,j+1},v)dv}\\
&<&\frac13\delta.
\end{eqnarray*}
Thus the assumptions of the previous proposition applies to these
rectangles.

As for those $\mathcal{X}_{ijk}\subset
\mathcal{G}(U_1)\setminus\mathcal{T}$,
we have similarly that
\begin{eqnarray*}
\sup_{\hat{v}_{i,k}\le v\le \hat{v}_{i,k+1}}
\int_{\hat{u}_{i,j}}^{\hat{u}_{i,j+1}}{|\nu|r^{-1}(u,v)du}
&=&-\sup_{\hat{v}_{i,k}\le v\le \hat{v}_{i,k+1}}
\int_{\hat{u}_{i,j}}^{\hat{u}_{i,j+1}}{\nu r^{-1}(u,v)du}\\
&\le& V_{ijk}(r),
\end{eqnarray*}
and
\begin{eqnarray*}
\sup_{\hat{u}_{i,j}\le u\le \hat{u}_{i,j+1}}
\int_{\hat{v}_{i,k}}^{\hat{v}_{i,k+1}}{|\lambda|r^{-1}(u,v)dv}
&=&\sup_{\hat{u}_{i,j}\le u\le \hat{u}_{i,j+1}}
\int_{\hat{v}_{i,k}}^{\hat{v}_{i,k+1}}{\lambda r^{-1}(u,v)dv}\\
&\le& V_{ijk}(r),
\end{eqnarray*}
while
\[
V_{ijk} (r)=\log r(\hat{u}_{i,j},\hat{v}_{i,k+1})
-\log r(\hat{u}_{i,j+1},\hat{u}_{i,k}).
\]
Now again by construction of $\mathcal{X}_{ijk}$, we have
\begin{eqnarray*}
\log r(\hat{u}_{i,j},\hat{v}_{i,k+1})-\log r(\hat{u}_{i,j+1},\hat{v}_{i,k})&=&
\int_{\hat{u}_{i,j}}^{\hat{u}_{i,j+1}}{|\nu|r^{-1}(u,\hat{v}_{i,k})du}\\
&&\hbox{}+
\int_{\hat{v}_{i,k}}^{\hat{v}_{i,k+1}}{|\lambda|r^{-1}(\hat{u}_{i,j},v)dv}\\
&<&\frac13\delta.
\end{eqnarray*}

Finally, for those $\mathcal{X}_{ijk}$ which are ``bisected'' by
$\mathcal{A}$,
\[
\sup_{\hat{v}_{i,k}\le v\le \hat{v}_{i,k+1}}
\int_{\hat{u}_{i,j}}^{\hat{u}_{i,j+1}}
{|\nu|r^{-1}(u,v)du}\le V_{ijk}(r)
\]
and
\[
\sup_{\hat{u}_{i,j}\le u\le \hat{u}_{i,j+1}}
\int_{\hat{v}_{i,k}}^{\hat{v}_{i,k+1}}
{|\lambda|r^{-1}(u,v)dv}\le 2V_{ijk}(r)
\]
as there can be at most one change of sign of $\lambda$.
On the other hand
\begin{eqnarray*}
V_{ijk} (r)
&=&
\log r(\hat{u}_{i,j},\hat{v}_{i,k+1})-
\log r (\hat{u}_{i,j+1},\hat{v}_{i,k+1})\\
&\le&
\int_{\hat{v}_{i,k}}^{\hat{v}_{i,k+1}}{(-\lambda)r^{-1}(\hat{u}_{i,j+1},v)dv}+
\int_{\hat{u}_{i,j}}^{\hat{u}_{i,j+1}}
{(-\nu)r^{-1}(u,\hat{v}_{i,k})du}\\
&&\hbox{}+
\int_{\hat{v}_{i,k}}^{\hat{v}_{i,k+1}}
{\lambda r^{-1}(\hat{u}_{i,j},v)dv}\\
&<&
\frac\delta2.
\end{eqnarray*}
Thus, all $\mathcal{X}_{ijk}$ 
satisfy the assumptions of Proposition~\ref{or9ogwvio}.
As the boundaries match up, by 
iterating the result of Proposition~\ref{or9ogwvio} at most
$N^3$ times, one obtains
the present proposition. \end{proof}

\appendix                               

\section{Two causal constructions}
Let $\tilde{\gamma}$ be a \emph{spacelike} curve in $\mathcal{K}(u_0)$ terminating
at $i^+$. Suppose $(u',v')\in\tilde{\gamma}$. 
Refer to the Penrose diagram below:
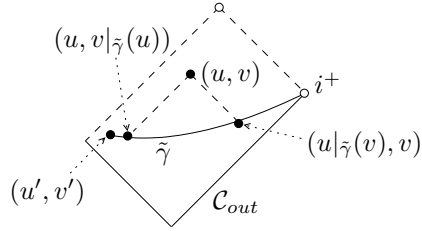
\[
\input{tomes.pstex_t}
\]
It follows that for
all $v\ge v'$, there is a unique $u$ such that $(u,v)\in\tilde{\gamma}$.
We will use the notation $u=u|_{\tilde{\gamma}(v)}$.
Similarly, for $u\le u'$, there is a unique $v$ such that
$(u,v)\in\tilde{\gamma}$, and we will write
$v=v|_{\tilde\gamma}(u)$. The latter can clearly be defined
even if $\tilde\gamma$ is assumed only to be \emph{achronal}.

We shall make use several times of the following construction:
Given two spacelike curves $\tilde\gamma$, $\tilde\gamma'$, terminating
at $i^+$, with
$\tilde\gamma'\subset I^+(\tilde\gamma)$, and a point $(u,v)$ 
such that $v|_{\tilde\gamma(u)}$, $v|_{\tilde\gamma'(u)}$, 
$u|_{\tilde\gamma(v)}$, and $v|_{\tilde\gamma'(u)}$ are defined,
the $(\tilde\gamma,\tilde\gamma',u,v)$-zigzag will be defined
as the union
\[
\bigcup_{i=1}^{I-1}\{u_i\}\times[v_i,v_{i+1}]\cup[u_{i+1},u_i]\times\{v_{i+1}\}
\]
where $u_1=\bar{u}$, $v_1=v|_{\tilde\gamma(\bar{u})}$,
and $u_i$, $v_i$ are defined inductively by
\[
v_{i+1}=\min\{v|_{\tilde\gamma'(\bar{u})},\bar{v}\}
\]
\[
u_{i+1}=u|_{\tilde\gamma(v_{i+1})},
\]
and $I$ is defined to be the first $i$ such that $v_i=\bar{v}$.
Refer to the Penrose diagram below:
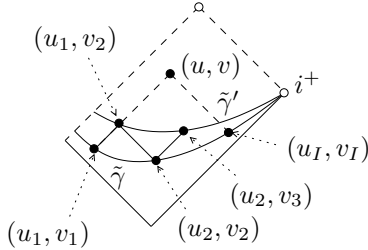
\[
\input{zigzag.pstex_t}
\]
By a compactness argument, it can be easily shown that $I<\infty$.

\vskip1pc
\noindent
{\bf Acknowledgement}
The author thanks Demetrios Christodoulou for his helpful comments
on a preliminary version of this paper, and for many interesting
discussions. This work is supported in part by
NSF grant  DMS-0302748.

\frenchspacing
\bibliographystyle{plain}

\end{document}

%% file: RN2.pstex_t
\begin{picture}(0,0)%
\includegraphics{RN2.pstex}%
\end{picture}%
\setlength{\unitlength}{3158sp}%
\begingroup\makeatletter\ifx\SetFigFont\undefined%
\gdef\SetFigFont#1#2#3#4#5{%
  \reset@font\fontsize{#1}{#2pt}%
  \fontfamily{#3}\fontseries{#4}\fontshape{#5}%
  \selectfont}%
\fi\endgroup%
\begin{picture}(2780,1913)(4161,-4919)
\put(6226,-3886){\makebox(0,0)[lb]{\smash{{\SetFigFont{10}{12.0}{\familydefault}{\mddefault}{\updefault}{\color[rgb]{0,0,0}$i^+$}%
}}}}
\put(4876,-4861){\makebox(0,0)[lb]{\smash{{\SetFigFont{10}{12.0}{\familydefault}{\mddefault}{\updefault}{\color[rgb]{0,0,0}$\mathcal{S}$}%
}}}}
\put(6032,-3162){\makebox(0,0)[lb]{\smash{{\SetFigFont{10}{12.0}{\familydefault}{\mddefault}{\updefault}{\color[rgb]{0,0,0}$\gamma$}%
}}}}
\put(6116,-4526){\makebox(0,0)[lb]{\smash{{\SetFigFont{10}{12.0}{\familydefault}{\mddefault}{\updefault}{\color[rgb]{0,0,0}$\mathcal{H}^+$}%
}}}}
\put(6517,-4209){\makebox(0,0)[lb]{\smash{{\SetFigFont{10}{12.0}{\familydefault}{\mddefault}{\updefault}{\color[rgb]{0,0,0}$\mathcal{I}^+$}%
}}}}
\put(5891,-3586){\makebox(0,0)[lb]{\smash{{\SetFigFont{10}{12.0}{\familydefault}{\mddefault}{\updefault}{\color[rgb]{0,0,0}$\mathcal{CH}^+$}%
}}}}
\end{picture}%

%% file: arxikoedw00.pstex_t
\begin{picture}(0,0)%
\includegraphics{arxikoedw00.pstex}%
\end{picture}%
\setlength{\unitlength}{3158sp}%
\begingroup\makeatletter\ifx\SetFigFont\undefined%
\gdef\SetFigFont#1#2#3#4#5{%
  \reset@font\fontsize{#1}{#2pt}%
  \fontfamily{#3}\fontseries{#4}\fontshape{#5}%
  \selectfont}%
\fi\endgroup%
\begin{picture}(1555,1339)(5101,-4844)
\put(6226,-3886){\makebox(0,0)[lb]{\smash{{\SetFigFont{10}{12.0}{\familydefault}{\mddefault}{\updefault}{\color[rgb]{0,0,0}$i^+$}%
}}}}
\put(5101,-4561){\makebox(0,0)[lb]{\smash{{\SetFigFont{10}{12.0}{\familydefault}{\mddefault}{\updefault}{\color[rgb]{0,0,0}$\mathcal{C}'_{in}$}%
}}}}
\put(5776,-4561){\makebox(0,0)[lb]{\smash{{\SetFigFont{10}{12.0}{\familydefault}{\mddefault}{\updefault}{\color[rgb]{0,0,0}$\mathcal{C}_{out}$}%
}}}}
\put(5401,-4786){\makebox(0,0)[lb]{\smash{{\SetFigFont{10}{12.0}{\familydefault}{\mddefault}{\updefault}{\color[rgb]{0,0,0}$p$}%
}}}}
\put(6001,-3661){\makebox(0,0)[lb]{\smash{{\SetFigFont{10}{12.0}{\rmdefault}{\mddefault}{\updefault}{\color[rgb]{0,0,0}$\mathcal{CH}^+$}%
}}}}
\put(5591,-4180){\makebox(0,0)[lb]{\smash{{\SetFigFont{10}{12.0}{\rmdefault}{\mddefault}{\updefault}{\color[rgb]{0,0,0}$\mathcal{G}$}%
}}}}
\end{picture}%

%% file: arxikoedw0.pstex_t
\begin{picture}(0,0)%
\includegraphics{arxikoedw0.pstex}%
\end{picture}%
\setlength{\unitlength}{3158sp}%
\begingroup\makeatletter\ifx\SetFigFont\undefined%
\gdef\SetFigFont#1#2#3#4#5{%
  \reset@font\fontsize{#1}{#2pt}%
  \fontfamily{#3}\fontseries{#4}\fontshape{#5}%
  \selectfont}%
\fi\endgroup%
\begin{picture}(1555,1495)(5101,-4844)
\put(6226,-3886){\makebox(0,0)[lb]{\smash{{\SetFigFont{10}{12.0}{\familydefault}{\mddefault}{\updefault}{\color[rgb]{0,0,0}$i^+$}%
}}}}
\put(5101,-4561){\makebox(0,0)[lb]{\smash{{\SetFigFont{10}{12.0}{\familydefault}{\mddefault}{\updefault}{\color[rgb]{0,0,0}$\mathcal{C}'_{in}$}%
}}}}
\put(5776,-4561){\makebox(0,0)[lb]{\smash{{\SetFigFont{10}{12.0}{\familydefault}{\mddefault}{\updefault}{\color[rgb]{0,0,0}$\mathcal{C}_{out}$}%
}}}}
\put(5401,-4786){\makebox(0,0)[lb]{\smash{{\SetFigFont{10}{12.0}{\familydefault}{\mddefault}{\updefault}{\color[rgb]{0,0,0}$p$}%
}}}}
\put(5626,-4186){\makebox(0,0)[lb]{\smash{{\SetFigFont{10}{12.0}{\rmdefault}{\mddefault}{\updefault}{\color[rgb]{0,0,0}$\mathcal{G}$}%
}}}}
\put(5946,-3661){\makebox(0,0)[lb]{\smash{{\SetFigFont{10}{12.0}{\rmdefault}{\mddefault}{\updefault}{\color[rgb]{0,0,0}$\mathcal{X}$}%
}}}}
\end{picture}%

%% file: arxikoedw.pstex_t
\begin{picture}(0,0)%
\includegraphics{arxikoedw.pstex}%
\end{picture}%
\setlength{\unitlength}{3158sp}%
\begingroup\makeatletter\ifx\SetFigFont\undefined%
\gdef\SetFigFont#1#2#3#4#5{%
  \reset@font\fontsize{#1}{#2pt}%
  \fontfamily{#3}\fontseries{#4}\fontshape{#5}%
  \selectfont}%
\fi\endgroup%
\begin{picture}(1840,2349)(5101,-5698)
\put(6608,-4271){\makebox(0,0)[lb]{\smash{\SetFigFont{10}{12.0}{\familydefault}{\mddefault}{\updefault}{\color[rgb]{0,0,0}$\mathcal{I}^+$}%
}}}
\put(6226,-3886){\makebox(0,0)[lb]{\smash{\SetFigFont{10}{12.0}{\familydefault}{\mddefault}{\updefault}{\color[rgb]{0,0,0}$i^+$}%
}}}
\put(5101,-4561){\makebox(0,0)[lb]{\smash{\SetFigFont{10}{12.0}{\familydefault}{\mddefault}{\updefault}{\color[rgb]{0,0,0}$\mathcal{C}'_{in}$}%
}}}
\put(6151,-4636){\makebox(0,0)[lb]{\smash{\SetFigFont{10}{12.0}{\familydefault}{\mddefault}{\updefault}{\color[rgb]{0,0,0}$\mathcal{H}^+$}%
}}}
\put(6301,-3586){\makebox(0,0)[lb]{\smash{\SetFigFont{10}{12.0}{\familydefault}{\mddefault}{\updefault}{\color[rgb]{0,0,0}$\mathcal{CH}^+$}%
}}}
\put(5476,-4820){\makebox(0,0)[lb]{\smash{\SetFigFont{10}{12.0}{\familydefault}{\mddefault}{\updefault}{\color[rgb]{0,0,0}$p$}%
}}}
\put(5926,-3661){\makebox(0,0)[lb]{\smash{\SetFigFont{10}{12.0}{\rmdefault}{\mddefault}{\updefault}{\color[rgb]{0,0,0}$\mathcal{X}$}%
}}}
\put(5926,-4861){\makebox(0,0)[lb]{\smash{\SetFigFont{10}{12.0}{\rmdefault}{\mddefault}{\updefault}{\color[rgb]{0,0,0}$\mathcal{D}$}%
}}}
\put(5626,-4186){\makebox(0,0)[lb]{\smash{\SetFigFont{10}{12.0}{\rmdefault}{\mddefault}{\updefault}{\color[rgb]{0,0,0}$\mathcal{G}$}%
}}}
\end{picture}

%% file: RNshiftveo.pstex_t
\begin{picture}(0,0)%
\includegraphics{RNshiftveo.pstex}%
\end{picture}%
\setlength{\unitlength}{3158sp}%
\begingroup\makeatletter\ifx\SetFigFont\undefined%
\gdef\SetFigFont#1#2#3#4#5{%
  \reset@font\fontsize{#1}{#2pt}%
  \fontfamily{#3}\fontseries{#4}\fontshape{#5}%
  \selectfont}%
\fi\endgroup%
\begin{picture}(2242,1964)(5164,-5886)
\put(5742,-5828){\makebox(0,0)[lb]{\smash{{\SetFigFont{10}{12.0}{\familydefault}{\mddefault}{\updefault}{\color[rgb]{0,0,0}$p$}%
}}}}
\put(6536,-4216){\makebox(0,0)[lb]{\smash{{\SetFigFont{10}{12.0}{\familydefault}{\mddefault}{\updefault}{\color[rgb]{0,0,0}$\mathcal{CH}^+$}%
}}}}
\put(6976,-4636){\makebox(0,0)[lb]{\smash{{\SetFigFont{10}{12.0}{\familydefault}{\mddefault}{\updefault}{\color[rgb]{0,0,0}$i^+$}%
}}}}
\put(5279,-5474){\makebox(0,0)[lb]{\smash{{\SetFigFont{10}{12.0}{\familydefault}{\mddefault}{\updefault}{\color[rgb]{0,0,0}$\mathcal{C}'_{in}$}%
}}}}
\put(6386,-5311){\makebox(0,0)[lb]{\smash{{\SetFigFont{10}{12.0}{\familydefault}{\mddefault}{\updefault}{\color[rgb]{0,0,0}$\mathcal{H}^+$}%
}}}}
\put(5729,-5337){\makebox(0,0)[lb]{\smash{{\SetFigFont{10}{12.0}{\familydefault}{\mddefault}{\updefault}{\color[rgb]{0,0,0}$\mathcal{R}$}%
}}}}
\put(6151,-4486){\makebox(0,0)[lb]{\smash{{\SetFigFont{10}{12.0}{\familydefault}{\mddefault}{\updefault}{\color[rgb]{0,0,0}$\mathcal{B}$}%
}}}}
\put(5587,-4896){\makebox(0,0)[lb]{\smash{{\SetFigFont{10}{12.0}{\familydefault}{\mddefault}{\updefault}{\color[rgb]{0,0,0}$\mathcal{N}$}%
}}}}
\end{picture}%

%% file: gevshiftveo.pstex_t
\begin{picture}(0,0)%
\includegraphics{gevshiftveo.pstex}%
\end{picture}%
\setlength{\unitlength}{3158sp}%
\begingroup\makeatletter\ifx\SetFigFont\undefined%
\gdef\SetFigFont#1#2#3#4#5{%
  \reset@font\fontsize{#1}{#2pt}%
  \fontfamily{#3}\fontseries{#4}\fontshape{#5}%
  \selectfont}%
\fi\endgroup%
\begin{picture}(2242,1964)(5164,-5886)
\put(5742,-5828){\makebox(0,0)[lb]{\smash{{\SetFigFont{10}{12.0}{\familydefault}{\mddefault}{\updefault}{\color[rgb]{0,0,0}$p$}%
}}}}
\put(6976,-4636){\makebox(0,0)[lb]{\smash{{\SetFigFont{10}{12.0}{\familydefault}{\mddefault}{\updefault}{\color[rgb]{0,0,0}$i^+$}%
}}}}
\put(6536,-4216){\makebox(0,0)[lb]{\smash{{\SetFigFont{10}{12.0}{\familydefault}{\mddefault}{\updefault}{\color[rgb]{0,0,0}$\mathcal{CH}^+$}%
}}}}
\put(5301,-5466){\makebox(0,0)[lb]{\smash{{\SetFigFont{10}{12.0}{\familydefault}{\mddefault}{\updefault}{\color[rgb]{0,0,0}$\mathcal{C}'_{in}$}%
}}}}
\put(5727,-5439){\makebox(0,0)[lb]{\smash{{\SetFigFont{10}{12.0}{\familydefault}{\mddefault}{\updefault}{\color[rgb]{0,0,0}$\mathcal{R}$}%
}}}}
\put(5716,-5122){\makebox(0,0)[lb]{\smash{{\SetFigFont{10}{12.0}{\familydefault}{\mddefault}{\updefault}{\color[rgb]{0,0,0}$\mathcal{R}$}%
}}}}
\put(5701,-4861){\makebox(0,0)[lb]{\smash{{\SetFigFont{10}{12.0}{\familydefault}{\mddefault}{\updefault}{\color[rgb]{0,0,0}$\mathcal{N}$}%
}}}}
\put(6134,-4265){\makebox(0,0)[lb]{\smash{{\SetFigFont{10}{12.0}{\familydefault}{\mddefault}{\updefault}{\color[rgb]{0,0,0}$\mathcal{N}$}%
}}}}
\put(5723,-4600){\makebox(0,0)[lb]{\smash{{\SetFigFont{10}{12.0}{\familydefault}{\mddefault}{\updefault}{\color[rgb]{0,0,0}$\mathcal{B}_\gamma$}%
}}}}
\put(5913,-4431){\makebox(0,0)[lb]{\smash{{\SetFigFont{10}{12.0}{\familydefault}{\mddefault}{\updefault}{\color[rgb]{0,0,0}$\mathcal{B}$}%
}}}}
\put(6859,-4378){\makebox(0,0)[lb]{\smash{{\SetFigFont{10}{12.0}{\familydefault}{\mddefault}{\updefault}{\color[rgb]{0,0,0}$\gamma$}%
}}}}
\put(6124,-5764){\makebox(0,0)[lb]{\smash{{\SetFigFont{10}{12.0}{\familydefault}{\mddefault}{\updefault}{\color[rgb]{0,0,0}$\mathcal{A}$}%
}}}}
\put(6557,-5446){\makebox(0,0)[lb]{\smash{{\SetFigFont{10}{12.0}{\familydefault}{\mddefault}{\updefault}{\color[rgb]{0,0,0}$\Gamma_E$}%
}}}}
\put(6830,-5085){\makebox(0,0)[lb]{\smash{{\SetFigFont{10}{12.0}{\familydefault}{\mddefault}{\updefault}{\color[rgb]{0,0,0}$\Gamma_{-\xi}$}%
}}}}
\end{picture}%

%% file: prop1veo.pstex_t
\begin{picture}(0,0)%
\includegraphics{prop1veo.pstex}%
\end{picture}%
\setlength{\unitlength}{3158sp}%
\begingroup\makeatletter\ifx\SetFigFont\undefined%
\gdef\SetFigFont#1#2#3#4#5{%
  \reset@font\fontsize{#1}{#2pt}%
  \fontfamily{#3}\fontseries{#4}\fontshape{#5}%
  \selectfont}%
\fi\endgroup%
\begin{picture}(2905,1799)(4501,-5721)
\put(6976,-4636){\makebox(0,0)[lb]{\smash{{\SetFigFont{10}{12.0}{\familydefault}{\mddefault}{\updefault}{\color[rgb]{0,0,0}$i^+$}%
}}}}
\put(6376,-5311){\makebox(0,0)[lb]{\smash{{\SetFigFont{10}{12.0}{\familydefault}{\mddefault}{\updefault}{\color[rgb]{0,0,0}$\mathcal{H}^+$}%
}}}}
\put(5776,-5386){\makebox(0,0)[lb]{\smash{{\SetFigFont{10}{12.0}{\familydefault}{\mddefault}{\updefault}{\color[rgb]{0,0,0}$\mathcal{R}$}%
}}}}
\put(5416,-5038){\makebox(0,0)[lb]{\smash{{\SetFigFont{10}{12.0}{\familydefault}{\mddefault}{\updefault}{\color[rgb]{0,0,0}$\mathcal{R}$}%
}}}}
\put(5101,-4636){\makebox(0,0)[lb]{\smash{{\SetFigFont{10}{12.0}{\familydefault}{\mddefault}{\updefault}{\color[rgb]{0,0,0}$\mathcal{A}$}%
}}}}
\put(6151,-4486){\makebox(0,0)[lb]{\smash{{\SetFigFont{10}{12.0}{\familydefault}{\mddefault}{\updefault}{\color[rgb]{0,0,0}$\Gamma_E$}%
}}}}
\put(4501,-5011){\makebox(0,0)[lb]{\smash{{\SetFigFont{10}{12.0}{\familydefault}{\mddefault}{\updefault}{\color[rgb]{0,0,0}$(V,U_1)$}%
}}}}
\end{picture}%

%% file: prop1bveo.pstex_t
\begin{picture}(0,0)%
\includegraphics{prop1bveo.pstex}%
\end{picture}%
\setlength{\unitlength}{3158sp}%
\begingroup\makeatletter\ifx\SetFigFont\undefined%
\gdef\SetFigFont#1#2#3#4#5{%
  \reset@font\fontsize{#1}{#2pt}%
  \fontfamily{#3}\fontseries{#4}\fontshape{#5}%
  \selectfont}%
\fi\endgroup%
\begin{picture}(2242,1799)(5164,-5721)
\put(6976,-4636){\makebox(0,0)[lb]{\smash{{\SetFigFont{10}{12.0}{\familydefault}{\mddefault}{\updefault}{\color[rgb]{0,0,0}$i^+$}%
}}}}
\put(6376,-5311){\makebox(0,0)[lb]{\smash{{\SetFigFont{10}{12.0}{\familydefault}{\mddefault}{\updefault}{\color[rgb]{0,0,0}$\mathcal{H}^+$}%
}}}}
\put(5776,-5386){\makebox(0,0)[lb]{\smash{{\SetFigFont{10}{12.0}{\familydefault}{\mddefault}{\updefault}{\color[rgb]{0,0,0}$\mathcal{R}$}%
}}}}
\put(6151,-4486){\makebox(0,0)[lb]{\smash{{\SetFigFont{10}{12.0}{\familydefault}{\mddefault}{\updefault}{\color[rgb]{0,0,0}$\Gamma_E$}%
}}}}
\put(6102,-5125){\makebox(0,0)[lb]{\smash{{\SetFigFont{10}{12.0}{\familydefault}{\mddefault}{\updefault}{\color[rgb]{0,0,0}$q$}%
}}}}
\end{picture}%

%% file: prop2veo.pstex_t
\begin{picture}(0,0)%
\includegraphics{prop2veo.pstex}%
\end{picture}%
\setlength{\unitlength}{3158sp}%
\begingroup\makeatletter\ifx\SetFigFont\undefined%
\gdef\SetFigFont#1#2#3#4#5{%
  \reset@font\fontsize{#1}{#2pt}%
  \fontfamily{#3}\fontseries{#4}\fontshape{#5}%
  \selectfont}%
\fi\endgroup%
\begin{picture}(3498,1799)(3908,-5721)
\put(6976,-4636){\makebox(0,0)[lb]{\smash{{\SetFigFont{10}{12.0}{\familydefault}{\mddefault}{\updefault}{\color[rgb]{0,0,0}$i^+$}%
}}}}
\put(6376,-5311){\makebox(0,0)[lb]{\smash{{\SetFigFont{10}{12.0}{\familydefault}{\mddefault}{\updefault}{\color[rgb]{0,0,0}$\mathcal{H}^+$}%
}}}}
\put(6015,-4347){\makebox(0,0)[lb]{\smash{{\SetFigFont{10}{12.0}{\familydefault}{\mddefault}{\updefault}{\color[rgb]{0,0,0}$(u,v)$}%
}}}}
\put(5756,-5486){\makebox(0,0)[lb]{\smash{{\SetFigFont{10}{12.0}{\familydefault}{\mddefault}{\updefault}{\color[rgb]{0,0,0}$\mathcal{R}$}%
}}}}
\put(3908,-5660){\makebox(0,0)[lb]{\smash{{\SetFigFont{10}{12.0}{\familydefault}{\mddefault}{\updefault}{\color[rgb]{0,0,0}$(u|_{\Gamma_E}(v),v|_{\Gamma_E}(u))$}%
}}}}
\put(5316,-4534){\makebox(0,0)[lb]{\smash{{\SetFigFont{10}{12.0}{\familydefault}{\mddefault}{\updefault}{\color[rgb]{0,0,0}$\Gamma_E$}%
}}}}
\put(6129,-4565){\makebox(0,0)[lb]{\smash{{\SetFigFont{10}{12.0}{\familydefault}{\mddefault}{\updefault}{\color[rgb]{0,0,0}$\Gamma_{-\xi}$}%
}}}}
\end{picture}%

%% file: prop3veo.pstex_t
\begin{picture}(0,0)%
\includegraphics{prop3veo.pstex}%
\end{picture}%
\setlength{\unitlength}{3158sp}%
\begingroup\makeatletter\ifx\SetFigFont\undefined%
\gdef\SetFigFont#1#2#3#4#5{%
  \reset@font\fontsize{#1}{#2pt}%
  \fontfamily{#3}\fontseries{#4}\fontshape{#5}%
  \selectfont}%
\fi\endgroup%
\begin{picture}(2242,1799)(5164,-5721)
\put(6976,-4636){\makebox(0,0)[lb]{\smash{{\SetFigFont{10}{12.0}{\familydefault}{\mddefault}{\updefault}{\color[rgb]{0,0,0}$i^+$}%
}}}}
\put(6376,-5311){\makebox(0,0)[lb]{\smash{{\SetFigFont{10}{12.0}{\familydefault}{\mddefault}{\updefault}{\color[rgb]{0,0,0}$\mathcal{H}^+$}%
}}}}
\put(6001,-4327){\makebox(0,0)[lb]{\smash{{\SetFigFont{10}{12.0}{\familydefault}{\mddefault}{\updefault}{\color[rgb]{0,0,0}$(u,v)$}%
}}}}
\put(5544,-5353){\makebox(0,0)[lb]{\smash{{\SetFigFont{10}{12.0}{\familydefault}{\mddefault}{\updefault}{\color[rgb]{0,0,0}$\mathcal{R}\cup\mathcal{N}$}%
}}}}
\put(5277,-4467){\makebox(0,0)[lb]{\smash{{\SetFigFont{10}{12.0}{\familydefault}{\mddefault}{\updefault}{\color[rgb]{0,0,0}$\Gamma_{-\xi}$}%
}}}}
\put(6267,-4570){\makebox(0,0)[lb]{\smash{{\SetFigFont{10}{12.0}{\familydefault}{\mddefault}{\updefault}{\color[rgb]{0,0,0}$\gamma$}%
}}}}
\end{picture}%

%% file: konta.pstex_t
\begin{picture}(0,0)%
\includegraphics{konta.pstex}%
\end{picture}%
\setlength{\unitlength}{3158sp}%
\begingroup\makeatletter\ifx\SetFigFont\undefined%
\gdef\SetFigFont#1#2#3#4#5{%
  \reset@font\fontsize{#1}{#2pt}%
  \fontfamily{#3}\fontseries{#4}\fontshape{#5}%
  \selectfont}%
\fi\endgroup%
\begin{picture}(1812,1964)(5164,-5886)
\put(5742,-5828){\makebox(0,0)[lb]{\smash{\SetFigFont{10}{12.0}{\familydefault}{\mddefault}{\updefault}{\color[rgb]{0,0,0}$p$}%
}}}
\put(6976,-4636){\makebox(0,0)[lb]{\smash{\SetFigFont{10}{12.0}{\familydefault}{\mddefault}{\updefault}{\color[rgb]{0,0,0}$i^+$}%
}}}
\put(6737,-5289){\makebox(0,0)[lb]{\smash{\SetFigFont{10}{12.0}{\familydefault}{\mddefault}{\updefault}{\color[rgb]{0,0,0}$(u_I,v_I)$}%
}}}
\put(5272,-5492){\makebox(0,0)[lb]{\smash{\SetFigFont{10}{12.0}{\familydefault}{\mddefault}{\updefault}{\color[rgb]{0,0,0}$\mathcal{C}'_{in}$}%
}}}
\put(5236,-4134){\makebox(0,0)[lb]{\smash{\SetFigFont{10}{12.0}{\familydefault}{\mddefault}{\updefault}{\color[rgb]{0,0,0}$(u,v)$}%
}}}
\put(5350,-5149){\makebox(0,0)[lb]{\smash{\SetFigFont{10}{12.0}{\familydefault}{\mddefault}{\updefault}{\color[rgb]{0,0,0}$(u_1,v_1)$}%
}}}
\put(5283,-4446){\makebox(0,0)[lb]{\smash{\SetFigFont{10}{12.0}{\familydefault}{\mddefault}{\updefault}{\color[rgb]{0,0,0}$\Gamma_{-\bar\epsilon}$}%
}}}
\put(6116,-4286){\makebox(0,0)[lb]{\smash{\SetFigFont{10}{12.0}{\familydefault}{\mddefault}{\updefault}{\color[rgb]{0,0,0}$\gamma$}%
}}}
\put(5808,-4554){\makebox(0,0)[lb]{\smash{\SetFigFont{10}{12.0}{\familydefault}{\mddefault}{\updefault}{\color[rgb]{0,0,0}$\Gamma_{-\xi}$}%
}}}
\put(6041,-5138){\makebox(0,0)[lb]{\smash{\SetFigFont{10}{12.0}{\familydefault}{\mddefault}{\updefault}{\color[rgb]{0,0,0}$\Gamma_{\bar\epsilon}$}%
}}}
\end{picture}

%% file: arxikoedw5.pstex_t
\begin{picture}(0,0)%
\includegraphics{arxikoedw5.pstex}%
\end{picture}%
\setlength{\unitlength}{3158sp}%
\begingroup\makeatletter\ifx\SetFigFont\undefined%
\gdef\SetFigFont#1#2#3#4#5{%
  \reset@font\fontsize{#1}{#2pt}%
  \fontfamily{#3}\fontseries{#4}\fontshape{#5}%
  \selectfont}%
\fi\endgroup%
\begin{picture}(1165,1495)(5101,-4844)
\put(6226,-3886){\makebox(0,0)[lb]{\smash{\SetFigFont{10}{12.0}{\familydefault}{\mddefault}{\updefault}{\color[rgb]{0,0,0}$i^+$}%
}}}
\put(5101,-4561){\makebox(0,0)[lb]{\smash{\SetFigFont{10}{12.0}{\familydefault}{\mddefault}{\updefault}{\color[rgb]{0,0,0}$\mathcal{C}'_{in}$}%
}}}
\put(5776,-4561){\makebox(0,0)[lb]{\smash{\SetFigFont{10}{12.0}{\familydefault}{\mddefault}{\updefault}{\color[rgb]{0,0,0}$\mathcal{C}_{out}$}%
}}}
\put(5401,-4786){\makebox(0,0)[lb]{\smash{\SetFigFont{10}{12.0}{\familydefault}{\mddefault}{\updefault}{\color[rgb]{0,0,0}$p$}%
}}}
\put(5946,-3661){\makebox(0,0)[lb]{\smash{\SetFigFont{10}{12.0}{\rmdefault}{\mddefault}{\updefault}{\color[rgb]{0,0,0}$\mathcal{X}$}%
}}}
\put(5412,-4193){\makebox(0,0)[lb]{\smash{\SetFigFont{10}{12.0}{\rmdefault}{\mddefault}{\updefault}{\color[rgb]{0,0,0}$\mathcal{G}(U_5)$}%
}}}
\end{picture}

%% file: step1veo.pstex_t
\begin{picture}(0,0)%
\includegraphics{step1veo.pstex}%
\end{picture}%
\setlength{\unitlength}{3158sp}%
\begingroup\makeatletter\ifx\SetFigFont\undefined%
\gdef\SetFigFont#1#2#3#4#5{%
  \reset@font\fontsize{#1}{#2pt}%
  \fontfamily{#3}\fontseries{#4}\fontshape{#5}%
  \selectfont}%
\fi\endgroup%
\begin{picture}(2375,1799)(5164,-5721)
\put(6536,-4216){\makebox(0,0)[lb]{\smash{{\SetFigFont{10}{12.0}{\familydefault}{\mddefault}{\updefault}{\color[rgb]{0,0,0}$\mathcal{CH}^+$}%
}}}}
\put(6976,-4636){\makebox(0,0)[lb]{\smash{{\SetFigFont{10}{12.0}{\familydefault}{\mddefault}{\updefault}{\color[rgb]{0,0,0}$i^+$}%
}}}}
\put(6360,-5365){\makebox(0,0)[lb]{\smash{{\SetFigFont{10}{12.0}{\familydefault}{\mddefault}{\updefault}{\color[rgb]{0,0,0}$\mathcal{H}^+$}%
}}}}
\put(5760,-4625){\makebox(0,0)[lb]{\smash{{\SetFigFont{10}{12.0}{\familydefault}{\mddefault}{\updefault}{\color[rgb]{0,0,0}$\gamma$}%
}}}}
\put(5294,-5117){\makebox(0,0)[lb]{\smash{{\SetFigFont{10}{12.0}{\familydefault}{\mddefault}{\updefault}{\color[rgb]{0,0,0}$(u_0,v_0)$}%
}}}}
\put(5206,-4163){\makebox(0,0)[lb]{\smash{{\SetFigFont{10}{12.0}{\familydefault}{\mddefault}{\updefault}{\color[rgb]{0,0,0}$(u_0,v_1)$}%
}}}}
\put(6601,-5086){\makebox(0,0)[lb]{\smash{{\SetFigFont{10}{12.0}{\familydefault}{\mddefault}{\updefault}{\color[rgb]{0,0,0}$(u_1,v_1)$}%
}}}}
\end{picture}%

%% file: konta2.pstex_t
\begin{picture}(0,0)%
\includegraphics{konta2.pstex}%
\end{picture}%
\setlength{\unitlength}{3158sp}%
\begingroup\makeatletter\ifx\SetFigFont\undefined%
\gdef\SetFigFont#1#2#3#4#5{%
  \reset@font\fontsize{#1}{#2pt}%
  \fontfamily{#3}\fontseries{#4}\fontshape{#5}%
  \selectfont}%
\fi\endgroup%
\begin{picture}(1812,1981)(5164,-5886)
\put(5742,-5828){\makebox(0,0)[lb]{\smash{\SetFigFont{10}{12.0}{\familydefault}{\mddefault}{\updefault}{\color[rgb]{0,0,0}$p$}%
}}}
\put(6976,-4636){\makebox(0,0)[lb]{\smash{\SetFigFont{10}{12.0}{\familydefault}{\mddefault}{\updefault}{\color[rgb]{0,0,0}$i^+$}%
}}}
\put(5272,-5492){\makebox(0,0)[lb]{\smash{\SetFigFont{10}{12.0}{\familydefault}{\mddefault}{\updefault}{\color[rgb]{0,0,0}$\mathcal{C}'_{in}$}%
}}}
\put(5949,-4393){\makebox(0,0)[lb]{\smash{\SetFigFont{10}{12.0}{\familydefault}{\mddefault}{\updefault}{\color[rgb]{0,0,0}$(u,v)$}%
}}}
\put(6849,-4061){\makebox(0,0)[lb]{\smash{\SetFigFont{10}{12.0}{\familydefault}{\mddefault}{\updefault}{\color[rgb]{0,0,0}$(u|_{\gamma}(v),v)$}%
}}}
\put(5283,-4446){\makebox(0,0)[lb]{\smash{\SetFigFont{10}{12.0}{\familydefault}{\mddefault}{\updefault}{\color[rgb]{0,0,0}$\Gamma_{-\bar\epsilon}$}%
}}}
\put(5996,-5141){\makebox(0,0)[lb]{\smash{\SetFigFont{10}{12.0}{\familydefault}{\mddefault}{\updefault}{\color[rgb]{0,0,0}$\Gamma_{\bar\epsilon}$}%
}}}
\put(6509,-4627){\makebox(0,0)[lb]{\smash{\SetFigFont{10}{12.0}{\familydefault}{\mddefault}{\updefault}{\color[rgb]{0,0,0}$\gamma$}%
}}}
\end{picture}

%% file: tomes.pstex_t
\begin{picture}(0,0)%
\includegraphics{tomes.pstex}%
\end{picture}%
\setlength{\unitlength}{3158sp}%
\begingroup\makeatletter\ifx\SetFigFont\undefined%
\gdef\SetFigFont#1#2#3#4#5{%
  \reset@font\fontsize{#1}{#2pt}%
  \fontfamily{#3}\fontseries{#4}\fontshape{#5}%
  \selectfont}%
\fi\endgroup%
\begin{picture}(2396,1776)(4580,-5698)
\put(6976,-4636){\makebox(0,0)[lb]{\smash{\SetFigFont{10}{12.0}{\familydefault}{\mddefault}{\updefault}{\color[rgb]{0,0,0}$i^+$}%
}}}
\put(4927,-4271){\makebox(0,0)[lb]{\smash{\SetFigFont{10}{12.0}{\familydefault}{\mddefault}{\updefault}{\color[rgb]{0,0,0}$(u,v|_{\tilde\gamma}(u))$}%
}}}
\put(6913,-5086){\makebox(0,0)[lb]{\smash{\SetFigFont{10}{12.0}{\familydefault}{\mddefault}{\updefault}{\color[rgb]{0,0,0}$(u|_{\tilde\gamma}(v),v)$}%
}}}
\put(6180,-5544){\makebox(0,0)[lb]{\smash{\SetFigFont{10}{12.0}{\familydefault}{\mddefault}{\updefault}{\color[rgb]{0,0,0}$\mathcal{C}_{out}$}%
}}}
\put(5731,-5159){\makebox(0,0)[lb]{\smash{\SetFigFont{10}{12.0}{\familydefault}{\mddefault}{\updefault}{\color[rgb]{0,0,0}$\tilde\gamma$}%
}}}
\put(4580,-5466){\makebox(0,0)[lb]{\smash{\SetFigFont{10}{12.0}{\familydefault}{\mddefault}{\updefault}{\color[rgb]{0,0,0}$(u',v')$}%
}}}
\put(6079,-4538){\makebox(0,0)[lb]{\smash{\SetFigFont{10}{12.0}{\familydefault}{\mddefault}{\updefault}{\color[rgb]{0,0,0}$(u,v)$}%
}}}
\end{picture}

%% file: zigzag.pstex_t
\begin{picture}(0,0)%
\includegraphics{zigzag.pstex}%
\end{picture}%
\setlength{\unitlength}{3158sp}%
\begingroup\makeatletter\ifx\SetFigFont\undefined%
\gdef\SetFigFont#1#2#3#4#5{%
  \reset@font\fontsize{#1}{#2pt}%
  \fontfamily{#3}\fontseries{#4}\fontshape{#5}%
  \selectfont}%
\fi\endgroup%
\begin{picture}(3076,1934)(4713,-5856)
\put(6976,-4636){\makebox(0,0)[lb]{\smash{{\SetFigFont{10}{12.0}{\familydefault}{\mddefault}{\updefault}{\color[rgb]{0,0,0}$i^+$}%
}}}}
\put(6053,-5758){\makebox(0,0)[lb]{\smash{{\SetFigFont{10}{12.0}{\familydefault}{\mddefault}{\updefault}{\color[rgb]{0,0,0}$(u_2,v_2)$}%
}}}}
\put(4927,-4271){\makebox(0,0)[lb]{\smash{{\SetFigFont{10}{12.0}{\familydefault}{\mddefault}{\updefault}{\color[rgb]{0,0,0}$(u_1,v_2)$}%
}}}}
\put(6467,-5498){\makebox(0,0)[lb]{\smash{{\SetFigFont{10}{12.0}{\familydefault}{\mddefault}{\updefault}{\color[rgb]{0,0,0}$(u_2,v_3)$}%
}}}}
\put(5534,-5349){\makebox(0,0)[lb]{\smash{{\SetFigFont{10}{12.0}{\familydefault}{\mddefault}{\updefault}{\color[rgb]{0,0,0}$\tilde\gamma$}%
}}}}
\put(6371,-4772){\makebox(0,0)[lb]{\smash{{\SetFigFont{10}{12.0}{\familydefault}{\mddefault}{\updefault}{\color[rgb]{0,0,0}$\tilde\gamma'$}%
}}}}
\put(4713,-5798){\makebox(0,0)[lb]{\smash{{\SetFigFont{10}{12.0}{\familydefault}{\mddefault}{\updefault}{\color[rgb]{0,0,0}$(u_1,v_1)$}%
}}}}
\put(6913,-5132){\makebox(0,0)[lb]{\smash{{\SetFigFont{10}{12.0}{\familydefault}{\mddefault}{\updefault}{\color[rgb]{0,0,0}$(u_I,v_I)$}%
}}}}
\put(6079,-4478){\makebox(0,0)[lb]{\smash{{\SetFigFont{10}{12.0}{\familydefault}{\mddefault}{\updefault}{\color[rgb]{0,0,0}$(u,v)$}%
}}}}
\end{picture}%

%% file: interior.bbl
\begin{thebibliography}{99}
\bibitem{lb:sbh} Leor Barack \emph{Late time dynamics of scalar perturbations
outside black holes. II. Schwarzschild geometry} Phys. Rev. D. {\bf 59}
(1999)

\bibitem{bo:lt} Leor Barack and Amos Ori \emph{Late-time decay
of scalar perturbations outside rotating black holes}
Phys. Rev. Lett {\bf 82} (1999), 4388-4391

\bibitem{jb:gcc} Jiri Bicak \emph{Gravitational collapse
with charge and small asymmetries I. Scalar perturbations}
General Relativity and Gravitation, {\bf 3} (1972), no. 4, 331--349

\bibitem{bdim:scsbhi} A. Bonanno, S. Droz, W. Israel and
S. M. Morsink \emph{Structure of the charged black hole
interior} Proc. Roy. Soc. London Ser. A {\bf 450} 
(1995), no. 1940, 553--567

\bibitem{brsm:bhs} Patrick Brady and John Smith
\emph{Black hole singularities: a numerical approach}
Phys. Rev. Lett. {\bf 75} (1995), no. 7, 1256--1259

\bibitem{lb:sb} Lior Burko \emph{Structure of
the black hole's Cauchy-horizon singularity}
Phys. Rev. Lett. {\bf 79} (1997), 4958--4961

\bibitem{lb:nc} Lior Burko \emph{Black hole singularities:
a new critical phenomenon} gr-qc/0209084

\bibitem{bo:lte} Lior Burko and Amos Ori
\emph{Late-time evolution of non-linear gravitational collapse}
Phys. Rev. D {\bf 56} (1997), 7828--7832

\bibitem{ch:cchRN} S. Chandrasekhar and J. B. Hartle
\emph{On crossing the Cauchy horizon of a Reissner-Nordstr\"om black-hole}
Proc. Roy. Soc. London Ser. A {\bf 384} (1982), no. 1787, 301--315

\bibitem{chge:givp} Yvonne Choquet-Bruhat and Robert Geroch
\emph{Global aspects of the Cauchy problem in general relativity}
Comm. Math. Phys. {\bf 14} 1969, 329--335

\bibitem{chr:ins} Demetrios Christodoulou \emph{The instability
of naked singularities in the gravitational collapse of
a scalar field} Ann. of Math. {\bf 149} (1999), no 1,
183--217

\bibitem{chr:mt} Demetrios Christodoulou \emph{A mathematical
theory of gravitational collapse} Comm. Math. Phys.
{\bf 109} (1987), no. 4, 613-647

\bibitem{chr:givp} Demetrios Christodoulou \emph{On the global
initial value problem and the issue of singularities}
Classical Quantum Gravity {\bf 16} (1999), no. 12A, A23--A35

\bibitem{chr:sgrf} Demetrios Christodoulou \emph {Self-gravitating
relativistic fluids: a two-phase model} Arch. Rational Mech. Anal. 
{\bf 130} (1995), no. 4, 343--400

\bibitem{chr:bv} Demetrios Christodoulou \emph{Bounded variation 
solutions of the spherically symmetric Einstein-scalar field equations}
Comm. Pure Appl. Math {\bf 46} (1992), no. 8, 1131--1220

\bibitem{chr:fbh} Demetrios Christodoulou \emph{The formation
of black holes and singularities in spherically symmetric gravitational
collapse} Comm. Pure Appl. Math. {\bf 44} (1991), no. 3, 339--373

\bibitem{chr:sgsf} Demetrios Christodoulou \emph{The problem
of a self-gravitating scalar field} Comm. Math. Phys. {\bf 105}
(1986), no. 3, 337--361

\bibitem{ck:sms} D. Christodoulou and S. Klainerman \emph{The 
global nonlinear stability of the Minkowski space}
Princeton University Press, Princeton, 1993

\bibitem{chru:uitl} Piotr Chru\'sciel \emph{On the uniqueness in
the large of solutions of the Einstein's equations (``strong
cosmic censorship'')} Australian National
University, Centre for Mathematics and its Applications,
Canberra, 1991

\bibitem{cla:asts} C. J. S. Clarke \emph{The analysis of
space-time singularities} Cambridge University Press,
Cambridge, 1993.

\bibitem{md:sssts} Mihalis Dafermos \emph{Spherically
symmetric spacetimes with a trapped surface} gr-qc/0403032, 
preprint, 2004


\bibitem{md:si} Mihalis Dafermos \emph{Stability and Instability
of the Cauchy horizon for the spherically-symmetric Einstein-Maxwell-Scalar
Field equations} Ann. of Math. {\bf 158} (2003), no 3,
875--928

\bibitem{md:bb} Mihalis Dafermos \emph{Stability and Instability
of the Reissner-Nordstr\"om Cauchy Horizon and the Problem of Uniqueness
in General relativity} gr-qc/0209052, Proceedings
of the Conference on Non-compact Variational Problems and
General Relativity in honor of Haim Brezis and Felix Browder
Contemp. Math. {\bf 350} (1994), 99--113

\bibitem{mi:mazi} Mihalis Dafermos and Igor Rodnianski
\emph{A proof of Price's law for the collapse
of a self-gravitating scalar field} gr-qc/0309115, preprint, 2003

\bibitem{rigid} Helmut Friedrich, Istv\'an R\'acz, and
Robert M. Wald \emph{On the rigidity theorem for spacetimes
with a stationary event horizon or a compact cauchy horizon}
Comm. Math. Phys. {\bf 204} (1999) 691--707

\bibitem{gpp:de1} C. Gundlach, R. H. Price, and J. Pullin
\emph{Late-time behavior of stellar collapse and explosions. I.
Linearized perturbations} Phys. Rev. D {\bf 49} (1994), 883--889

\bibitem{gpp:de} C. Gundlach, R. H. Price, and J. Pullin
\emph{Late-time behavior of stellar collapse and explosions. II.
Nonlinear evolution} Phys. Rev. D {\bf 49} (1994), 890--899

\bibitem{he:lssst} S. W. Hawking and G. F. R. Ellis
\emph{The large scale structure of space-time} Cambridge
Monographs on Mathematical Physics, No. 1. Cambridge
University Press, London-New York, 1973

\bibitem{kr:rs} S. Klainerman and I. Rodnianski
\emph{Rough solutions for the Einstein vacuum equations}
preprint, 2001

\bibitem{st:pc} Matei Machedon and John Stalker, 
\emph{Decay of solutions to the wave equation on a spherically symmetric background},
preprint


\bibitem{mc:bhsf} R. L. Marsa and M. W. Choptuik
\emph{Black-hole--scalar-field interactions in spherical symmetry}
Phys. Rev. D {\bf 54} (1996), 4929--4943

\bibitem{ao:pa} Amos Ori \emph{Perturbative approach to
the inner structure of a rotating black hole}
Gen. Relativity Gravitation {\bf 29} (1997), 881-929

\bibitem{rp:gcsts} Roger Penrose \emph{Gravitational collapse
and space-time singularities} Phys. Rev. Lett. {\bf 14} (1965), 57--59

\bibitem{ispo:isbh} Eric Poisson and Werner Israel
\emph{Internal structure of black holes}
Phys. Rev. D (3) {\bf 41} (1990), no. 6, 1796--1809

\bibitem{rpr:ns} Richard Price \emph{Nonspherical perturbations
of relativistic gravitational collapse. I. Scalar and gravitational
perturbations} Phys. Rev. D (3) {\bf 5} (1972), 2419-2438

\bibitem{sp:RN} Michael Simpson and Roger Penrose \emph{Internal
instability in a Reissner-Nordstr\"om Black Hole} Int. Journ.
Theor. Phys. {\bf 7} (1973), no. 3, 183--197

\bibitem{yau:ps} S. T. Yau \emph{Problem Section} in Seminar on
Differential Geometry, edited by S. T. Yau, Annals of
Math. Studies, Princeton, N. J., 1982
\end{thebibliography}
